# From Multi-site Kaplan–Meier Curves to Evidence Synthesis: An Empirical Evaluation of IPD Reconstruction and Reconstruction-Assisted Meta-Analysis

Hui Yao[1], Nazmul Islam[2], Yicheng Shen[1], Lang Lang[3], Yanxun Xu[3], and Jiayi Tong[1,*]

[1] Department of Biostatistics, Johns Hopkins Bloomberg School of Public Health, Baltimore, MD, USA
[2] RefinedScience, Aurora, CO, USA
[3] Department of Applied Mathematics and Statistics, Johns Hopkins University, Baltimore, MD, USA
[*] Corresponding author: Jiayi Tong, jtong20@jh.edu

## Abstract

**Objective.** To compare methods for reconstructing individual participant data (IPD) from Kaplan–Meier (KM) curves and evaluate their use when reported trial estimates are unavailable for meta-analysis.

**Materials and Methods.** We compared iterative inverse-KM (iKM), IPDfromKM, KMtoIPD, and quadratic programming (QP) using their specified packages and auxiliary inputs. The single-trial benchmark evaluated curve agreement, event and risk-set accuracy, and treatment-effect accuracy across different survival settings. The meta-analysis benchmark compared omission of trials with unavailable estimates against four hybrid strategies that retained reported estimates and reconstructed missing estimates under missing-completely-at-random reporting.

**Results.** Relative method performance varied by setting and accuracy measure. Accurate reproduction of curves or supplied risk tables did not consistently identify the smallest treatment-effect errors. In 30-trial meta-analyses with 80% of reported hazard ratios unavailable, hybrids retained 27.9–30.0 trials, compared with 6.0 for omission, and achieved pooled log-hazard-ratio root-mean-squared error of 0.0356–0.0371 versus 0.0852.

**Discussion.** The comparisons reflect method-specific inputs, including accurately supplied auxiliary information, rather than algorithms receiving identical information. Controlled coordinate perturbation did not reproduce full image-digitization error.

**Conclusion.** Reconstruction should be selected and evaluated for its intended analytical use. Under the tested conditions, hybrid strategies supported meta-analysis by supplementing unavailable compatible trial estimates.

## 1 BACKGROUND AND SIGNIFICANCE

Kaplan–Meier (KM) curves are widely used to present time-to-event outcomes in clinical research, particularly in clinical trials [1]. Trial reports may also provide hazard ratios (HRs), numbers at risk, and total event counts, although reporting completeness varies across publications [2,3]. Even though these summaries may be sufficient for the primary question addressed in the original report, they may not contain the information required for alternative estimands or subsequent secondary analyses. At the single-trial level, incomplete reporting may prevent investigators from estimating alternative treatment-effect measures, such as an unreported restricted mean survival time (RMST) contrast. This contrast expresses the between-group difference in average survival time over a specified follow-up period, providing a time-based interpretation of the treatment effect [4]. At the evidence-synthesis level, the absence of such study-level effect estimates or measures of uncertainty may lead to the exclusion of otherwise eligible studies from meta-analysis, reducing precision and potentially introducing selection bias [3,5,6]. Original individual participant data (IPD) contain participant-level observations, including treatment group, follow-up time, and event status. Access to these data could address many of these limitations [7]. However, privacy safeguards and data-sharing agreements can limit access [8]. Consequently, limited access to IPD can constrain both trial-level secondary analysis and comprehensive synthesis of time-to-event evidence.

These practical constraints have motivated a range of methods for reconstructing pseudo-IPD from published KM curves [9]. Guyot et al. introduced iterative inverse-KM reconstruction (iKM), which uses digitized KM coordinates and reported numbers at risk; total event counts provide an additional constraint when available [10]. Building on this, Liu et al. developed IPDfromKM, a workflow that integrates coordinate extraction, preprocessing, reconstruction, accuracy assessment, and downstream survival analysis [11]. Other methods use different reconstruction strategies. Rogula et al. developed KMtoIPD to incorporate marked censoring times directly [12], whereas Titman formulated reconstruction as a constrained quadratic-programming (QP) problem that can combine multiple forms of auxiliary information [13]. More recent frameworks extend beyond numerical reconstruction: KM-GPT combines image preprocessing, multimodal artificial intelligence, and iterative reconstruction [14], while RESOLVE-IPD combines extraction of KM coordinates and censoring marks with uncertainty-aware subgroup recovery and meta-analysis [15]. Together, these advances have expanded the range of methods that differ in their mathematical formulation, required inputs, censoring assumptions, automation, and analytical scope.

We conceptualize the pathway from a published KM figure to meta-analysis as an error-propagation pipeline (Figure 1). At the digitization stage, extracting curve coordinates and censoring marks can introduce numerical errors [10–12,16]. At reconstruction, methods infer event and censoring records using different constraints and allocation assumptions [10–13]. At trial-level analysis, discrepancies in these records may affect Cox log-hazard ratios (logHRs), RMST contrasts, and standard errors [13,17]. At meta-analysis, reconstruction failure can leave a trial without an estimate, while errors in estimates or standard errors can alter study weights and pooled results [3,5]. Thus, this sequence motivates evaluating reconstruction together with its consequences for subsequent analyses.

Previous evaluations have addressed several important questions about KM reconstruction. Wan et al. compared least-squares and graphical approaches with the Hoyle–Henley and Guyot methods for estimating mean survival across sample sizes, censoring levels, and survival distributions [18]. Saluja et al. evaluated the Guyot, Williamson, Parmar, and Hoyle–Henley methods for estimating HRs from published oncology trial curves [17]. Meanwhile, researchers have already used reconstructed IPD in survival meta-analyses: Amacher et al. pooled reconstructed survival data to examine long-term outcomes after out-of-hospital cardiac arrest [19], and Fu et al. developed a Bayesian framework that combines reconstructed survival information with other aggregate summaries [20]. However, these evaluations and applications provide limited evidence on how method- and scenario-dependent reconstruction errors affect trial-level estimates and subsequent meta-analysis. A coordinated benchmark is therefore needed to compare reconstruction accuracy across analytical uses and assess how combining reconstructed and reported estimates affects trial inclusion, pooled effects, and uncertainty.

We therefore conducted a two-part simulation benchmark. Part I compared iKM, IPDfromKM, KMtoIPD, and QP at the single-trial level across survival patterns, censoring mechanisms, coordinate error, sample size, and arm allocation. Part II compared strategies for combining reported and reconstruction-derived Cox estimates in random-effects meta-analysis when reported estimates were missing completely at random. Together, the two benchmarks examine which aspects of reconstruction performance matter for trial-level analysis and whether reconstruction can supplement otherwise unavailable evidence in meta-analysis.

## 2 MATERIALS AND METHODS

### 2.1 Overall design and error-propagation framework

We evaluated single-trial reconstruction and reconstruction-assisted meta-analysis within the framework in Figure 1. Table 1 summarizes the simulation settings for both benchmarks. Part I compared iKM, IPDfromKM, KMtoIPD, and QP [10–13]. The comparisons covered curve agreement, event and risk-set accuracy, and treatment-effect accuracy across simulation settings. Part II compared strategies for combining reported and reconstructed trial estimates.

For Part I, we generated IPD under survival and censoring mechanisms representing a range of survival and follow-up patterns. These reference IPD were converted into method-specific reconstruction inputs representing information that may be obtained from a published KM report, including KM coordinates, numbers at risk, total event counts, and marked censoring times. Because the reconstruction methods rely on different auxiliary information, each method was evaluated as a complete workflow using its prespecified input configuration. We then compared the reconstructed datasets with the simulated reference IPD using the measures described in Section 2.3.2.

For Part II, we generated collections of trials with varying treatment effects and randomly withheld their reported logHRs and standard errors. We compared omission of trials with unavailable estimates against strategies that recovered those estimates from reconstructed IPD. We assessed pooled-effect error and confidence-interval coverage against the mean logHR used to generate the trials, and examined trial retention and analysis success as explanatory outcomes. We modeled one component of manual or automated digitization error by perturbing KM survival probabilities while retaining their time coordinates. We did not simulate image resolution, overlapping curves, incorrect arm labels, or extraction errors in risk tables and censor marks. Sections 2.3 and 2.4 describe the two benchmarks.

### 2.2 Reconstruction methods

Table 2 summarizes the principles and inputs of iKM [10], IPDfromKM [11], KMtoIPD [12], and QP [13]. Each method reconstructed the two trial arms separately, after which we combined the reconstructed arms for analysis. The descriptions below specify the software versions and input configurations used in this benchmark.

1. Iterative inverse-KM (iKM) [10]. This method iteratively inverts the KM product-limit calculation to infer event and censoring records from curve coordinates and numbers at risk. Preprocessing enforced nonincreasing survival probabilities, restricted coordinates to the risk-table horizon, and added the step-function survival value at each risk-table time when needed. The resulting coordinates and risk table served as inputs to the method. The iKM implementation used reconstructKM::KM_reconstruct() (version 0.4.0) [21].

2. IPDfromKM [11]. This method reconstructs interval-specific event and censoring counts from KM coordinates and risk-table information and can incorporate a reported total event count. The implementation used preprocess() and getIPD() from IPDfromKM (version 0.1.10) [22]. Inputs comprised KM coordinates, the risk table, the initial sample size, and the observed total number of events in each simulated arm.
3. KMtoIPD [12]. This method uses KM decreases and marked censoring times to reconstruct participant records. We used getIPD() from KMtoIPD (version 1.0.0), supplying initial sample size, nonincreasing post-drop coordinates, and observed censoring times without extraction error, but no follow-up risk table or event total.
4. Quadratic-programming approach (QP) [13]. This method estimates event and censoring counts by constrained quadratic programming. We used CIFresolve::KM_resolve() and CIFresolve::make_data() (version 0.1.1), supplying nonincreasing post-drop coordinates, the risk table, and observed censor marks without extraction error, but no event total. The primary configuration enforced supplied censor marks and used the package's approximate optimization and integer-rounding procedure.

## 2.3 Part I: single-trial reconstruction benchmark

### *2.3.1 Design and data-generation mechanisms*

Part I varied four features of time-to-event studies: survival and treatment-effect patterns; censoring amount, timing, and informativeness; participant numbers and allocation; and accuracy of extracted survival probabilities. The main benchmark combined factors, whereas one-factor experiments changed one feature while holding others fixed (Supplementary Methods S2) [23]. Table 1, Panel A defines four scenario categories: Weibull and piecewise-exponential settings, each with either 30% censoring and survival-probability error standard deviation (SD) 0.01 or 60% censoring and SD 0.04. The latter combination represents less complete follow-up and less precise digitization; the 60%-censoring Weibull settings additionally use smaller, unequal arms. Each category contains multiple survival and censoring scenarios. Within each survival-model family, these paired categories allowed us to examine how reconstruction performance changed when greater censoring and larger digitization errors occurred together, with smaller and unequal arms also contributing to the comparison in the Weibull settings. The complementary one-factor experiments helped distinguish the effects of individual features from the combined differences observed in the main benchmark.

Weibull models generated proportional hazards with decreasing, constant, or increasing baseline risk, representing risk that attenuates, remains constant, or accumulates during follow-up. Piecewise-exponential models allowed delayed treatment effects and crossing curves, where relative prognosis changes over time. Crossing curves have been reported between prognostic subgroups of acute myeloid leukemia patients treated with venetoclax and azacitidine [24]. These models represented general survival patterns, not particular diseases.

Increasing censoring represented less complete follow-up. Uniform censoring distributed observation endings throughout follow-up; truncated-exponential and front-loaded censoring concentrated them earlier; back-loaded censoring concentrated them later (Figure 2). These patterns resemble distributed losses, early withdrawal, and late administrative cutoff. Informative censoring linked censoring time to latent event time; censoring at a prognosis-related postbaseline intervention provides a clinical analogy [25,26]. Supplementary Methods S1 gives the mathematical definitions.

Smaller or imbalanced arms represented less or unequal information. To model imprecision when reading vertical coordinates during manual or automated digitization, we added mean-zero Gaussian error to KM survival probabilities. Its SD, termed survival-probability error SD, quantified the imposed imprecision; larger values indicated noisier inputs.

Each participant's earlier event or censoring time determined observed follow-up and event status. These observations generated KM-drop times, risk tables, and auxiliary inputs; only survival probabilities were perturbed. Exact inputs therefore meant quantities derived from observed simulated data without extraction error.

#### *2.3.2 Performance measures*

The evaluation covered three aspects of reconstruction because users may need to reproduce a survival curve, recover event counts and numbers at risk, or estimate treatment effects. We assessed each aspect separately because accuracy for one use may not imply accuracy for another.

KM root-mean-squared error (RMSE) measured curve agreement by comparing reconstructed and reference survival probabilities at 200 equally spaced times over the evaluation period. Smaller RMSE values indicated closer agreement between the curves.

Relative absolute event-count deviation expressed the difference in total events as a percentage of the reference count. Maximum absolute number-at-risk deviation captured the largest discrepancy at scheduled risk-table times. Smaller values indicated closer recovery. Agreement with quantities supplied as inputs demonstrated reproduction of those constraints rather than independent recovery.

Treatment-effect accuracy was assessed using absolute errors in the unadjusted Cox logHR and the treatment-minus-control RMST contrast. Cox logHR error assesses recovery of a commonly reported relative treatment effect that also enters inverse-variance meta-analysis. RMST-contrast error assesses recovery of an absolute difference in survival time and remains interpretable under nonproportional hazards [4]. For both measures, smaller errors indicate closer agreement with estimates calculated from the simulated reference IPD.

For each method and setting, we calculated median errors across valid reconstructed datasets. Supplementary Methods S2.3 gives the equations and implementation details; Supplementary Methods S2 and S4 describe the one-factor experiments and auxiliary-input analyses.

### 2.4 Part II: reconstruction-assisted meta-analysis benchmark

#### *2.4.1 Meta-analysis strategies and estimand*

We used two-stage inverse-variance random-effects meta-analysis for all principal strategies, estimating between-study variance by restricted maximum likelihood. The estimand was the mean generating logHR. Aggregate data (AD) comprised a trial-level Cox logHR and standard error. We compared six principal strategies:

1. Full AD oracle. We used Cox logHRs and model-based standard errors calculated from the simulated reference IPD for every trial. This benchmark represents complete reporting of all required trial-level estimates.
2. AD only. We used available reported estimates and omitted trials whose estimates were unavailable.
3. Hybrid–iKM. We recovered unavailable trial estimates using iKM.
4. Hybrid–IPDfromKM. We recovered unavailable trial estimates using IPDfromKM.
5. Hybrid–KMtoIPD. We recovered unavailable trial estimates using KMtoIPD.
6. Hybrid–QP. We recovered unavailable trial estimates using QP.

Each hybrid retained available reported estimates and used reconstruction only when an estimate was unavailable. Four additional reconstruction-only strategies used reconstructed estimates for every trial. We also evaluated a pooled true-IPD comparator (Supplementary Methods S3.1). The reference IPD record observed follow-up times and event indicators, with event times after censoring unobserved, so even exact reconstruction cannot correct any bias due to informative censoring.

### *2.4.2 Trial generation, report availability, and reconstruction*

Part II varied trial count, participants per arm, censoring, between-study heterogeneity, coordinate accuracy, and availability of reported estimates to test whether reconstruction could retain otherwise omitted trials (Table 1, Panel B).

The 30-trial benchmark used 250 participants per arm, 30% censoring, and survival-probability error SD 0.01. The 10- and 15-trial benchmarks used 100 participants per arm, 60% censoring, and SD 0.04, representing less available information. We generated exponential event times under proportional hazards, allowing the logHR to vary across trials. We used the mean of the generating logHR distribution as the target for meta-analysis.

For each trial, we calculated a reference-IPD Cox estimate and standard error, then withheld them randomly and independently of trial outcomes. All strategies used the same trials and reporting-availability patterns. This design isolated missing-completely-at-random reporting from selective reporting.

For trials with unavailable reported estimates, we omitted a trial if reconstruction failed or the subsequent Cox analysis could not provide a valid logHR and standard error. For example, some QP runs failed because the optimization procedure could not find a solution satisfying the specified input constraints. Available reported estimates remained usable regardless of reconstruction failure. When a reported estimate was unavailable, the hybrid strategies used the logHR and model-based standard error from a Cox model fitted to the reconstructed pseudo-IPD. These standard errors did not account for uncertainty in the reconstruction (Supplementary Methods S3.1–S3.2).

### *2.4.3 Meta-analysis performance measures*

We used two primary measures. Pooled-logHR RMSE measured error relative to the generating mean treatment effect; smaller values indicated greater accuracy. Confidence-interval coverage was the proportion of simulated meta-analyses whose 95% confidence interval contained that mean effect. We counted failed analyses as noncoverage so that coverage reflected both interval performance and the ability to produce an estimate.

Bias, mean trials retained, and analysis success distinguished systematic error, trial omission, and failure to obtain a pooled estimate. Supplementary Methods S3.3 provides definitions.

# 3 RESULTS

## 3.1 Single-trial reconstruction

Figure 3 compares four errors in ten matched Weibull scenarios under each censoring–coordinate-error configuration. Table 3 summarizes all 61 settings. Accuracy summaries used valid reconstructed datasets; Supplementary Results S1.2 reports completion separately.

### *3.1.1 Curve, event-count, risk-set, and treatment-effect accuracy*

Across 25 Weibull settings with 30% censoring, median setting-specific KM RMSEs ranged from 0.0091 for iKM to 0.0096 for IPDfromKM (Table 3). Event-count deviation ranged from 0.0% for KMtoIPD to 2.6% for iKM and IPDfromKM, and maximum number-at-risk deviation from 0 for IPDfromKM and QP to 3 for KMtoIPD. These comparisons reflected differing inputs: IPDfromKM received event totals and risk tables, KMtoIPD exact censor marks, and QP risk tables and exact marks.

Treatment-effect errors showed a different ordering: Cox logHR error ranged from 0.0080 for KMtoIPD to 0.0118 for QP, and RMST-contrast error from 0.112 for iKM and KMtoIPD to 0.164 for QP. Curve accuracy therefore did not consistently identify the smallest treatment-effect error. Figure 3 shows the matched Weibull comparison.

Within reconstructed datasets, changes in KM RMSE did not consistently correspond to changes in event-count, risk-set, Cox, or RMST error (Figure 4). These patterns suggest that the measures capture different aspects of performance; they do not establish statistical independence.

### *3.1.2 Method-specific performance profiles*

The method profiles below summarize how curve, event-count, risk-set, and treatment-effect accuracy varied across Weibull and piecewise-exponential settings with 30% or 60% target censoring (Table 3). The corresponding coordinate-error SDs were 0.01 and 0.04, respectively.

iKM. This method had the lowest KM RMSE in Weibull settings at 30% and 60% censoring and the lowest RMST-contrast error in both survival models at 60% censoring. Event-count deviation increased with 60% censoring and coordinate-error SD 0.04.

IPDfromKM. This method closely reproduced supplied risk tables. In both survival models at 30% censoring, its curve, event-count, and treatment-effect errors were generally within the range of the other methods. Agreement with event totals and risk tables requires cautious interpretation because both were supplied inputs.

KMtoIPD. With exact censor marks, it had the smallest event-count deviation in Weibull and piecewise-exponential settings at 30% censoring. It also had the lowest Cox error in those settings and in piecewise-exponential settings at 60% censoring. Median maximum number-at-risk deviation ranged from 3 to 13.5; unlike the other methods, it received no follow-up risk table.

QP. Median maximum number-at-risk deviation was zero in all four scenario categories, reflecting supplied risk-table constraints. QP had the smallest Cox error in Weibull settings at 60% censoring and the smallest KM RMSE and event-count deviation in piecewise-exponential settings at 60% censoring. At 30% censoring, its Cox error exceeded KMtoIPD's, and its RMST error exceeded iKM's and KMtoIPD's.

Under the evaluated input configurations, iKM generally showed favorable curve and RMST accuracy, whereas IPDfromKM closely reproduced supplied risk tables without a consistent advantage in treatment-effect accuracy. KMtoIPD showed favorable event-count and Cox accuracy with exact censor marks. QP reproduced supplied risk sets and had the smallest Cox error in the 60%-censoring Weibull settings and the smallest curve and event-count errors in the 60%-censoring piecewise-exponential settings. Method selection should therefore consider the analytical target, available auxiliary inputs, and similarity to the evaluated scenarios.

### *3.1.3 Performance under combined challenges and individual factor changes*

The matched Weibull comparison changed censoring, coordinate error, sample size, and arm balance together, so it could not isolate individual factors. Errors increased under 60% censoring, coordinate-error SD 0.04, and 100/200 participants compared with 30% censoring, SD 0.01, and 250/250 participants (Figure 3). Method ordering also changed: QP had the lowest Cox error in the former configuration but the largest in the latter. Figure S1 reports the piecewise-exponential comparisons.

The dedicated one-factor experiments isolated individual design changes (Figures S2A–S2D). Increasing survival-probability error consistently increased KM RMSE, event-count deviation, Cox logHR error, and RMST-contrast error. At a fixed total sample size of 500, changing allocation from 250/250 to 100/400 increased average method-specific median Cox and RMST errors by approximately 91% and 51%, respectively, while KM RMSE changed by less than 2%. Arm imbalance therefore affected treatment-effect recovery more than curve agreement.

Censoring effects depended on the method and accuracy measure (Figure 5). The mechanism with the largest event-count error did not necessarily yield the largest RMST error. Increasing the censoring proportion generally increased KM RMSE, whereas Cox and RMST errors rose and fell across successive levels (Figure S2A). Thus, the relationship between censoring proportion and reconstruction error depended on the method and analytical quantity.

#### *3.1.4 Performance with alternative auxiliary inputs*

QP performance varied with the auxiliary information supplied (Figure S7 and Table S5). In the scenario with informative censoring and a 60% censoring target, using only the initial sample size and exact censor marks alongside the curve yielded a median relative event-count deviation of 129.9%. The median of the maximum discrepancies in the number at risk was 59 participants. When event totals were supplied, the median event-count deviation among valid reconstructions was zero in all four scenarios. Supplying event totals did not consistently reduce errors in the Cox logHR or the RMST contrast. Across scenarios and input configurations, 86.5% to 100% of QP reconstructions were valid.

For KMtoIPD, exact censor marks substantially reduced event-count error, with smaller and less consistent changes in curve and RMST error (Table S5). In these simulations, each censored participant contributed one mark, so the observed event count equaled the initial sample size minus the number of marks.

### 3.2 Reconstruction-assisted meta-analysis

Every hybrid strategy produced a valid pooled analysis in every replicate because at least two usable trial estimates remained, even when individual reconstructions failed. Supplementary Results S2.1 distinguishes reconstruction completion from meta-analysis success.

#### *3.2.1 Thirty-trial benchmark*

With complete reporting, AD only, the Full AD oracle, and all hybrids used identical estimates, retaining 30 trials with RMSE 0.0355 and coverage 94.2%. As reporting gaps increased, AD only retained fewer trials, while hybrids remained close to the oracle across the three heterogeneity levels (Figure 6; Table 4).

At 40% unavailable HRs, AD only retained 17.9 trials with RMSE 0.0457, whereas hybrids retained 28.9–30.0 with RMSE 0.0355–0.0363. Table 4 reports all method-specific measures.

At 80% unavailable HRs, AD only retained 6.0 trials, succeeded in 98.7% of replicates, and achieved RMSE 0.0852 and coverage 90.4%. Hybrids retained 27.9–30.0 trials, with RMSE 0.0356–0.0371 and coverage 93.7%–94.3%. AD-only RMSE was 2.40 times the oracle value; hybrid RMSE remained within approximately 5%. Bias stayed close to zero, indicating that differences mainly concerned information retention and variability.

#### *3.2.2 Benchmarks with 10 or 15 trials*

With 10 trials and 80% unavailable HRs, AD only retained 2.0 trials, produced valid analyses in 62.8% of replicates, and achieved RMSE 0.2224 and coverage 55.0%. All hybrids produced valid analyses, with RMSE 0.1056–0.1202 and coverage 92.5%–92.8%, compared with oracle coverage 92.0% (Table 4; Figure S8).

With 15 trials and 80% unavailable HRs, AD-only success was 84.2%, RMSE 0.2054, and coverage 75.7%. Hybrids retained 14.1–15.0 trials, with RMSE 0.0883–0.1019 and coverage 92.0%–92.7%. A similar ordering occurred at 60% missingness. Hybrid–QP improved on omission but had modestly higher RMSE than the other hybrids.

### *3.2.3 Secondary strategy comparisons*

Differences between AD only and hybrids generally exceeded those among Hybrid–iKM, Hybrid–IPDfromKM, and Hybrid–KMtoIPD. AD-only bias remained small despite increased RMSE and failures.

Reconstruction-only strategies sometimes resembled hybrids when reconstruction error was small, but discarded available reference-IPD estimates that contained no reconstruction error. They therefore exposed reliable available information to avoidable reconstruction error or failure. The pooled true-IPD stratified Cox comparator showed undercoverage and estimated a common logHR rather than the mean of heterogeneous trial effects; it was a secondary comparator, not the oracle (Table S4; Figures S4–S6).

These comparisons support supplementing unavailable compatible estimates under the evaluated conditions while retaining available reported estimates.

# 4 DISCUSSION

The two benchmarks identified method-specific performance differences and their consequences for evidence synthesis. In Part I, relative accuracy depended on the analytical quantity and setting. Agreement with supplied risk tables or event totals reflected those constraints rather than establishing treatment-effect accuracy. Digitization imprecision increased errors across measures, while arm imbalance affected treatment-effect recovery more than curve agreement. Method selection therefore requires consideration of both the analytical purpose and available inputs.

Part II extended the evaluation to evidence synthesis. Under missing-completely-at-random reporting, hybrids retained more trials and achieved RMSE and coverage closer to the Full AD oracle than omission. Benefits persisted with 10 or 15 trials and 60% censoring, where omission often left insufficient information for a valid analysis. These results support reconstruction to supplement unavailable compatible estimates while preserving reported information.

Earlier clinical validation supported RMST recovery from reconstructed IPD [27]. Our benchmark connects evaluation across methods, scenarios, and analytical targets with reconstruction-assisted meta-analysis. The findings concern the tested software and input configurations, without establishing an input-independent method ranking.

Combining IPD and aggregate information is an established evidence-synthesis approach [28]. In practice, analysts should first identify the treatment-effect measure they need and the auxiliary information they can reliably extract. They can then consider performance in settings resembling their data and assess reconstructed outputs relevant to that measure, rather than relying on curve agreement alone. Analysts should document the software, preprocessing, follow-up horizon, supplied inputs, and failures. For meta-analysis, compatible reported estimates should be retained, with reconstruction used to supplement unavailable estimates; results should be compared with and without reconstructed trials. Unadjusted reconstruction-derived HRs should not automatically be treated as equivalent to adjusted reported HRs.

Several limitations qualify these findings. Coordinate perturbations did not reproduce time-axis, curve-overlap, arm-label, or auxiliary-input extraction errors, and methods received unequal, accurately supplied auxiliary information. Part II assumed missing-completely-at-random reporting, unadjusted Cox estimates, proportional hazards, and normally distributed trial effects. Reconstructed standard errors included no additional reconstruction uncertainty; observed coverage therefore does not establish complete uncertainty propagation. We included neither a real-image benchmark nor an empirical meta-analysis application, and runtime was not compared under common hardware.

Future work should evaluate the full image-to-analysis pathway, separating errors introduced by manual or automated extraction from reconstruction discrepancies. SurvdigitizeR, KM-GPT, and RESOLVE-IPD offer candidate workflows for this evaluation [14–16]. KMDATA could broaden the range of published trial figures considered, although reconstructed datasets should not replace original IPD as validation truth [29]. Repeated extraction by different operators could quantify human contributions to error.

Further benchmarks should examine imperfect auxiliary inputs, selective reporting, and nonproportional effects, and assess extensions to competing risks or network meta-analysis. Resampling or hierarchical measurement-error models could carry extraction and reconstruction uncertainty into trial estimates and pooled intervals, but their coverage requires validation. Common-hardware timing and empirical meta-analysis applications would clarify practical costs and whether reconstructed estimates change clinical conclusions.

## 5 CONCLUSION

Across the evaluated scenarios, reconstruction methods differed in curve, event-count, risk-set, and treatment-effect accuracy. These differences support selecting and evaluating methods for the intended estimand and available inputs. Under missing-completely-at-random reporting, hybrid meta-analysis retained more trials and produced results closer to the Full AD oracle than trial omission, supporting reconstruction to supplement compatible reported evidence under the conditions studied.

## ACKNOWLEDGMENTS

The authors used OpenAI Codex to assist with manuscript formatting and figure preparation.

## COMPETING INTERESTS

The authors declare no competing interests.

## FUNDING

This research received no specific funding.

## AUTHOR CONTRIBUTIONS

Hui Yao — simulation implementation, data analysis, figures and tables, and manuscript drafting

Nazmul Islam — input on the study's practical motivation and interpretation of the findings

Yicheng Shen — assistance with the simulation work and review of the results

Lang Lang — methodological input and interpretation of the method comparisons

Yanxun Xu — input on the study design and statistical interpretation

Jiayi Tong — study conception and design, supervision, and manuscript revision

## DATA AVAILABILITY

This study used simulated data. The analysis code and supporting results are available in the KM2IPD repository (release v1.0.0): https://github.com/HuiYao-Amber/KM2IPD-arxiv/releases/tag/v1.0.0

**Table 1.** Simulation scenario settings. Target censoring proportion is the intended overall proportion censored. Panel A arm sizes are control/treatment, and design pairs are target censoring proportion / survival-probability error SD. Part I included 61 settings with 1,000 replicates each. Part II scenarios used 300 replicates each. Heterogeneity SD is on the logHR scale. SD, standard deviation; PH, proportional hazards; NPH, nonproportional hazards; HR, hazard ratio; K, number of trials. k, Weibull shape parameter.

**Panel A. Single-trial reconstruction benchmark**

| Survival model | Design pair | Treatment-effect settings | Arm size | Censoring | Survival-probability error SD | Censoring pattern | Settings |
|---|---|---|---|---|---|---|---|
| Weibull PH | 30% / 0.01 | k=1: HR 0.67, 0.85, or 1.00; k=0.6: HR 0.67; k=2: HR 0.67 | 250/250 | 0.30 | 0.01 | Uniform, truncated-exponential, front-loaded, back-loaded, informative | 25 |
| | 60% / 0.04 | k=1: HR 0.67 or 1.00 | 100/200 | 0.60 | 0.04 | Uniform, truncated-exponential, front-loaded, back-loaded, informative | 10 |
| Piecewise exponential | 30% / 0.01 | 3 PH, 2 delayed-effect NPH, and 1 crossing scenario | 250/250 | 0.30 | 0.01 | Uniform, front-loaded, back-loaded | 18 |
| | 60% / 0.04 | 1 delayed-effect NPH and 1 crossing scenario | 250/250 | 0.60 | 0.04 | Uniform, front-loaded, back-loaded, informative | 8 |

**Panel B. Reconstruction-assisted meta-analysis benchmark**

| Trial set | Survival model | Trials (K); participants/arm | Censoring | Survival-probability error SD | Censoring pattern | Missing HRs, % | Between-trial SD | Scenarios |
|---|---|---|---|---|---|---|---|---|
| 30 trials | Weibull PH; median trial HR 0.75 | 30; 250 | 0.30 | 0.01 | Uniform, front-loaded, back-loaded | 0, 20, 40, 60, 80 | 0.05, 0.10, 0.25 | 45 |
| 10 or 15 trials | Weibull PH; median trial HR 0.75 | 10 or 15; 100 | 0.60 | 0.04 | Uniform, front-loaded, back-loaded, informative | 60, 80 | 0.25 | 16 |

**Table 2.** Core principles and input requirements of the evaluated IPD reconstruction methods

| Method | Core Idea | Digitized KM curve coordinates | Initial number at risk | Follow-up risk table | Total event count | Censor marks |
|---|---|---|---|---|---|---|
| iKM | Iterative inverse-KM reconstruction with censoring allocated between risk-table times | Required | Required | Required | Not required | Not required |
| IPDfromKM | Modified iKM reconstruction using interval-specific event and censoring estimates | Required | Required | Optional | Optional | Not required |
| KMtoIPD | Reconstruction from KM decreases with optional incorporation of marked censoring times | Required (KM drop-bottom coordinates) | Required | Not required | Not required | Optional |
| QP | Constrained quadratic-programming estimation of event and censoring allocations | Required | Required | Optional | Optional | Optional |

**Table 3.** Single-trial reconstruction accuracy. Medians and 10th–90th percentiles (P10–P90) summarize setting-specific medians, weighting settings equally. The four categories pair Weibull or piecewise-exponential survival with a target censoring proportion of 30% (error SD 0.01) or 60% (error SD 0.04). Error SD refers to Gaussian error added to survival probabilities. Cells show median (P10–P90), except risk-set deviation, which shows the median maximum absolute number-at-risk difference. Cox and RMST errors are absolute; RMST refers to the treatment-minus-control contrast. Piecewise denotes piecewise exponential.

| Scenario category | Method | Settings | KM RMSE ($\times 10^{-3}$) | Event deviation (%) | Risk-set deviation | Cox logHR error ($\times 10^{-3}$) | RMST error ($\times 10^{-2}$) |
|---|---|---|---|---|---|---|---|
| Weibull 30% / 0.01 | iKM | 25 | 9.1 (8.3–9.5) | 2.6 (1.3–5.3) | 1 | 8.4 (6.8–12) | 11.2 (5.5–19.8) |
| Weibull 30% / 0.01 | IPDfromKM | 25 | 9.6 (8.6–11.6) | 2.6 (0.6–5.3) | 0 | 8.3 (7.1–12.1) | 14.1 (7.1–23.9) |
| Weibull 30% / 0.01 | KMtoIPD | 25 | 9.3 (8.2–10) | 0.0 (0.0–0.3) | 3 | 8 (6.4–11.8) | 11.2 (5.2–22.6) |
| Weibull 30% / 0.01 | QP | 25 | 9.5 (8.4–12.5) | 1.4 (1.1–2.0) | 0 | 11.8 (10.4–16.5) | 16.4 (9.1–26.9) |
| Weibull 60% / 0.04 | iKM | 10 | 39 (36.6–40.2) | 12.9 (11.6–16.8) | 0 | 53 (46–63.9) | 35.4 (19.9–60.5) |
| Weibull 60% / 0.04 | IPDfromKM | 10 | 41.1 (38.2–43.3) | 10.8 (9.9–15.9) | 0 | 53.9 (46.5–68.7) | 40.3 (22.8–66.6) |
| Weibull 60% / 0.04 | KMtoIPD | 10 | 43.2 (38.1–49.4) | 2.5 (0.8–6.8) | 8.5 | 58.3 (47.5–70) | 49 (23.5–72.7) |
| Weibull 60% / 0.04 | QP | 10 | 44.3 (36–47.3) | 5.4 (5.0–9.6) | 0 | 47.4 (40.8–61.2) | 39.3 (26–62.3) |
| Piecewise 30% / 0.01 | iKM | 18 | 9.1 (9–9.3) | 1.7 (1.4–2.0) | 1 | 7.8 (6.8–8.7) | 3.1 (2.8–3.8) |
| Piecewise 30% / 0.01 | IPDfromKM | 18 | 9.1 (9–10.2) | 1.7 (0.6–1.7) | 1 | 8 (7.3–8.8) | 3.1 (2.8–5) |
| Piecewise 30% / 0.01 | KMtoIPD | 18 | 9.1 (9–9.4) | 0.6 (0.3–0.9) | 4 | 7.4 (6.2–7.9) | 3.1 (2.8–4.3) |
| Piecewise 30% / 0.01 | QP | 18 | 11.8 (11.3–12.5) | 2.6 (2.3–2.7) | 0 | 13.1 (12.3–15) | 9 (8.2–9.9) |
| Piecewise 60% / 0.04 | iKM | 8 | 41.3 (38.7–42.1) | 16.0 (13.0–34.8) | 0 | 40.5 (29–86.5) | 35.1 (19.7–48.6) |
| Piecewise 60% / 0.04 | IPDfromKM | 8 | 42.3 (37–43) | 14.2 (9.8–34.5) | 0 | 38.7 (29.4–82.5) | 42.2 (20.1–57.5) |
| Piecewise 60% / 0.04 | KMtoIPD | 8 | 42.7 (39.5–45.6) | 5.3 (3.5–8.2) | 13.5 | 36.6 (33–50.1) | 44.7 (20.5–57.5) |
| Piecewise 60% / 0.04 | QP | 8 | 37.6 (25.5–44.8) | 3.8 (2.1–5.2) | 0 | 39.8 (27.9–45.3) | 40.9 (23.3–46.8) |

**Table 4.** Reconstruction-assisted meta-analysis results at selected reported-estimate missingness levels.

| Trial set | K | Missing HRs (%) | Strategy | Success (%) | Trials retained | Bias (×10⁻³) | RMSE (×10⁻²) | Coverage (%) | Mean τ² (×10⁻²) |
|---|---|---|---|---|---|---|---|---|---|
| 30 trials | 30 | 0 | Full AD oracle | 100 | 30 | 1.5 | 3.55 | 94.2 | 2.49 |
| 30 trials | 30 | 0 | AD only | 100 | 30 | 1.5 | 3.55 | 94.2 | 2.49 |
| 30 trials | 30 | 0 | Hybrid—iKM | 100 | 30 | 1.5 | 3.55 | 94.2 | 2.49 |
| 30 trials | 30 | 0 | Hybrid—IPDfromKM | 100 | 30 | 1.5 | 3.55 | 94.2 | 2.49 |
| 30 trials | 30 | 0 | Hybrid—KMtoIPD | 100 | 30 | 1.5 | 3.55 | 94.2 | 2.49 |
| 30 trials | 30 | 0 | Hybrid—QP | 100 | 30 | 1.5 | 3.55 | 94.2 | 2.49 |
| 30 trials | 30 | 40 | Full AD oracle | 100 | 30 | 1.5 | 3.55 | 94.2 | 2.49 |
| 30 trials | 30 | 40 | AD only | 100 | 17.9 | 1.6 | 4.57 | 93.3 | 2.48 |
| 30 trials | 30 | 40 | Hybrid—iKM | 100 | 29.6 | 0.8 | 3.58 | 94.2 | 2.53 |
| 30 trials | 30 | 40 | Hybrid—IPDfromKM | 100 | 30 | 0.7 | 3.56 | 94.1 | 2.53 |
| 30 trials | 30 | 40 | Hybrid—KMtoIPD | 100 | 30 | 1 | 3.55 | 94.3 | 2.51 |
| 30 trials | 30 | 40 | Hybrid—QP | 100 | 28.9 | 1.1 | 3.63 | 94 | 2.51 |
| 30 trials | 30 | 80 | Full AD oracle | 100 | 30 | 1.5 | 3.55 | 94.2 | 2.49 |
| 30 trials | 30 | 80 | AD only | 98.7 | 6 | 0.8 | 8.52 | 90.4 | 2.58 |
| 30 trials | 30 | 80 | Hybrid—iKM | 100 | 29.3 | 0.3 | 3.6 | 94.2 | 2.57 |
| 30 trials | 30 | 80 | Hybrid—IPDfromKM | 100 | 30 | 0.1 | 3.57 | 94 | 2.57 |
| 30 trials | 30 | 80 | Hybrid—KMtoIPD | 100 | 30 | 0.5 | 3.56 | 94.3 | 2.53 |
| 30 trials | 30 | 80 | Hybrid—QP | 100 | 27.9 | 0.7 | 3.71 | 93.7 | 2.52 |
| 10/15 trials | 10 | 60 | Full AD oracle | 100 | 10 | -1.6 | 10.74 | 92 | 6.6 |
| 10/15 trials | 10 | 60 | AD only | 95.7 | 4 | -4.2 | 17.81 | 87.1 | 7.27 |
| 10/15 trials | 10 | 60 | Hybrid—iKM | 100 | 9.8 | 6.8 | 10.63 | 92.2 | 6.59 |
| 10/15 trials | 10 | 60 | Hybrid—IPDfromKM | 100 | 10 | 6.2 | 10.61 | 92.8 | 6.5 |
| 10/15 trials | 10 | 60 | Hybrid—KMtoIPD | 100 | 10 | 7.3 | 10.57 | 93 | 6.28 |
| 10/15 trials | 10 | 60 | Hybrid—QP | 100 | 9.5 | -8.3 | 11.75 | 91.9 | 7.57 |
| 10/15 trials | 10 | 80 | Full AD oracle | 100 | 10 | -1.6 | 10.74 | 92 | 6.6 |
| 10/15 trials | 10 | 80 | AD only | 62.8 | 2 | 1.6 | 22.24 | 55 | 8.16 |
| 10/15 trials | 10 | 80 | Hybrid—iKM | 100 | 9.8 | 8.8 | 10.67 | 92.5 | 6.58 |
| 10/15 trials | 10 | 80 | Hybrid—IPDfromKM | 100 | 10 | 8.8 | 10.63 | 92.6 | 6.46 |
| 10/15 trials | 10 | 80 | Hybrid—KMtoIPD | 100 | 10 | 10.3 | 10.56 | 92.8 | 6.17 |

| Trial set | K | Missing HRs (%) | Strategy | Success (%) | Trials retained | Bias (×10⁻³) | RMSE (×10⁻²) | Coverage (%) | Mean τ² (×10⁻²) |
|---|---|---|---|---|---|---|---|---|---|
| 10/15 trials | 10 | 80 | Hybrid—QP | 100 | 9.4 | -10.9 | 12.02 | 92.6 | 7.86 |
| 10/15 trials | 15 | 60 | Full AD oracle | 100 | 15 | -7.1 | 9.13 | 92.6 | 6.75 |
| 10/15 trials | 15 | 60 | AD only | 99.5 | 6 | -7.9 | 15.08 | 90.4 | 7.45 |
| 10/15 trials | 15 | 60 | Hybrid—iKM | 100 | 14.8 | 1.8 | 8.99 | 92.8 | 6.76 |
| 10/15 trials | 15 | 60 | Hybrid—IPDfromKM | 100 | 15 | 1.2 | 8.91 | 92.3 | 6.69 |
| 10/15 trials | 15 | 60 | Hybrid—KMtoIPD | 100 | 15 | 1.9 | 8.89 | 92.4 | 6.42 |
| 10/15 trials | 15 | 60 | Hybrid—QP | 100 | 14.3 | -14.1 | 9.87 | 92.7 | 7.5 |
| 10/15 trials | 15 | 80 | Full AD oracle | 100 | 15 | -7.1 | 9.13 | 92.6 | 6.75 |
| 10/15 trials | 15 | 80 | AD only | 84.2 | 3.1 | -7.5 | 20.54 | 75.7 | 8.2 |
| 10/15 trials | 15 | 80 | Hybrid—iKM | 100 | 14.7 | 4.3 | 8.98 | 92.7 | 6.76 |
| 10/15 trials | 15 | 80 | Hybrid—IPDfromKM | 100 | 15 | 3.6 | 8.87 | 92 | 6.67 |
| 10/15 trials | 15 | 80 | Hybrid—KMtoIPD | 100 | 15 | 4.9 | 8.83 | 92.2 | 6.31 |
| 10/15 trials | 15 | 80 | Hybrid—QP | 100 | 14.1 | -17.1 | 10.19 | 92.4 | 7.78 |

Note: Thirty-trial summaries pool three censoring mechanisms and three heterogeneity SDs. Summaries for the 10- or 15-trial, 60%-censoring benchmark pool four censoring mechanisms within each trial-count and missingness combination. Bias and RMSE are on the logHR scale. Coverage is unconditional, with failed fits counted as noncoverage. Shared scaling factors are shown in the column headings.

## FIGURES

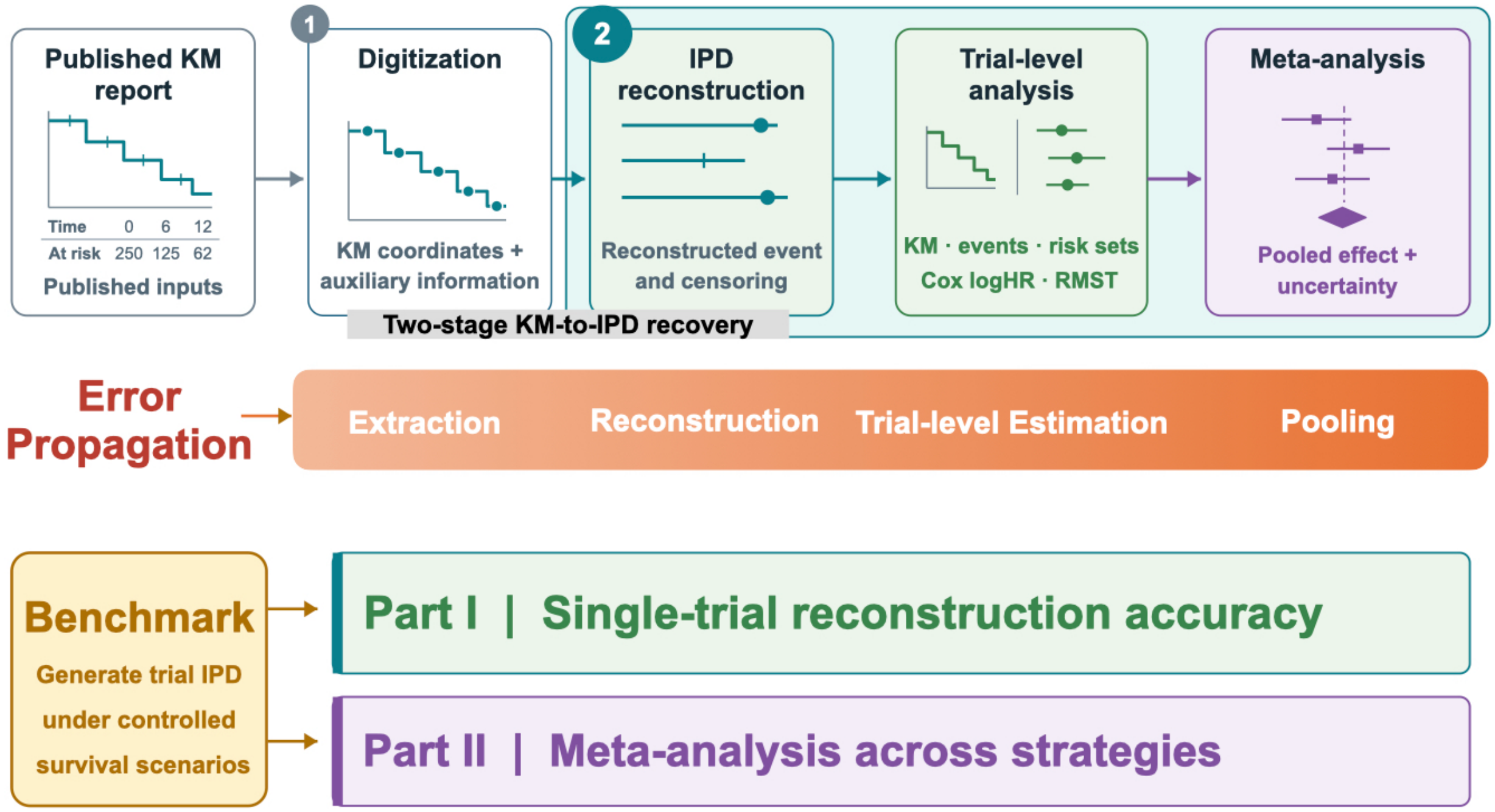

**Figure 1.** IPD reconstruction pipeline and the two simulation benchmarks. The upper pathway distinguishes published information, digitized inputs, reconstructed participant records, trial-level estimates, and meta-analysis estimates. Errors introduced at one stage can affect subsequent stages. The two numbered steps denote digitization and IPD reconstruction. The shaded area covers the stages evaluated with simulated inputs: reconstruction, trial-level analysis, and meta-analysis. Part I evaluates reconstruction and trial-level accuracy; Part II evaluates strategies that combine available reported estimates with reconstruction-derived estimates. The simulations perturb survival probabilities rather than extracting inputs from raster images.

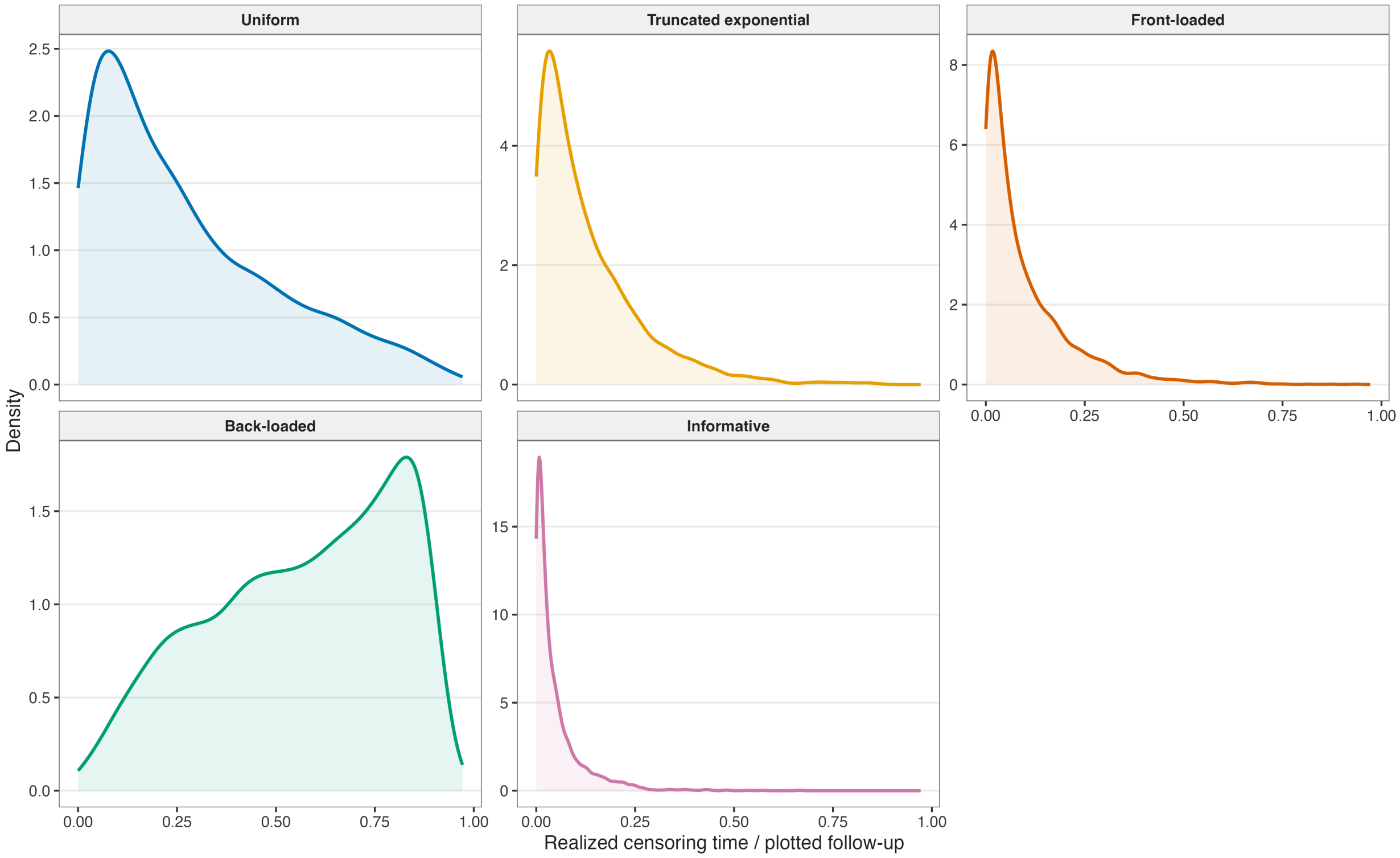

**Figure 2.** Simulated temporal patterns of censoring. Distributions of realized censoring times under five mechanisms are shown on a normalized follow-up scale. The mechanisms were calibrated to comparable target censoring proportions. The panels show how the timing of observed censoring varied across mechanisms.

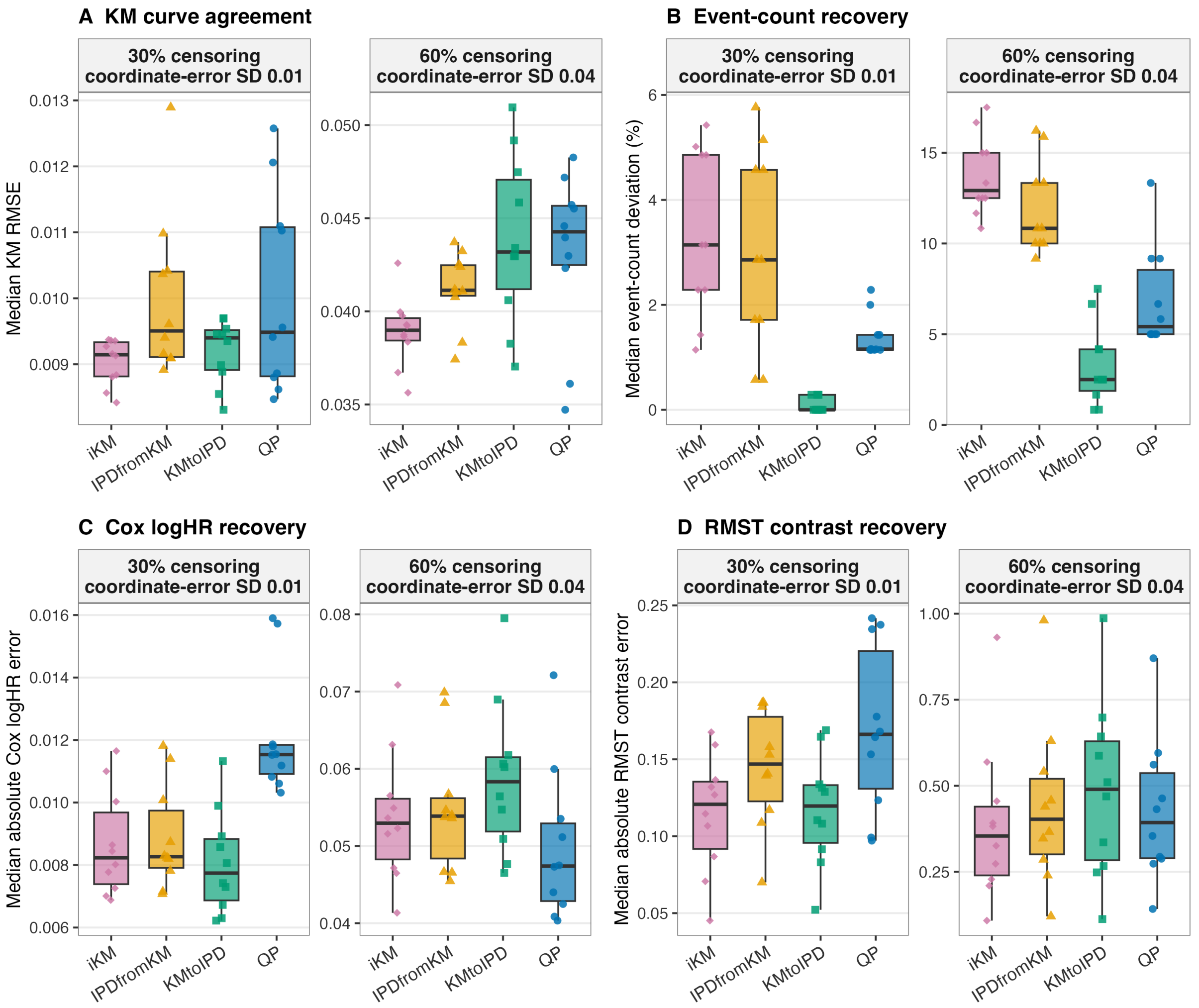

**Figure 3.** Setting-level median reconstruction errors for (A) KM RMSE, (B) relative absolute event-count deviation, (C) absolute Cox logHR error, and (D) absolute RMST-contrast error. Each configuration includes ten shape-1 Weibull scenarios formed by two HRs (0.67 and 1.00) and five censoring mechanisms. The 30%-censoring configuration used 250/250 participants and survival-probability error SD 0.01; the 60%-censoring configuration used 100/200 participants and error SD 0.04. Each point summarizes valid reconstructions from 1,000 replicates in one setting. Boxes show the median and interquartile range across settings; whiskers extend to the most extreme points within 1.5 interquartile ranges of the box.

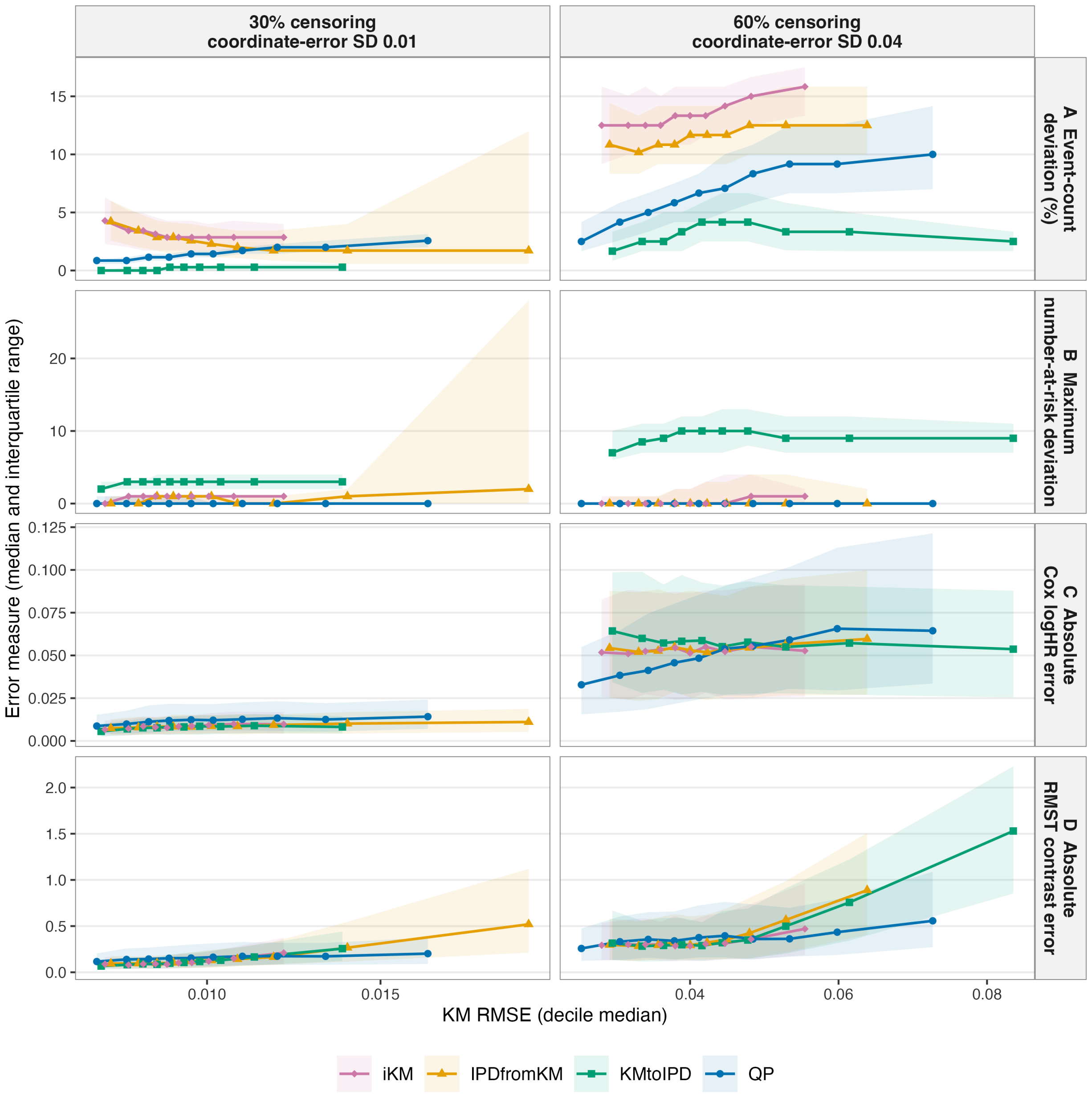

**Figure 4.** Replicate-level relationships between KM RMSE and (A) relative event-count deviation, (B) maximum number-at-risk deviation, (C) absolute Cox logHR error, and (D) absolute RMST-contrast error in the matched Weibull subset. Within each method and censoring–coordinate-error configuration, datasets were divided into KM-RMSE deciles; lines connect decile medians and bands show interquartile ranges.

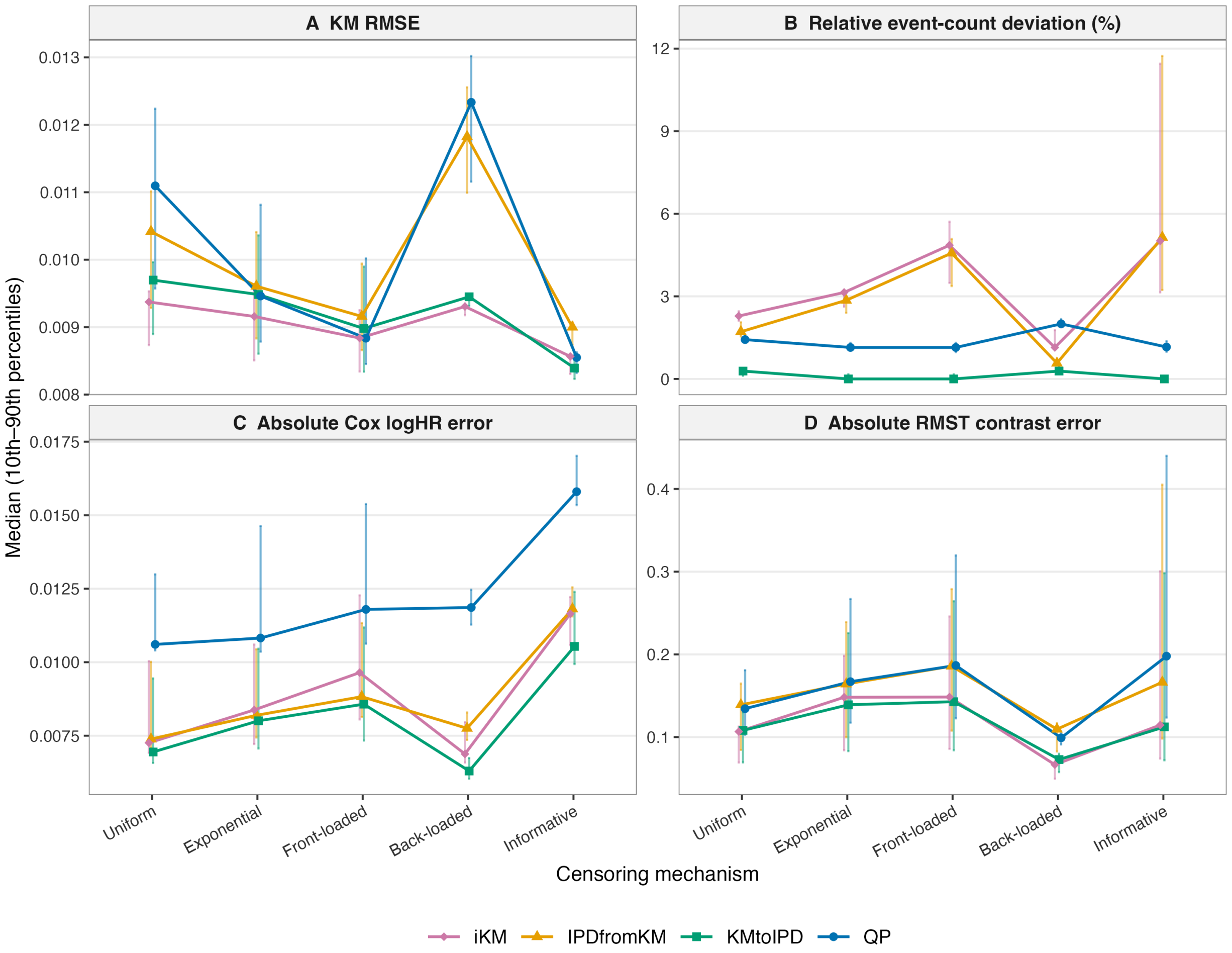

**Figure 5.** Reconstruction performance across five censoring mechanisms in the matched 30%-censoring Weibull settings. For each mechanism, points show method-specific medians across the five survival scenarios and bars show P10–P90 across setting-level medians. Panels report (A) KM RMSE, (B) relative event-count deviation, (C) absolute Cox logHR error, and (D) absolute RMST-contrast error; lower values indicate better recovery.

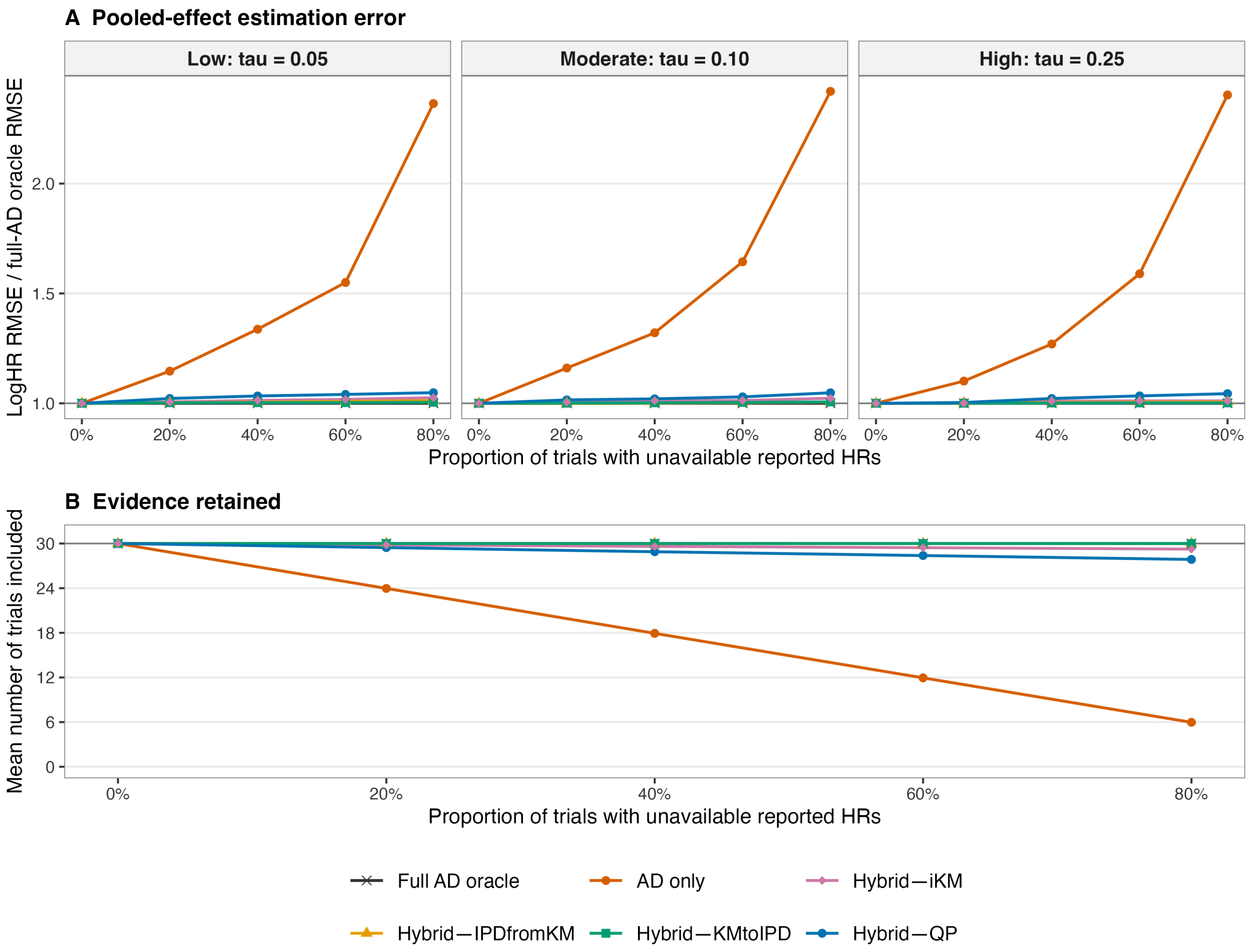

**Figure 6.** Performance of principal meta-analysis strategies in the larger 30-trial benchmark. (A) Pooled-logHR RMSE relative to the full-aggregate-data oracle across proportions of missing reported HRs, stratified by heterogeneity SD (τ); (B) mean number of trials retained across heterogeneity and censoring conditions. Each scenario contained 30 trials and used 300 replicates. Hybrid strategies retained reported estimates when available and substituted reconstruction-derived Cox estimates only when the reported estimate was missing.

# Supplementary material

From Multi-site Kaplan–Meier Curves to Evidence Synthesis: An Empirical Evaluation of IPD Reconstruction and Reconstruction-Assisted Meta-Analysis

Hui Yao, Nazmul Islam, Yicheng Shen, Lang Lang, Yanxun Xu, and Jiayi Tong

## Contents

## SUPPLEMENTARY METHODS S1. PART I DATA-GENERATING MECHANISMS AND CONTROLLED KM INPUTS

### S1.1 Weibull proportional-hazards scenarios

We generated independent latent event times from Weibull proportional-hazards (PH) models using R's shape–scale parameterization. For shape k, the control-arm scale was

$$\lambda_0 = \frac{10}{\Gamma(1 + 1/k)}.$$

This fixed the mean control event time at 10. For the target hazard ratio (HR), the treatment-arm scale was

$$\lambda_1 = \lambda_0 \, \mathrm{HR}^{-1/k}.$$

Table S1 lists the complete 61-setting registry, including survival patterns, sample sizes, censoring mechanisms, and survival-probability error standard deviations (SDs).

### S1.2 Piecewise-exponential scenarios

Piecewise-exponential event times used intervals [0,6), [6,12), and [12,∞). We inverted a unit-rate exponential variate through the cumulative hazard defined by the arm-specific hazard vectors in Table S1, Panel A.

The administrative follow-up limit was 24 time units. Table S1, Panel B specifies the combinations of survival patterns and censoring mechanisms.

### S1.3 Censoring mechanisms

Let c denote the calibrated censoring scale and U a standard uniform variate. Noninformative censoring times were generated as follows:

- uniform censoring: $C = cU$;
- front-loaded censoring: $C = cB$, where $B \sim \mathrm{Beta}(0.7, 2.0)$;
- back-loaded censoring: $C = cB$, where $B \sim \mathrm{Beta}(2.0, 0.7)$;
- truncated-exponential censoring: $C = F^{-1}(U(1 - \exp(-2)))$, where F is an exponential distribution with rate 2/c.

For informative censoring, an independent uniform indicator below $p_{\mathrm{inf}} = \min(\max(p_c, 0.01), 0.85)$ selected participants for censoring at $U \min(T, c)$, where $p_c$ was the target censoring proportion and T the latent event time. Other participants were subject only to the administrative limit. Event status was $\mathrm{I}(T \leq C, T \leq A)$, where A was c for Weibull settings and 24 for piecewise-exponential settings. Within each replicate, we selected c by root finding, with a grid search when the target could not be attained. In the piecewise-exponential scenarios with informative censoring, the fixed administrative limit and the selected participants determined the censoring proportion. At a 60% target, mean achieved censoring was 67.2% in the delayed-effect scenario and 66.6% in the crossing-curves scenario.

### S1.4 Controlled KM-derived inputs

The number at risk at time t was defined as the number of participants whose observed follow-up was at least t. Nine risk-table times defined eight equal intervals from zero to the display limit. For piecewise-exponential scenarios, the display limit was 24; for Weibull scenarios, it was the largest observed follow-up rounded upward to the next multiple of eight.

For each arm, coordinates included baseline, every observed Kaplan–Meier (KM) decrease, each risk-table time, and the display limit. For empirical survival probability $S(t_i)$ from the simulated individual participant data (IPD), the supplied coordinate was

$$S^*(t_i) = \min[1, \max(0, S(t_i) + \varepsilon_i)], \quad \varepsilon_i \sim N(0, \sigma_S^2).$$

We fixed baseline survival at one and applied cumulative minima to enforce nonincreasing survival probabilities. Time coordinates remained unchanged. The imposed SDs are listed in Table S1.

All four methods used the same simulated IPD and controlled coordinates within each setting and replicate. Their auxiliary inputs are specified in main-text Section 2.2 and Table 2.

## SUPPLEMENTARY METHODS S2. CONTROLLED ONE-FACTOR EXPERIMENTS

### S2.1 Dedicated one-factor design

The dedicated one-factor experiments used a Weibull PH reference design with shape 1, HR 0.67, 250 participants per arm, uniform censoring, target censoring proportion 0.30, and survival-probability error SD 0.01. One component was changed at a time while all other components remained at these reference values:

- target censoring proportion: 0.10, 0.30, 0.45, 0.60, or 0.70;
- survival-probability error SD: 0, 0.005, 0.01, 0.02, or 0.04;
- balanced sample size: 100/100, 250/250, or 500/500 participants in the control/treatment arms;
- arm allocation with total sample size fixed at 500: 250/250, 200/300, 150/350, or 100/400.

Each of the 17 factor levels used 300 replicates, with the same IPD and coordinates supplied to all methods within a replicate. Figures S2A–S2D show medians and P10–P90 intervals; Table S3 gives the numerical summaries.

### S2.2 Matched factor comparisons from the 30%-censoring Weibull settings

Matched comparisons within the 25 Weibull settings with 30% censoring used 1,000 replicates per setting. We compared shapes 0.6, 1, and 2 at HR 0.67, and HRs 0.67, 0.85, and 1.00 at shape 1, within each censoring mechanism (Figure S3). Censoring-mechanism comparisons are shown in main-text Figure 5.

### S2.3 Technical definitions and calculation details

The equations below specify the measures introduced in main-text Section 2.3.2. Subscripts rec and true denote reconstructed and reference datasets; a denotes trial arm.

We calculated arm-specific KM root mean squared error (RMSE) at G = 200 equally spaced times up to the final scheduled risk-table time and averaged the two arm-specific values:

$$\mathrm{KM-RMSE}(a) = \left[G^{-1} \sum_{g=1}^{G} \left\{\hat{S}_{\mathrm{rec},a}[t(g)] - \hat{S}_{\mathrm{true},a}[t(g)]\right\}^2\right]^{1/2}.$$

For total observed events D across both arms and arm-specific risk counts Y at scheduled times $r_j$, the structural errors were

$$\text{Relative event deviation} = 100 \times \frac{|D_{\mathrm{rec}} - D_{\mathrm{true}}|}{D_{\mathrm{true}}},$$

$$\text{Maximum risk-set deviation} = \max_{a,j} \left|Y_{\mathrm{rec},a}(r_j) - Y_{\mathrm{true},a}(r_j)\right|,$$

Figure S5 additionally reports absolute event-count deviation, $|D_{\mathrm{rec}} - D_{\mathrm{true}}|$.

Let β denote the unadjusted Cox log hazard ratio (logHR). For dataset x, the treatment-minus-control restricted mean survival time (RMST) contrast and its reconstruction error were

$$\text{Cox logHR error} = \left|\hat{\beta}_{\mathrm{rec}} - \hat{\beta}_{\mathrm{true}}\right|,$$

$$\Delta\text{RMST}(x;\tau) = \int_0^{\tau}\{\hat{S}_{x,1}(t) - \hat{S}_{x,0}(t)\}\,dt.$$

$$RMST\ contrast\ error\ =\ |\Delta RMST(rec;\ \tau)\ -\ \Delta RMST(true;\ \tau)|$$

RMST used the final risk-table time as τ and a 1,000-point integration grid. Cox models used Efron handling for tied events. Figure S3 reports signed Cox and RMST differences.

## SUPPLEMENTARY METHODS S3. PART II META-ANALYSIS STRATEGIES, SCENARIOS, AND IMPLEMENTATION

### S3.1 Meta-analysis strategies and estimand

For valid trial Cox logHRs $y_j$ with standard errors $s_j$, the random-effects model was

$$y_j \mid \theta_j \sim N(\theta_j, s_j^2), \quad \theta_j \sim N(\mu, \tau^2).$$

We estimated μ and τ² by restricted maximum likelihood, with generating mean μ = log(0.75). The principal strategies using aggregate data (AD) and reconstructed estimates are defined in main-text Section 2.4.1.

The pooled true-IPD comparator fitted a Cox model with trial-specific baseline-hazard strata. Its common conditional coefficient differed from the mean generating trial effect μ, so it served as a secondary comparator.

A meta-analysis failed when fewer than two valid trial estimates remained or the random-effects model returned a nonfinite pooled estimate or standard error.

### S3.2 Trial generation, reported-estimate availability, and reconstruction

For trial j in meta-analysis replicate r, the generating logHR was

$$\theta_{jr} \sim N(\log(0.75), \tau^2), \quad j = 1, \ldots, K.$$

The combinations in main-text Table 1, Panel B yielded 45 scenarios for the 30-trial benchmark and 16 for the 10- or 15-trial benchmark. Each scenario used 300 replicates.

We drew one U(0,1) variate per trial and withheld its reference-IPD Cox estimate and standard error at missingness proportion p when U < p. Thus, trials withheld at a lower p remained withheld at higher p. Every strategy used the same trial collection and availability pattern.

Trials were converted to the controlled inputs in S1.4 and reconstructed using the configurations in main-text Section 2.2. KMtoIPD used regenerated copies of the frozen trials with matching seeds and availability patterns; checks confirmed agreement with archived trial-level truth.

### S3.3 Meta-analysis performance summaries

Bias and pooled-logHR RMSE summarized valid analyses relative to μ = log(0.75). Coverage counted nominal 95% Wald intervals, $\hat{\mu} \pm 1.96\,\text{SE}(\hat{\mu})$, containing μ across all assigned replicates; failed analyses counted as noncoverage. Analysis success was the proportion yielding a valid pooled estimate. We also summarized mean trials retained and mean estimated heterogeneity variance. Table S4 reports all 11 strategies; Figures S4–S6 provide extended comparisons.

## SUPPLEMENTARY METHODS S4. AUXILIARY-INPUT SENSITIVITY ANALYSES

Four Weibull designs with shape 1 and HR 0.67 crossed uniform or informative censoring with two configurations: 30% censoring, 250/250 participants, and survival-probability error SD 0.01; or 60% censoring, 100/200 participants, and SD 0.04. Each design used 200 replicates. Input states shared the same IPD and coordinates within a replicate.

Seven quadratic-programming (QP) configurations were evaluated:

1. risk table plus exact censor marks, with every supplied mark enforced (the primary QP configuration);
2. risk table plus exact censor marks treated as candidate censoring locations;
3. risk table, exact censor marks, and true total events;
4. risk table only;
5. risk table plus true total events;
6. initial sample size plus exact censor marks;
7. initial sample size, exact censor marks, and true total events.

When marks were supplied, nonprimary QP configurations used strict_tick = FALSE, treating marks as candidate locations without requiring a positive censor count at every mark. All configurations used strict_dec = FALSE and no fallback reconstruction. Cross-method comparisons removed censor marks from KMtoIPD or total events from IPDfromKM, with iKM as a contextual comparator. Figure S7 shows the seven QP configurations, and Table S5 reports all configurations.

## SUPPLEMENTARY METHODS S5. SOFTWARE CONTROLS

Software versions, functions, and primary inputs are specified in main-text Section 2.2. For QP, we used optmethod = "approx", strict_dec = FALSE, and strict_tick = TRUE. The approximate option solves the continuous quadratic program before integer rounding; strict_dec = FALSE does not require an event at every apparent decrease after coordinate perturbation. Infeasible solutions remained failures. Because software and validation procedures differed, completion proportions were retained as diagnostics. Runtime was omitted because implementations were not timed under common hardware and execution conditions.

## SUPPLEMENTARY RESULTS S1. EXTENDED PART I RESULTS

### S1.1 Survival patterns, censoring, and controlled factors

Figure S1 extends the comparisons to proportional-hazards, delayed-effect, and crossing-survival patterns. Nonproportional hazards or crossing curves alone did not consistently yield the largest errors across methods and analytical targets.

Figures S2A–S2D and Table S3 provide complete one-factor results. Figure S3 separately examines Weibull shape and HR in matched settings, including signed Cox and RMST errors.

### S1.2 Reconstruction completion

Across 61,000 assigned replicates per method, valid-reconstruction proportions were 98.0% for iKM, 100% for IPDfromKM and KMtoIPD, and 93.5% for QP. The 3,937 QP failures were infeasible optimization results under the fixed primary controls. Table S2 provides the complete setting-level results.

## SUPPLEMENTARY RESULTS S2. EXTENDED PART II RESULTS

### S2.1 Trial-level reconstructions entering meta-analysis

For iKM, IPDfromKM, KMtoIPD, and QP, respectively, reconstruction completion was 96.9%, 100%, 100%, and 91.2% in the 30-trial benchmark, and 97.6%, 100%, 100%, and 92.5% in the 10- or 15-trial benchmark. Corresponding median absolute Cox logHR errors were 0.0078, 0.0082, 0.0072, and 0.0116, and 0.0437, 0.0454, 0.0462, and 0.0553 (Figure S5). Failed reconstructions reduced the trial estimates available for synthesis.

### S2.2 Principal and secondary meta-analysis strategies

Figure S4 shows how interval coverage and estimated heterogeneity varied with report availability and generating heterogeneity. AD-only coverage deteriorated most clearly when both missingness and heterogeneity were high. Figure S8 provides the corresponding pooled-effect, coverage, and success comparisons for 10 or 15 trials.

Figure S6 and Table S4 extend the numerical comparison to all reconstruction-only strategies and the pooled true-IPD comparator. Their values complement the principal-strategy results in main-text Table 4.

## SUPPLEMENTARY RESULTS S3. AUXILIARY-INPUT SENSITIVITY RESULTS

Of 5,600 QP sensitivity runs, 5,226 produced valid reconstructed datasets. Figure S7 and Table S5 show that completion and accuracy varied across input configurations.

In the 60%-censoring informative design, risk-table-plus-total-events inputs yielded median Cox logHR error 0.0969 and RMST-contrast error 1.073, compared with 0.0725 and 0.850 for risk table alone. Strict-versus-relaxed mark comparisons did not consistently attribute failure to enforcement of every mark. In the 30%-censoring uniform design, the primary strict-mark configuration succeeded in 94.0% of replicates with median Cox error 0.0104; risk table alone succeeded in 99.0% with error 0.0064. Under 60% uniform censoring, the primary configuration instead had lower Cox error (0.0487 versus 0.0566).

Across the four scenarios, median relative event-count deviation ranged from 1.2% to 8.3% when QP used the risk table and enforced every censor mark, and from 2.3% to 15.8% when QP used the risk table alone. For KMtoIPD, median relative event-count deviation ranged from 0.0% to 4.2% with exact censor marks and from 36.3% to 151.3% without them. Removing total events from IPDfromKM produced no paired median change in the four displayed accuracy measures, although upper-tail differences appeared under informative censoring. These findings apply to the tested inputs and implementations.

**Table S1.** Full registry of 61 Part I design settings

**Panel A. Piecewise-exponential survival patterns and hazard vectors**

| Scenario ID | Pattern | Control hazards | Treatment hazards |
|---|---|---|---|
| PH_H1_HR067 | Constant-baseline PH, HR 0.67 | 0.080, 0.080, 0.080 | 0.0536, 0.0536, 0.0536 |
| PH_H2_HR085 | Increasing-baseline PH, HR 0.85 | 0.045, 0.080, 0.120 | 0.03825, 0.068, 0.102 |
| PH_H2_HR100 | Increasing-baseline PH, HR 1.00 | 0.045, 0.080, 0.120 | 0.045, 0.080, 0.120 |
| NPHdelayed_H1_D1 | Delayed benefit | 0.080, 0.080, 0.080 | 0.095, 0.075, 0.045 |
| NPHdelayed_H2_D2 | Delayed benefit with increasing baseline hazard | 0.045, 0.080, 0.120 | 0.055, 0.075, 0.080 |
| NPHcross_H1_C1 | Early harm followed by late benefit | 0.060, 0.080, 0.100 | 0.130, 0.075, 0.035 |

**Panel B. Complete setting registry**

| Design ID | Event-time model | Scenario ID | Shape k | HR | $n_0$ | $n_1$ | Censoring | Target censoring | Error SD | Follow-up cap |
|---|---|---|---|---|---|---|---|---|---|---|
| 1 | Weibull PH | W_shape1_HR067 | 1 | 0.67 | 250 | 250 | random | 0.3 | 0.01 | |
| 2 | Weibull PH | W_shape1_HR067 | 1 | 0.67 | 250 | 250 | exp | 0.3 | 0.01 | |
| 3 | Weibull PH | W_shape1_HR067 | 1 | 0.67 | 250 | 250 | front | 0.3 | 0.01 | |
| 4 | Weibull PH | W_shape1_HR067 | 1 | 0.67 | 250 | 250 | back | 0.3 | 0.01 | |
| 5 | Weibull PH | W_shape1_HR067 | 1 | 0.67 | 250 | 250 | informative | 0.3 | 0.01 | |
| 6 | Weibull PH | W_shape1_HR085 | 1 | 0.85 | 250 | 250 | random | 0.3 | 0.01 | |
| 7 | Weibull PH | W_shape1_HR085 | 1 | 0.85 | 250 | 250 | exp | 0.3 | 0.01 | |
| 8 | Weibull PH | W_shape1_HR085 | 1 | 0.85 | 250 | 250 | front | 0.3 | 0.01 | |
| 9 | Weibull PH | W_shape1_HR085 | 1 | 0.85 | 250 | 250 | back | 0.3 | 0.01 | |
| 10 | Weibull PH | W_shape1_HR085 | 1 | 0.85 | 250 | 250 | informative | 0.3 | 0.01 | |
| 11 | Weibull PH | W_shape1_HR100 | 1 | 1 | 250 | 250 | random | 0.3 | 0.01 | |
| 12 | Weibull PH | W_shape1_HR100 | 1 | 1 | 250 | 250 | exp | 0.3 | 0.01 | |
| 13 | Weibull PH | W_shape1_HR100 | 1 | 1 | 250 | 250 | front | 0.3 | 0.01 | |
| 14 | Weibull PH | W_shape1_HR100 | 1 | 1 | 250 | 250 | back | 0.3 | 0.01 | |
| 15 | Weibull PH | W_shape1_HR100 | 1 | 1 | 250 | 250 | informative | 0.3 | 0.01 | |
| 16 | Weibull PH | W_shape0p6_HR067 | 0.6 | 0.67 | 250 | 250 | random | 0.3 | 0.01 | |
| 17 | Weibull PH | W_shape0p6_HR067 | 0.6 | 0.67 | 250 | 250 | exp | 0.3 | 0.01 | |
| 18 | Weibull PH | W_shape0p6_HR067 | 0.6 | 0.67 | 250 | 250 | front | 0.3 | 0.01 | |

| Design ID | Event-time model | Scenario ID | Shape k | HR | $n_0$ | $n_1$ | Censoring | Target censoring | Error SD | Follow-up cap |
|---|---|---|---|---|---|---|---|---|---|---|
| 19 | Weibull PH | W_shape0p6_HR067 | 0.6 | 0.67 | 250 | 250 | back | 0.3 | 0.01 | |
| 20 | Weibull PH | W_shape0p6_HR067 | 0.6 | 0.67 | 250 | 250 | informative | 0.3 | 0.01 | |
| 21 | Weibull PH | W_shape2_HR067 | 2 | 0.67 | 250 | 250 | random | 0.3 | 0.01 | |
| 22 | Weibull PH | W_shape2_HR067 | 2 | 0.67 | 250 | 250 | exp | 0.3 | 0.01 | |
| 23 | Weibull PH | W_shape2_HR067 | 2 | 0.67 | 250 | 250 | front | 0.3 | 0.01 | |
| 24 | Weibull PH | W_shape2_HR067 | 2 | 0.67 | 250 | 250 | back | 0.3 | 0.01 | |
| 25 | Weibull PH | W_shape2_HR067 | 2 | 0.67 | 250 | 250 | informative | 0.3 | 0.01 | |
| 26 | Weibull PH | W_shape1_HR067 | 1 | 0.67 | 100 | 200 | random | 0.6 | 0.04 | |
| 27 | Weibull PH | W_shape1_HR067 | 1 | 0.67 | 100 | 200 | exp | 0.6 | 0.04 | |
| 28 | Weibull PH | W_shape1_HR067 | 1 | 0.67 | 100 | 200 | front | 0.6 | 0.04 | |
| 29 | Weibull PH | W_shape1_HR067 | 1 | 0.67 | 100 | 200 | back | 0.6 | 0.04 | |
| 30 | Weibull PH | W_shape1_HR067 | 1 | 0.67 | 100 | 200 | informative | 0.6 | 0.04 | |
| 31 | Weibull PH | W_shape1_HR100 | 1 | 1 | 100 | 200 | random | 0.6 | 0.04 | |
| 32 | Weibull PH | W_shape1_HR100 | 1 | 1 | 100 | 200 | exp | 0.6 | 0.04 | |
| 33 | Weibull PH | W_shape1_HR100 | 1 | 1 | 100 | 200 | front | 0.6 | 0.04 | |
| 34 | Weibull PH | W_shape1_HR100 | 1 | 1 | 100 | 200 | back | 0.6 | 0.04 | |
| 35 | Weibull PH | W_shape1_HR100 | 1 | 1 | 100 | 200 | informative | 0.6 | 0.04 | |
| 36 | Piecewise exponential | PH_H1_HR067 | | 0.67 | 250 | 250 | random | 0.3 | 0.01 | 24 |
| 37 | Piecewise exponential | PH_H1_HR067 | | 0.67 | 250 | 250 | front | 0.3 | 0.01 | 24 |
| 38 | Piecewise exponential | PH_H1_HR067 | | 0.67 | 250 | 250 | back | 0.3 | 0.01 | 24 |
| 39 | Piecewise exponential | PH_H2_HR085 | | 0.85 | 250 | 250 | random | 0.3 | 0.01 | 24 |
| 40 | Piecewise exponential | PH_H2_HR085 | | 0.85 | 250 | 250 | front | 0.3 | 0.01 | 24 |
| 41 | Piecewise exponential | PH_H2_HR085 | | 0.85 | 250 | 250 | back | 0.3 | 0.01 | 24 |
| 42 | Piecewise exponential | PH_H2_HR100 | | 1 | 250 | 250 | random | 0.3 | 0.01 | 24 |
| 43 | Piecewise exponential | PH_H2_HR100 | | 1 | 250 | 250 | front | 0.3 | 0.01 | 24 |
| 44 | Piecewise exponential | PH_H2_HR100 | | 1 | 250 | 250 | back | 0.3 | 0.01 | 24 |
| 45 | Piecewise exponential | NPHdelayed_H1_D1 | | | 250 | 250 | random | 0.3 | 0.01 | 24 |
| 46 | Piecewise exponential | NPHdelayed_H1_D1 | | | 250 | 250 | front | 0.3 | 0.01 | 24 |
| 47 | Piecewise exponential | NPHdelayed_H1_D1 | | | 250 | 250 | back | 0.3 | 0.01 | 24 |

| Design ID | Event-time model | Scenario ID | Shape k | HR | $n_0$ | $n_1$ | Censoring | Target censoring | Error SD | Follow-up cap |
|---|---|---|---|---|---|---|---|---|---|---|
| 48 | Piecewise exponential | NPHdelayed_H2_D2 | | | 250 | 250 | random | 0.3 | 0.01 | 24 |
| 49 | Piecewise exponential | NPHdelayed_H2_D2 | | | 250 | 250 | front | 0.3 | 0.01 | 24 |
| 50 | Piecewise exponential | NPHdelayed_H2_D2 | | | 250 | 250 | back | 0.3 | 0.01 | 24 |
| 51 | Piecewise exponential | NPHcross_H1_C1 | | | 250 | 250 | random | 0.3 | 0.01 | 24 |
| 52 | Piecewise exponential | NPHcross_H1_C1 | | | 250 | 250 | front | 0.3 | 0.01 | 24 |
| 53 | Piecewise exponential | NPHcross_H1_C1 | | | 250 | 250 | back | 0.3 | 0.01 | 24 |
| 54 | Piecewise exponential | NPHdelayed_H1_D1 | | | 250 | 250 | random | 0.6 | 0.04 | 24 |
| 55 | Piecewise exponential | NPHdelayed_H1_D1 | | | 250 | 250 | front | 0.6 | 0.04 | 24 |
| 56 | Piecewise exponential | NPHdelayed_H1_D1 | | | 250 | 250 | back | 0.6 | 0.04 | 24 |
| 57 | Piecewise exponential | NPHdelayed_H1_D1 | | | 250 | 250 | informative | 0.6 | 0.04 | 24 |
| 58 | Piecewise exponential | NPHcross_H1_C1 | | | 250 | 250 | random | 0.6 | 0.04 | 24 |
| 59 | Piecewise exponential | NPHcross_H1_C1 | | | 250 | 250 | front | 0.6 | 0.04 | 24 |
| 60 | Piecewise exponential | NPHcross_H1_C1 | | | 250 | 250 | back | 0.6 | 0.04 | 24 |
| 61 | Piecewise exponential | NPHcross_H1_C1 | | | 250 | 250 | informative | 0.6 | 0.04 | 24 |

**Note:** Every setting used 1,000 replicates and nine risk-table points. $n_0$ and $n_1$ are control and treatment arm sizes. Target censoring is the intended proportion; error SD applies to survival probabilities; follow-up cap is the administrative limit. Codes random, exp, front, and back denote uniform, truncated-exponential, front-loaded, and back-loaded censoring. Weibull scenarios are identified by shape and HR; piecewise hazard vectors and pattern labels are given in Panel A.

**Table S2.** Setting-level Part I accuracy and valid-reconstruction summaries

| Scenario category | Scenario ID | Censoring | Method | Valid IPD (%) | KM RMSE | Relative event deviation (%) | Maximum risk-set deviation | Absolute Cox logHR error | Absolute RMST contrast error |
|---|---|---|---|---|---|---|---|---|---|
| Weibull—30% | W_shape1_HR067 | random | iKM | 96.7 | 0.0094 | 2.3 | 0.0 | 0.0078 | 0.107 |
| Weibull—30% | W_shape1_HR067 | random | IPDfromKM | 100.0 | 0.0104 | 1.7 | 0.0 | 0.0082 | 0.139 |
| Weibull—30% | W_shape1_HR067 | random | KMtoIPD | 100.0 | 0.0097 | 0.3 | 3.0 | 0.0074 | 0.108 |
| Weibull—30% | W_shape1_HR067 | random | QP | 92.2 | 0.0110 | 1.4 | 0.0 | 0.0103 | 0.168 |
| Weibull—30% | W_shape1_HR067 | exp | iKM | 94.0 | 0.0092 | 3.1 | 1.0 | 0.0084 | 0.168 |
| Weibull—30% | W_shape1_HR067 | exp | IPDfromKM | 100.0 | 0.0094 | 2.9 | 0.0 | 0.0083 | 0.187 |
| Weibull—30% | W_shape1_HR067 | exp | KMtoIPD | 100.0 | 0.0093 | 0.0 | 2.0 | 0.0081 | 0.169 |
| Weibull—30% | W_shape1_HR067 | exp | QP | 90.5 | 0.0096 | 1.1 | 0.0 | 0.0112 | 0.237 |
| Weibull—30% | W_shape1_HR067 | front | iKM | 92.0 | 0.0088 | 4.9 | 1.0 | 0.0100 | 0.159 |
| Weibull—30% | W_shape1_HR067 | front | IPDfromKM | 100.0 | 0.0092 | 4.6 | 1.0 | 0.0101 | 0.187 |
| Weibull—30% | W_shape1_HR067 | front | KMtoIPD | 100.0 | 0.0089 | 0.0 | 2.0 | 0.0089 | 0.165 |
| Weibull—30% | W_shape1_HR067 | front | QP | 88.7 | 0.0088 | 1.1 | 0.0 | 0.0118 | 0.235 |
| Weibull—30% | W_shape1_HR067 | back | iKM | 100.0 | 0.0094 | 1.4 | 1.0 | 0.0069 | 0.045 |
| Weibull—30% | W_shape1_HR067 | back | IPDfromKM | 100.0 | 0.0110 | 0.6 | 1.0 | 0.0071 | 0.070 |
| Weibull—30% | W_shape1_HR067 | back | KMtoIPD | 100.0 | 0.0095 | 0.3 | 4.0 | 0.0062 | 0.052 |
| Weibull—30% | W_shape1_HR067 | back | QP | 93.2 | 0.0126 | 2.3 | 0.0 | 0.0119 | 0.097 |
| Weibull—30% | W_shape1_HR067 | informative | iKM | 95.8 | 0.0084 | 5.4 | 1.0 | 0.0116 | 0.137 |
| Weibull—30% | W_shape1_HR067 | informative | IPDfromKM | 100.0 | 0.0089 | 5.8 | 2.0 | 0.0118 | 0.184 |
| Weibull—30% | W_shape1_HR067 | informative | KMtoIPD | 100.0 | 0.0083 | 0.0 | 3.0 | 0.0113 | 0.132 |
| Weibull—30% | W_shape1_HR067 | informative | QP | 88.4 | 0.0085 | 1.2 | 0.0 | 0.0159 | 0.242 |
| Weibull—30% | W_shape1_HR085 | random | iKM | 97.8 | 0.0094 | 2.3 | 0.0 | 0.0072 | 0.107 |
| Weibull—30% | W_shape1_HR085 | random | IPDfromKM | 100.0 | 0.0106 | 1.7 | 0.0 | 0.0074 | 0.139 |
| Weibull—30% | W_shape1_HR085 | random | KMtoIPD | 100.0 | 0.0098 | 0.3 | 3.0 | 0.0069 | 0.112 |
| Weibull—30% | W_shape1_HR085 | random | QP | 91.2 | 0.0112 | 1.4 | 0.0 | 0.0105 | 0.134 |
| Weibull—30% | W_shape1_HR085 | exp | iKM | 93.5 | 0.0093 | 3.1 | 1.0 | 0.0084 | 0.148 |
| Weibull—30% | W_shape1_HR085 | exp | IPDfromKM | 100.0 | 0.0097 | 2.9 | 0.0 | 0.0081 | 0.164 |
| Weibull—30% | W_shape1_HR085 | exp | KMtoIPD | 100.0 | 0.0095 | 0.0 | 3.0 | 0.0080 | 0.139 |

| Scenario category | Scenario ID | Censoring | Method | Valid IPD (%) | KM RMSE | Relative event deviation (%) | Maximum risk-set deviation | Absolute Cox logHR error | Absolute RMST contrast error |
|---|---|---|---|---|---|---|---|---|---|
| Weibull—30% | W_shape1_HR085 | exp | QP | 90.8 | 0.0095 | 1.1 | 0.0 | 0.0103 | 0.167 |
| Weibull—30% | W_shape1_HR085 | front | iKM | 93.3 | 0.0089 | 4.9 | 1.0 | 0.0096 | 0.148 |
| Weibull—30% | W_shape1_HR085 | front | IPDfromKM | 100.0 | 0.0092 | 4.6 | 1.0 | 0.0088 | 0.186 |
| Weibull—30% | W_shape1_HR085 | front | KMtoIPD | 100.0 | 0.0090 | 0.0 | 2.0 | 0.0084 | 0.143 |
| Weibull—30% | W_shape1_HR085 | front | QP | 90.6 | 0.0088 | 1.1 | 0.0 | 0.0123 | 0.187 |
| Weibull—30% | W_shape1_HR085 | back | iKM | 100.0 | 0.0093 | 1.1 | 0.0 | 0.0064 | 0.068 |
| Weibull—30% | W_shape1_HR085 | back | IPDfromKM | 100.0 | 0.0118 | 0.6 | 0.0 | 0.0077 | 0.110 |
| Weibull—30% | W_shape1_HR085 | back | KMtoIPD | 100.0 | 0.0095 | 0.3 | 4.0 | 0.0059 | 0.073 |
| Weibull—30% | W_shape1_HR085 | back | QP | 93.6 | 0.0123 | 2.0 | 0.0 | 0.0111 | 0.105 |
| Weibull—30% | W_shape1_HR085 | informative | iKM | 95.2 | 0.0086 | 4.9 | 1.0 | 0.0103 | 0.112 |
| Weibull—30% | W_shape1_HR085 | informative | IPDfromKM | 100.0 | 0.0091 | 5.1 | 2.0 | 0.0118 | 0.166 |
| Weibull—30% | W_shape1_HR085 | informative | KMtoIPD | 100.0 | 0.0085 | 0.0 | 3.0 | 0.0100 | 0.112 |
| Weibull—30% | W_shape1_HR085 | informative | QP | 88.4 | 0.0086 | 1.2 | 0.0 | 0.0158 | 0.198 |
| Weibull—30% | W_shape1_HR100 | random | iKM | 97.9 | 0.0094 | 2.3 | 0.0 | 0.0073 | 0.087 |
| Weibull—30% | W_shape1_HR100 | random | IPDfromKM | 100.0 | 0.0104 | 1.7 | 0.0 | 0.0071 | 0.109 |
| Weibull—30% | W_shape1_HR100 | random | KMtoIPD | 100.0 | 0.0097 | 0.3 | 3.0 | 0.0067 | 0.092 |
| Weibull—30% | W_shape1_HR100 | random | QP | 92.0 | 0.0111 | 1.4 | 0.0 | 0.0106 | 0.123 |
| Weibull—30% | W_shape1_HR100 | exp | iKM | 92.9 | 0.0091 | 3.1 | 1.0 | 0.0080 | 0.127 |
| Weibull—30% | W_shape1_HR100 | exp | IPDfromKM | 100.0 | 0.0096 | 2.9 | 0.0 | 0.0082 | 0.141 |
| Weibull—30% | W_shape1_HR100 | exp | KMtoIPD | 100.0 | 0.0095 | 0.0 | 3.0 | 0.0073 | 0.129 |
| Weibull—30% | W_shape1_HR100 | exp | QP | 90.5 | 0.0094 | 1.1 | 0.0 | 0.0108 | 0.153 |
| Weibull—30% | W_shape1_HR100 | front | iKM | 92.0 | 0.0088 | 4.9 | 1.0 | 0.0086 | 0.132 |
| Weibull—30% | W_shape1_HR100 | front | IPDfromKM | 100.0 | 0.0091 | 4.6 | 1.0 | 0.0087 | 0.158 |
| Weibull—30% | W_shape1_HR100 | front | KMtoIPD | 100.0 | 0.0090 | 0.0 | 2.0 | 0.0086 | 0.134 |
| Weibull—30% | W_shape1_HR100 | front | QP | 89.8 | 0.0089 | 1.1 | 0.0 | 0.0115 | 0.164 |
| Weibull—30% | W_shape1_HR100 | back | iKM | 100.0 | 0.0093 | 1.1 | 0.0 | 0.0070 | 0.071 |
| Weibull—30% | W_shape1_HR100 | back | IPDfromKM | 100.0 | 0.0129 | 0.6 | 0.0 | 0.0078 | 0.117 |
| Weibull—30% | W_shape1_HR100 | back | KMtoIPD | 100.0 | 0.0095 | 0.3 | 4.0 | 0.0063 | 0.083 |

| Scenario category | Scenario ID | Censoring | Method | Valid IPD (%) | KM RMSE | Relative event deviation (%) | Maximum risk-set deviation | Absolute Cox logHR error | Absolute RMST contrast error |
|---|---|---|---|---|---|---|---|---|---|
| Weibull—30% | W_shape1_HR100 | back | QP | 93.5 | 0.0121 | 2.0 | 0.0 | 0.0115 | 0.099 |
| Weibull—30% | W_shape1_HR100 | informative | iKM | 95.0 | 0.0086 | 5.0 | 1.0 | 0.0110 | 0.115 |
| Weibull—30% | W_shape1_HR100 | informative | IPDfromKM | 100.0 | 0.0091 | 5.1 | 2.0 | 0.0114 | 0.153 |
| Weibull—30% | W_shape1_HR100 | informative | KMtoIPD | 100.0 | 0.0085 | 0.0 | 3.0 | 0.0099 | 0.110 |
| Weibull—30% | W_shape1_HR100 | informative | QP | 87.3 | 0.0086 | 1.2 | 0.0 | 0.0157 | 0.178 |
| Weibull—30% | W_shape0p6_HR067 | random | iKM | 99.3 | 0.0096 | 2.0 | 0.0 | 0.0067 | 0.116 |
| Weibull—30% | W_shape0p6_HR067 | random | IPDfromKM | 100.0 | 0.0113 | 1.4 | 0.0 | 0.0071 | 0.182 |
| Weibull—30% | W_shape0p6_HR067 | random | KMtoIPD | 100.0 | 0.0101 | 0.3 | 3.0 | 0.0065 | 0.142 |
| Weibull—30% | W_shape0p6_HR067 | random | QP | 93.9 | 0.0130 | 1.7 | 0.0 | 0.0112 | 0.190 |
| Weibull—30% | W_shape0p6_HR067 | exp | iKM | 97.6 | 0.0097 | 3.1 | 0.0 | 0.0067 | 0.219 |
| Weibull—30% | W_shape0p6_HR067 | exp | IPDfromKM | 100.0 | 0.0109 | 2.3 | 0.0 | 0.0070 | 0.274 |
| Weibull—30% | W_shape0p6_HR067 | exp | KMtoIPD | 100.0 | 0.0109 | 0.3 | 2.0 | 0.0069 | 0.264 |
| Weibull—30% | W_shape0p6_HR067 | exp | QP | 94.8 | 0.0117 | 1.1 | 0.0 | 0.0104 | 0.287 |
| Weibull—30% | W_shape0p6_HR067 | front | iKM | 95.5 | 0.0095 | 6.3 | 0.0 | 0.0077 | 0.304 |
| Weibull—30% | W_shape0p6_HR067 | front | IPDfromKM | 100.0 | 0.0104 | 5.4 | 0.0 | 0.0077 | 0.340 |
| Weibull—30% | W_shape0p6_HR067 | front | KMtoIPD | 100.0 | 0.0105 | 0.3 | 2.0 | 0.0067 | 0.331 |
| Weibull—30% | W_shape0p6_HR067 | front | QP | 94.6 | 0.0108 | 0.9 | 0.0 | 0.0100 | 0.376 |
| Weibull—30% | W_shape0p6_HR067 | back | iKM | 100.0 | 0.0093 | 1.1 | 1.0 | 0.0069 | 0.067 |
| Weibull—30% | W_shape0p6_HR067 | back | IPDfromKM | 100.0 | 0.0110 | 0.6 | 1.0 | 0.0078 | 0.101 |
| Weibull—30% | W_shape0p6_HR067 | back | KMtoIPD | 100.0 | 0.0094 | 0.6 | 4.0 | 0.0068 | 0.075 |
| Weibull—30% | W_shape0p6_HR067 | back | QP | 96.3 | 0.0133 | 2.0 | 0.0 | 0.0128 | 0.109 |
| Weibull—30% | W_shape0p6_HR067 | informative | iKM | 96.5 | 0.0086 | 15.5 | 1.0 | 0.0120 | 0.410 |
| Weibull—30% | W_shape0p6_HR067 | informative | IPDfromKM | 100.0 | 0.0090 | 15.7 | 2.0 | 0.0127 | 0.553 |
| Weibull—30% | W_shape0p6_HR067 | informative | KMtoIPD | 100.0 | 0.0084 | 0.0 | 2.0 | 0.0105 | 0.409 |
| Weibull—30% | W_shape0p6_HR067 | informative | QP | 89.0 | 0.0082 | 0.9 | 1.0 | 0.0151 | 0.572 |
| Weibull—30% | W_shape2_HR067 | random | iKM | 94.0 | 0.0083 | 2.6 | 1.0 | 0.0115 | 0.058 |
| Weibull—30% | W_shape2_HR067 | random | IPDfromKM | 100.0 | 0.0086 | 2.3 | 0.0 | 0.0112 | 0.068 |
| Weibull—30% | W_shape2_HR067 | random | KMtoIPD | 100.0 | 0.0084 | 0.0 | 3.0 | 0.0108 | 0.055 |

| Scenario category | Scenario ID | Censoring | Method | Valid IPD (%) | KM RMSE | Relative event deviation (%) | Maximum risk-set deviation | Absolute Cox logHR error | Absolute RMST contrast error |
|---|---|---|---|---|---|---|---|---|---|
| Weibull—30% | W_shape2_HR067 | random | QP | 88.6 | 0.0086 | 1.4 | 0.0 | 0.0142 | 0.089 |
| Weibull—30% | W_shape2_HR067 | exp | iKM | 94.7 | 0.0081 | 2.3 | 1.0 | 0.0120 | 0.056 |
| Weibull—30% | W_shape2_HR067 | exp | IPDfromKM | 100.0 | 0.0085 | 2.6 | 1.0 | 0.0119 | 0.073 |
| Weibull—30% | W_shape2_HR067 | exp | KMtoIPD | 100.0 | 0.0081 | 0.0 | 3.0 | 0.0121 | 0.052 |
| Weibull—30% | W_shape2_HR067 | exp | QP | 91.9 | 0.0084 | 1.4 | 0.0 | 0.0169 | 0.094 |
| Weibull—30% | W_shape2_HR067 | front | iKM | 96.0 | 0.0080 | 2.6 | 1.0 | 0.0138 | 0.055 |
| Weibull—30% | W_shape2_HR067 | front | IPDfromKM | 100.0 | 0.0084 | 2.6 | 1.0 | 0.0122 | 0.074 |
| Weibull—30% | W_shape2_HR067 | front | KMtoIPD | 100.0 | 0.0080 | 0.0 | 3.0 | 0.0127 | 0.051 |
| Weibull—30% | W_shape2_HR067 | front | QP | 89.1 | 0.0082 | 1.4 | 0.0 | 0.0174 | 0.095 |
| Weibull—30% | W_shape2_HR067 | back | iKM | 99.8 | 0.0091 | 2.0 | 0.0 | 0.0086 | 0.056 |
| Weibull—30% | W_shape2_HR067 | back | IPDfromKM | 100.0 | 0.0120 | 0.9 | 0.0 | 0.0086 | 0.117 |
| Weibull—30% | W_shape2_HR067 | back | KMtoIPD | 100.0 | 0.0092 | 0.3 | 4.0 | 0.0067 | 0.066 |
| Weibull—30% | W_shape2_HR067 | back | QP | 90.1 | 0.0106 | 2.0 | 0.0 | 0.0120 | 0.087 |
| Weibull—30% | W_shape2_HR067 | informative | iKM | 97.4 | 0.0082 | 2.0 | 1.0 | 0.0123 | 0.049 |
| Weibull—30% | W_shape2_HR067 | informative | IPDfromKM | 100.0 | 0.0086 | 2.0 | 1.0 | 0.0124 | 0.062 |
| Weibull—30% | W_shape2_HR067 | informative | KMtoIPD | 100.0 | 0.0082 | 0.0 | 3.0 | 0.0131 | 0.046 |
| Weibull—30% | W_shape2_HR067 | informative | QP | 86.9 | 0.0085 | 1.5 | 0.0 | 0.0178 | 0.088 |
| Weibull—60% | W_shape1_HR067 | random | iKM | 99.5 | 0.0400 | 12.5 | 0.0 | 0.0565 | 0.210 |
| Weibull—60% | W_shape1_HR067 | random | IPDfromKM | 100.0 | 0.0425 | 10.0 | 0.0 | 0.0567 | 0.239 |
| Weibull—60% | W_shape1_HR067 | random | KMtoIPD | 100.0 | 0.0434 | 4.2 | 10.0 | 0.0602 | 0.266 |
| Weibull—60% | W_shape1_HR067 | random | QP | 88.6 | 0.0483 | 9.2 | 0.0 | 0.0512 | 0.288 |
| Weibull—60% | W_shape1_HR067 | exp | iKM | 97.5 | 0.0387 | 12.5 | 0.0 | 0.0549 | 0.382 |
| Weibull—60% | W_shape1_HR067 | exp | IPDfromKM | 100.0 | 0.0410 | 10.8 | 0.0 | 0.0546 | 0.439 |
| Weibull—60% | W_shape1_HR067 | exp | KMtoIPD | 100.0 | 0.0458 | 2.5 | 9.0 | 0.0618 | 0.510 |
| Weibull—60% | W_shape1_HR067 | exp | QP | 90.2 | 0.0455 | 5.8 | 0.0 | 0.0425 | 0.463 |
| Weibull—60% | W_shape1_HR067 | front | iKM | 95.0 | 0.0384 | 15.0 | 0.0 | 0.0523 | 0.455 |
| Weibull—60% | W_shape1_HR067 | front | IPDfromKM | 100.0 | 0.0408 | 13.3 | 0.0 | 0.0537 | 0.541 |
| Weibull—60% | W_shape1_HR067 | front | KMtoIPD | 100.0 | 0.0492 | 2.5 | 8.0 | 0.0606 | 0.698 |

| Scenario category | Scenario ID | Censoring | Method | Valid IPD (%) | KM RMSE | Relative event deviation (%) | Maximum risk-set deviation | Absolute Cox logHR error | Absolute RMST contrast error |
|---|---|---|---|---|---|---|---|---|---|
| Weibull—60% | W_shape1_HR067 | front | QP | 90.8 | 0.0423 | 5.0 | 0.0 | 0.0409 | 0.561 |
| Weibull—60% | W_shape1_HR067 | back | iKM | 100.0 | 0.0398 | 10.8 | 5.0 | 0.0631 | 0.228 |
| Weibull—60% | W_shape1_HR067 | back | IPDfromKM | 100.0 | 0.0412 | 9.2 | 5.0 | 0.0699 | 0.285 |
| Weibull—60% | W_shape1_HR067 | back | KMtoIPD | 100.0 | 0.0406 | 6.7 | 12.0 | 0.0690 | 0.248 |
| Weibull—60% | W_shape1_HR067 | back | QP | 94.6 | 0.0440 | 9.2 | 0.0 | 0.0600 | 0.273 |
| Weibull—60% | W_shape1_HR067 | informative | iKM | 94.9 | 0.0356 | 17.5 | 1.0 | 0.0709 | 0.931 |
| Weibull—60% | W_shape1_HR067 | informative | IPDfromKM | 100.0 | 0.0374 | 16.2 | 1.0 | 0.0685 | 0.981 |
| Weibull—60% | W_shape1_HR067 | informative | KMtoIPD | 100.0 | 0.0370 | 0.8 | 7.0 | 0.0795 | 0.987 |
| Weibull—60% | W_shape1_HR067 | informative | QP | 88.8 | 0.0347 | 5.0 | 0.0 | 0.0473 | 0.871 |
| Weibull—60% | W_shape1_HR100 | random | iKM | 99.9 | 0.0393 | 12.5 | 0.0 | 0.0465 | 0.273 |
| Weibull—60% | W_shape1_HR100 | random | IPDfromKM | 100.0 | 0.0437 | 10.0 | 0.0 | 0.0466 | 0.347 |
| Weibull—60% | W_shape1_HR100 | random | KMtoIPD | 100.0 | 0.0430 | 4.2 | 10.0 | 0.0465 | 0.335 |
| Weibull—60% | W_shape1_HR100 | random | QP | 91.3 | 0.0457 | 6.7 | 0.0 | 0.0535 | 0.294 |
| Weibull—60% | W_shape1_HR100 | exp | iKM | 98.6 | 0.0392 | 13.3 | 0.0 | 0.0413 | 0.325 |
| Weibull—60% | W_shape1_HR100 | exp | IPDfromKM | 100.0 | 0.0424 | 10.8 | 0.0 | 0.0455 | 0.366 |
| Weibull—60% | W_shape1_HR100 | exp | KMtoIPD | 100.0 | 0.0475 | 2.5 | 8.0 | 0.0477 | 0.469 |
| Weibull—60% | W_shape1_HR100 | exp | QP | 92.2 | 0.0446 | 5.0 | 0.0 | 0.0440 | 0.354 |
| Weibull—60% | W_shape1_HR100 | front | iKM | 95.6 | 0.0386 | 15.0 | 0.0 | 0.0472 | 0.392 |
| Weibull—60% | W_shape1_HR100 | front | IPDfromKM | 100.0 | 0.0411 | 13.3 | 0.0 | 0.0466 | 0.457 |
| Weibull—60% | W_shape1_HR100 | front | KMtoIPD | 100.0 | 0.0509 | 1.7 | 8.0 | 0.0547 | 0.643 |
| Weibull—60% | W_shape1_HR100 | front | QP | 91.7 | 0.0430 | 5.0 | 0.0 | 0.0403 | 0.432 |
| Weibull—60% | W_shape1_HR100 | back | iKM | 100.0 | 0.0426 | 11.7 | 7.0 | 0.0516 | 0.108 |
| Weibull—60% | W_shape1_HR100 | back | IPDfromKM | 100.0 | 0.0432 | 10.0 | 7.0 | 0.0540 | 0.120 |
| Weibull—60% | W_shape1_HR100 | back | KMtoIPD | 100.0 | 0.0430 | 7.5 | 13.0 | 0.0509 | 0.112 |
| Weibull—60% | W_shape1_HR100 | back | QP | 90.8 | 0.0472 | 13.3 | 0.0 | 0.0721 | 0.142 |
| Weibull—60% | W_shape1_HR100 | informative | iKM | 94.0 | 0.0367 | 16.7 | 1.0 | 0.0536 | 0.569 |
| Weibull—60% | W_shape1_HR100 | informative | IPDfromKM | 100.0 | 0.0383 | 15.9 | 1.0 | 0.0536 | 0.631 |
| Weibull—60% | W_shape1_HR100 | informative | KMtoIPD | 100.0 | 0.0383 | 0.8 | 7.0 | 0.0564 | 0.587 |

| Scenario category | Scenario ID | Censoring | Method | Valid IPD (%) | KM RMSE | Relative event deviation (%) | Maximum risk-set deviation | Absolute Cox logHR error | Absolute RMST contrast error |
|---|---|---|---|---|---|---|---|---|---|
| Weibull—60% | W_shape1_HR100 | informative | QP | 86.3 | 0.0361 | 5.0 | 0.0 | 0.0475 | 0.595 |
| Piecewise—30% | PH_H1_HR067 | random | iKM | 100.0 | 0.0092 | 1.7 | 2.0 | 0.0085 | 0.032 |
| Piecewise—30% | PH_H1_HR067 | random | IPDfromKM | 100.0 | 0.0092 | 1.7 | 1.0 | 0.0083 | 0.030 |
| Piecewise—30% | PH_H1_HR067 | random | KMtoIPD | 100.0 | 0.0092 | 0.9 | 4.0 | 0.0077 | 0.030 |
| Piecewise—30% | PH_H1_HR067 | random | QP | 100.0 | 0.0121 | 2.9 | 0.0 | 0.0155 | 0.100 |
| Piecewise—30% | PH_H1_HR067 | front | iKM | 100.0 | 0.0092 | 1.7 | 3.0 | 0.0088 | 0.028 |
| Piecewise—30% | PH_H1_HR067 | front | IPDfromKM | 100.0 | 0.0092 | 1.7 | 2.0 | 0.0082 | 0.028 |
| Piecewise—30% | PH_H1_HR067 | front | KMtoIPD | 100.0 | 0.0092 | 0.9 | 4.0 | 0.0085 | 0.028 |
| Piecewise—30% | PH_H1_HR067 | front | QP | 100.0 | 0.0118 | 2.6 | 0.0 | 0.0165 | 0.098 |
| Piecewise—30% | PH_H1_HR067 | back | iKM | 100.0 | 0.0092 | 1.4 | 2.0 | 0.0069 | 0.029 |
| Piecewise—30% | PH_H1_HR067 | back | IPDfromKM | 100.0 | 0.0093 | 0.9 | 2.0 | 0.0071 | 0.030 |
| Piecewise—30% | PH_H1_HR067 | back | KMtoIPD | 100.0 | 0.0092 | 0.6 | 4.0 | 0.0068 | 0.030 |
| Piecewise—30% | PH_H1_HR067 | back | QP | 99.9 | 0.0124 | 2.9 | 0.0 | 0.0146 | 0.098 |
| Piecewise—30% | PH_H2_HR085 | random | iKM | 100.0 | 0.0091 | 1.7 | 1.0 | 0.0078 | 0.028 |
| Piecewise—30% | PH_H2_HR085 | random | IPDfromKM | 100.0 | 0.0091 | 1.7 | 0.0 | 0.0079 | 0.028 |
| Piecewise—30% | PH_H2_HR085 | random | KMtoIPD | 100.0 | 0.0091 | 0.6 | 4.0 | 0.0079 | 0.028 |
| Piecewise—30% | PH_H2_HR085 | random | QP | 100.0 | 0.0114 | 2.6 | 0.0 | 0.0135 | 0.082 |
| Piecewise—30% | PH_H2_HR085 | front | iKM | 100.0 | 0.0091 | 1.7 | 2.0 | 0.0083 | 0.028 |
| Piecewise—30% | PH_H2_HR085 | front | IPDfromKM | 100.0 | 0.0090 | 1.7 | 1.0 | 0.0086 | 0.028 |
| Piecewise—30% | PH_H2_HR085 | front | KMtoIPD | 100.0 | 0.0091 | 0.9 | 4.0 | 0.0077 | 0.028 |
| Piecewise—30% | PH_H2_HR085 | front | QP | 100.0 | 0.0113 | 2.6 | 0.0 | 0.0149 | 0.089 |
| Piecewise—30% | PH_H2_HR085 | back | iKM | 100.0 | 0.0092 | 1.4 | 1.0 | 0.0071 | 0.031 |
| Piecewise—30% | PH_H2_HR085 | back | IPDfromKM | 100.0 | 0.0098 | 0.6 | 1.0 | 0.0073 | 0.036 |
| Piecewise—30% | PH_H2_HR085 | back | KMtoIPD | 100.0 | 0.0092 | 0.3 | 4.0 | 0.0061 | 0.031 |
| Piecewise—30% | PH_H2_HR085 | back | QP | 92.8 | 0.0117 | 2.6 | 0.0 | 0.0128 | 0.088 |
| Piecewise—30% | PH_H2_HR100 | random | iKM | 100.0 | 0.0090 | 1.7 | 1.0 | 0.0076 | 0.028 |
| Piecewise—30% | PH_H2_HR100 | random | IPDfromKM | 100.0 | 0.0091 | 1.7 | 0.0 | 0.0075 | 0.028 |
| Piecewise—30% | PH_H2_HR100 | random | KMtoIPD | 100.0 | 0.0091 | 0.6 | 4.0 | 0.0069 | 0.028 |

| Scenario category | Scenario ID | Censoring | Method | Valid IPD (%) | KM RMSE | Relative event deviation (%) | Maximum risk-set deviation | Absolute Cox logHR error | Absolute RMST contrast error |
|---|---|---|---|---|---|---|---|---|---|
| Piecewise—30% | PH_H2_HR100 | random | QP | 100.0 | 0.0113 | 2.6 | 0.0 | 0.0131 | 0.083 |
| Piecewise—30% | PH_H2_HR100 | front | iKM | 100.0 | 0.0090 | 2.0 | 2.0 | 0.0083 | 0.028 |
| Piecewise—30% | PH_H2_HR100 | front | IPDfromKM | 100.0 | 0.0090 | 1.7 | 1.0 | 0.0083 | 0.028 |
| Piecewise—30% | PH_H2_HR100 | front | KMtoIPD | 100.0 | 0.0090 | 0.6 | 4.0 | 0.0079 | 0.028 |
| Piecewise—30% | PH_H2_HR100 | front | QP | 100.0 | 0.0110 | 2.3 | 0.0 | 0.0129 | 0.082 |
| Piecewise—30% | PH_H2_HR100 | back | iKM | 100.0 | 0.0093 | 1.4 | 1.0 | 0.0070 | 0.036 |
| Piecewise—30% | PH_H2_HR100 | back | IPDfromKM | 100.0 | 0.0103 | 0.6 | 1.0 | 0.0076 | 0.048 |
| Piecewise—30% | PH_H2_HR100 | back | KMtoIPD | 100.0 | 0.0093 | 0.3 | 4.0 | 0.0059 | 0.042 |
| Piecewise—30% | PH_H2_HR100 | back | QP | 87.6 | 0.0118 | 2.6 | 0.0 | 0.0131 | 0.092 |
| Piecewise—30% | NPHdelayed_H1_D1 | random | iKM | 100.0 | 0.0091 | 1.7 | 1.0 | 0.0080 | 0.036 |
| Piecewise—30% | NPHdelayed_H1_D1 | random | IPDfromKM | 100.0 | 0.0091 | 1.4 | 0.0 | 0.0081 | 0.034 |
| Piecewise—30% | NPHdelayed_H1_D1 | random | KMtoIPD | 100.0 | 0.0091 | 0.6 | 4.0 | 0.0073 | 0.035 |
| Piecewise—30% | NPHdelayed_H1_D1 | random | QP | 100.0 | 0.0123 | 2.6 | 0.0 | 0.0137 | 0.093 |
| Piecewise—30% | NPHdelayed_H1_D1 | front | iKM | 100.0 | 0.0090 | 2.0 | 2.0 | 0.0083 | 0.031 |
| Piecewise—30% | NPHdelayed_H1_D1 | front | IPDfromKM | 100.0 | 0.0090 | 1.7 | 1.0 | 0.0080 | 0.032 |
| Piecewise—30% | NPHdelayed_H1_D1 | front | KMtoIPD | 100.0 | 0.0090 | 0.9 | 4.0 | 0.0077 | 0.033 |
| Piecewise—30% | NPHdelayed_H1_D1 | front | QP | 100.0 | 0.0118 | 2.6 | 0.0 | 0.0132 | 0.088 |
| Piecewise—30% | NPHdelayed_H1_D1 | back | iKM | 100.0 | 0.0093 | 1.4 | 1.0 | 0.0067 | 0.043 |
| Piecewise—30% | NPHdelayed_H1_D1 | back | IPDfromKM | 100.0 | 0.0102 | 0.6 | 1.0 | 0.0078 | 0.055 |
| Piecewise—30% | NPHdelayed_H1_D1 | back | KMtoIPD | 100.0 | 0.0094 | 0.3 | 4.0 | 0.0063 | 0.046 |
| Piecewise—30% | NPHdelayed_H1_D1 | back | QP | 93.0 | 0.0127 | 2.6 | 0.0 | 0.0123 | 0.092 |
| Piecewise—30% | NPHdelayed_H2_D2 | random | iKM | 100.0 | 0.0091 | 1.7 | 1.0 | 0.0076 | 0.028 |
| Piecewise—30% | NPHdelayed_H2_D2 | random | IPDfromKM | 100.0 | 0.0091 | 1.7 | 0.0 | 0.0072 | 0.028 |
| Piecewise—30% | NPHdelayed_H2_D2 | random | KMtoIPD | 100.0 | 0.0091 | 0.6 | 4.0 | 0.0076 | 0.029 |
| Piecewise—30% | NPHdelayed_H2_D2 | random | QP | 100.0 | 0.0116 | 2.6 | 0.0 | 0.0127 | 0.082 |
| Piecewise—30% | NPHdelayed_H2_D2 | front | iKM | 100.0 | 0.0090 | 1.7 | 2.0 | 0.0087 | 0.028 |
| Piecewise—30% | NPHdelayed_H2_D2 | front | IPDfromKM | 100.0 | 0.0090 | 1.7 | 1.0 | 0.0087 | 0.028 |
| Piecewise—30% | NPHdelayed_H2_D2 | front | KMtoIPD | 100.0 | 0.0091 | 0.9 | 4.0 | 0.0079 | 0.028 |

| Scenario category | Scenario ID | Censoring | Method | Valid IPD (%) | KM RMSE | Relative event deviation (%) | Maximum risk-set deviation | Absolute Cox logHR error | Absolute RMST contrast error |
|---|---|---|---|---|---|---|---|---|---|
| Piecewise—30% | NPHdelayed_H2_D2 | front | QP | 100.0 | 0.0114 | 2.3 | 0.0 | 0.0136 | 0.088 |
| Piecewise—30% | NPHdelayed_H2_D2 | back | iKM | 100.0 | 0.0093 | 1.4 | 1.0 | 0.0068 | 0.034 |
| Piecewise—30% | NPHdelayed_H2_D2 | back | IPDfromKM | 100.0 | 0.0098 | 0.6 | 1.0 | 0.0073 | 0.040 |
| Piecewise—30% | NPHdelayed_H2_D2 | back | KMtoIPD | 100.0 | 0.0093 | 0.3 | 4.0 | 0.0065 | 0.035 |
| Piecewise—30% | NPHdelayed_H2_D2 | back | QP | 92.9 | 0.0121 | 2.6 | 0.0 | 0.0121 | 0.087 |
| Piecewise—30% | NPHcross_H1_C1 | random | iKM | 100.0 | 0.0091 | 1.7 | 0.0 | 0.0077 | 0.035 |
| Piecewise—30% | NPHcross_H1_C1 | random | IPDfromKM | 100.0 | 0.0091 | 1.4 | 0.0 | 0.0076 | 0.035 |
| Piecewise—30% | NPHcross_H1_C1 | random | KMtoIPD | 100.0 | 0.0090 | 0.6 | 4.0 | 0.0070 | 0.035 |
| Piecewise—30% | NPHcross_H1_C1 | random | QP | 100.0 | 0.0123 | 2.6 | 0.0 | 0.0121 | 0.091 |
| Piecewise—30% | NPHcross_H1_C1 | front | iKM | 100.0 | 0.0090 | 2.0 | 2.0 | 0.0091 | 0.035 |
| Piecewise—30% | NPHcross_H1_C1 | front | IPDfromKM | 100.0 | 0.0091 | 2.0 | 1.0 | 0.0093 | 0.036 |
| Piecewise—30% | NPHcross_H1_C1 | front | KMtoIPD | 100.0 | 0.0090 | 0.9 | 4.0 | 0.0078 | 0.034 |
| Piecewise—30% | NPHcross_H1_C1 | front | QP | 100.0 | 0.0119 | 2.3 | 0.0 | 0.0131 | 0.095 |
| Piecewise—30% | NPHcross_H1_C1 | back | iKM | 100.0 | 0.0093 | 1.4 | 1.0 | 0.0073 | 0.050 |
| Piecewise—30% | NPHcross_H1_C1 | back | IPDfromKM | 100.0 | 0.0106 | 0.6 | 1.0 | 0.0113 | 0.073 |
| Piecewise—30% | NPHcross_H1_C1 | back | KMtoIPD | 100.0 | 0.0094 | 0.6 | 4.0 | 0.0071 | 0.056 |
| Piecewise—30% | NPHcross_H1_C1 | back | QP | 94.0 | 0.0127 | 2.3 | 0.0 | 0.0125 | 0.100 |
| Piecewise—60% | NPHdelayed_H1_D1 | random | iKM | 100.0 | 0.0408 | 14.5 | 0.0 | 0.0301 | 0.423 |
| Piecewise—60% | NPHdelayed_H1_D1 | random | IPDfromKM | 100.0 | 0.0430 | 13.0 | 0.0 | 0.0306 | 0.554 |
| Piecewise—60% | NPHdelayed_H1_D1 | random | KMtoIPD | 100.0 | 0.0434 | 5.5 | 14.0 | 0.0315 | 0.560 |
| Piecewise—60% | NPHdelayed_H1_D1 | random | QP | 93.5 | 0.0423 | 4.5 | 0.0 | 0.0398 | 0.453 |
| Piecewise—60% | NPHdelayed_H1_D1 | front | iKM | 98.9 | 0.0418 | 17.0 | 0.0 | 0.0265 | 0.253 |
| Piecewise—60% | NPHdelayed_H1_D1 | front | IPDfromKM | 100.0 | 0.0429 | 15.5 | 0.0 | 0.0266 | 0.269 |
| Piecewise—60% | NPHdelayed_H1_D1 | front | KMtoIPD | 100.0 | 0.0454 | 3.5 | 13.0 | 0.0336 | 0.336 |
| Piecewise—60% | NPHdelayed_H1_D1 | front | QP | 94.1 | 0.0457 | 5.5 | 0.0 | 0.0398 | 0.343 |
| Piecewise—60% | NPHdelayed_H1_D1 | back | iKM | 100.0 | 0.0388 | 13.0 | 0.0 | 0.0374 | 0.490 |
| Piecewise—60% | NPHdelayed_H1_D1 | back | IPDfromKM | 100.0 | 0.0371 | 9.5 | 0.0 | 0.0390 | 0.551 |
| Piecewise—60% | NPHdelayed_H1_D1 | back | KMtoIPD | 100.0 | 0.0397 | 8.5 | 16.0 | 0.0363 | 0.508 |

| Scenario category | Scenario ID | Censoring | Method | Valid IPD (%) | KM RMSE | Relative event deviation (%) | Maximum risk-set deviation | Absolute Cox logHR error | Absolute RMST contrast error |
|---|---|---|---|---|---|---|---|---|---|
| Piecewise—60% | NPHdelayed_H1_D1 | back | QP | 95.9 | 0.0333 | 3.5 | 0.0 | 0.0443 | 0.457 |
| Piecewise—60% | NPHdelayed_H1_D1 | informative | iKM | 99.9 | 0.0423 | 34.6 | 8.0 | 0.0650 | 0.177 |
| Piecewise—60% | NPHdelayed_H1_D1 | informative | IPDfromKM | 100.0 | 0.0424 | 34.3 | 8.0 | 0.0632 | 0.180 |
| Piecewise—60% | NPHdelayed_H1_D1 | informative | KMtoIPD | 100.0 | 0.0425 | 5.1 | 8.0 | 0.0393 | 0.177 |
| Piecewise—60% | NPHdelayed_H1_D1 | informative | QP | 100.0 | 0.0256 | 2.0 | 3.0 | 0.0256 | 0.204 |
| Piecewise—60% | NPHcross_H1_C1 | random | iKM | 99.9 | 0.0398 | 15.0 | 0.0 | 0.0436 | 0.434 |
| Piecewise—60% | NPHcross_H1_C1 | random | IPDfromKM | 100.0 | 0.0421 | 13.0 | 0.0 | 0.0385 | 0.587 |
| Piecewise—60% | NPHcross_H1_C1 | random | KMtoIPD | 100.0 | 0.0429 | 5.5 | 14.0 | 0.0369 | 0.610 |
| Piecewise—60% | NPHcross_H1_C1 | random | QP | 93.0 | 0.0411 | 4.0 | 0.0 | 0.0399 | 0.459 |
| Piecewise—60% | NPHcross_H1_C1 | front | iKM | 98.3 | 0.0419 | 17.5 | 0.0 | 0.0338 | 0.279 |
| Piecewise—60% | NPHcross_H1_C1 | front | IPDfromKM | 100.0 | 0.0431 | 16.0 | 0.0 | 0.0318 | 0.292 |
| Piecewise—60% | NPHcross_H1_C1 | front | KMtoIPD | 100.0 | 0.0459 | 3.5 | 13.0 | 0.0356 | 0.386 |
| Piecewise—60% | NPHcross_H1_C1 | front | QP | 92.3 | 0.0444 | 5.0 | 0.0 | 0.0412 | 0.365 |
| Piecewise—60% | NPHcross_H1_C1 | back | iKM | 100.0 | 0.0384 | 13.0 | 0.0 | 0.0621 | 0.484 |
| Piecewise—60% | NPHcross_H1_C1 | back | IPDfromKM | 100.0 | 0.0367 | 10.0 | 0.0 | 0.0537 | 0.570 |
| Piecewise—60% | NPHcross_H1_C1 | back | KMtoIPD | 100.0 | 0.0389 | 8.0 | 15.0 | 0.0562 | 0.508 |
| Piecewise—60% | NPHcross_H1_C1 | back | QP | 96.1 | 0.0341 | 3.5 | 0.0 | 0.0474 | 0.489 |
| Piecewise—60% | NPHcross_H1_C1 | informative | iKM | 100.0 | 0.0421 | 35.3 | 8.0 | 0.1366 | 0.205 |
| Piecewise—60% | NPHcross_H1_C1 | informative | IPDfromKM | 100.0 | 0.0421 | 35.0 | 8.0 | 0.1276 | 0.210 |
| Piecewise—60% | NPHcross_H1_C1 | informative | KMtoIPD | 100.0 | 0.0420 | 4.8 | 8.0 | 0.0475 | 0.216 |
| Piecewise—60% | NPHcross_H1_C1 | informative | QP | 100.0 | 0.0254 | 2.2 | 3.0 | 0.0289 | 0.245 |

Note: Each row summarizes 1,000 assigned replicates for one setting-method combination. Censoring percentages in the scenario labels are target proportions. Valid reconstruction uses all assigned runs and is reported descriptively; accuracy outcomes summarize valid reconstructed datasets. Runtime is omitted because the four final implementations were not timed under one common hardware and execution protocol. Valid IPD denotes the percentage of assigned runs producing a valid reconstructed dataset. Risk-set deviation is the maximum absolute number-at-risk difference.

**Table S3.** Method-specific results from the dedicated one-factor experiments

| Factor | Level | Method | Valid runs | KM RMSE | Event deviation (%) | Cox logHR error | RMST contrast error |
|---|---|---|---|---|---|---|---|
| Target censoring proportion | 0.1 | iKM | 280 | 0.007956 [0.00946] | 2.222 [3.333] | 0.01396 [0.03324] | 0.1669 [0.4237] |
| Target censoring proportion | 0.1 | IPDfromKM | 300 | 0.008162 [0.01008] | 2.556 [4.444] | 0.01213 [0.03219] | 0.1883 [0.5199] |
| Target censoring proportion | 0.1 | KMtoIPD | 300 | 0.007865 [0.009406] | 0 [0.2222] | 0.01406 [0.0335] | 0.1699 [0.3968] |
| Target censoring proportion | 0.1 | QP | 273 | 0.00736 [0.009562] | 1.111 [1.956] | 0.01574 [0.05371] | 0.2619 [0.7228] |
| Target censoring proportion | 0.3 | iKM | 287 | 0.009359 [0.01194] | 2.286 [3.429] | 0.008146 [0.01986] | 0.1025 [0.2751] |
| Target censoring proportion | 0.3 | IPDfromKM | 300 | 0.01057 [0.01421] | 1.714 [2.857] | 0.008054 [0.02037] | 0.1482 [0.4001] |
| Target censoring proportion | 0.3 | KMtoIPD | 300 | 0.009627 [0.01346] | 0.2857 [0.5714] | 0.007204 [0.01796] | 0.1142 [0.3173] |
| Target censoring proportion | 0.3 | QP | 275 | 0.01109 [0.01507] | 1.714 [2.286] | 0.01045 [0.02827] | 0.1652 [0.4257] |
| Target censoring proportion | 0.45 | iKM | 300 | 0.009812 [0.01239] | 1.818 [2.909] | 0.007218 [0.01858] | 0.07696 [0.198] |
| Target censoring proportion | 0.45 | IPDfromKM | 300 | 0.01339 [0.02197] | 0.7273 [1.818] | 0.007546 [0.01997] | 0.1336 [0.3914] |
| Target censoring proportion | 0.45 | KMtoIPD | 300 | 0.01039 [0.01664] | 0.3636 [0.7273] | 0.007913 [0.01596] | 0.1031 [0.3107] |
| Target censoring proportion | 0.45 | QP | 279 | 0.01394 [0.01929] | 2.182 [3.273] | 0.0114 [0.03397] | 0.1131 [0.3491] |
| Target censoring proportion | 0.6 | iKM | 300 | 0.0103 [0.01371] | 2 [3] | 0.009305 [0.02338] | 0.06016 [0.1311] |
| Target censoring proportion | 0.6 | IPDfromKM | 300 | 0.01531 [0.03307] | 0.5 [1.5] | 0.01027 [0.02685] | 0.1253 [0.3449] |
| Target censoring proportion | 0.6 | KMtoIPD | 300 | 0.01061 [0.01683] | 0.5 [1] | 0.009001 [0.01936] | 0.07385 [0.1965] |
| Target censoring proportion | 0.6 | QP | 291 | 0.01563 [0.02213] | 3 [4.5] | 0.01468 [0.03787] | 0.08732 [0.2011] |
| Target censoring proportion | 0.7 | iKM | 300 | 0.01024 [0.01541] | 2 [4] | 0.01219 [0.03322] | 0.0909 [0.2409] |
| Target censoring proportion | 0.7 | IPDfromKM | 300 | 0.02246 [0.05331] | 0.6667 [2.667] | 0.01452 [0.0428] | 0.2549 [0.8512] |
| Target censoring proportion | 0.7 | KMtoIPD | 300 | 0.01057 [0.02089] | 0.6667 [1.333] | 0.0121 [0.02687] | 0.1237 [0.3605] |
| Target censoring proportion | 0.7 | QP | 295 | 0.01502 [0.02565] | 2.667 [4] | 0.01749 [0.04646] | 0.1015 [0.3042] |
| Survival-probability error SD | 0 | iKM | 300 | 0.002082 [0.003648] | 0.8571 [1.714] | 0.003731 [0.01054] | 0.047 [0.1156] |
| Survival-probability error SD | 0 | IPDfromKM | 300 | 0.004443 [0.01072] | 0.2857 [1.143] | 0.004202 [0.009934] | 0.09559 [0.3335] |
| Survival-probability error SD | 0 | KMtoIPD | 300 | 0 [0] | 0 [0] | 0 [0] | 0 [0] |
| Survival-probability error SD | 0 | QP | 300 | 0 [0] | 0 [0] | 0 [0] | 0 [0] |

| Factor | Level | Method | Valid runs | KM RMSE | Event deviation (%) | Cox logHR error | RMST contrast error |
|---|---|---|---|---|---|---|---|
| Survival-probability error SD | 0.005 | iKM | 299 | 0.005434 [0.007194] | 1.143 [2.286] | 0.00468 [0.01261] | 0.07004 [0.1793] |
| Survival-probability error SD | 0.005 | IPDfromKM | 300 | 0.006898 [0.01104] | 0.5714 [1.429] | 0.005665 [0.01311] | 0.1226 [0.326] |
| Survival-probability error SD | 0.005 | KMtoIPD | 300 | 0.004552 [0.006554] | 0 [0.2857] | 0.003144 [0.009405] | 0.02649 [0.163] |
| Survival-probability error SD | 0.005 | QP | 289 | 0.00672 [0.009991] | 0.8571 [1.429] | 0.006887 [0.01579] | 0.09658 [0.2556] |
| Survival-probability error SD | 0.01 | iKM | 289 | 0.009259 [0.01187] | 2.286 [3.714] | 0.007348 [0.02027] | 0.1032 [0.2711] |
| Survival-probability error SD | 0.01 | IPDfromKM | 300 | 0.01009 [0.01459] | 1.714 [3.143] | 0.008292 [0.02085] | 0.1323 [0.4044] |
| Survival-probability error SD | 0.01 | KMtoIPD | 300 | 0.009762 [0.01434] | 0.2857 [0.5714] | 0.006827 [0.0191] | 0.1138 [0.3295] |
| Survival-probability error SD | 0.01 | QP | 282 | 0.0111 [0.01498] | 1.429 [2.286] | 0.01062 [0.0284] | 0.1583 [0.4271] |
| Survival-probability error SD | 0.02 | iKM | 284 | 0.01892 [0.02259] | 5.714 [7.429] | 0.01393 [0.03981] | 0.1705 [0.4559] |
| Survival-probability error SD | 0.02 | IPDfromKM | 300 | 0.01964 [0.02492] | 5.143 [7.714] | 0.01617 [0.03854] | 0.1944 [0.5418] |
| Survival-probability error SD | 0.02 | KMtoIPD | 300 | 0.01968 [0.02509] | 0.5714 [1.143] | 0.01507 [0.03794] | 0.2071 [0.5247] |
| Survival-probability error SD | 0.02 | QP | 250 | 0.01877 [0.02625] | 2.286 [4] | 0.01992 [0.04662] | 0.298 [0.7344] |
| Survival-probability error SD | 0.04 | iKM | 280 | 0.04337 [0.05023] | 14.86 [18.29] | 0.03796 [0.08711] | 0.3488 [0.9071] |
| Survival-probability error SD | 0.04 | IPDfromKM | 300 | 0.04329 [0.05076] | 14 [17.46] | 0.03334 [0.07922] | 0.42 [1.136] |
| Survival-probability error SD | 0.04 | KMtoIPD | 300 | 0.04318 [0.05119] | 1.714 [2.857] | 0.02931 [0.07576] | 0.4386 [1.094] |
| Survival-probability error SD | 0.04 | QP | 252 | 0.03127 [0.04288] | 3.429 [6] | 0.03289 [0.08621] | 0.5586 [1.343] |

| Factor | Level | Method | Valid runs | KM RMSE | Event deviation (%) | Cox logHR error | RMST contrast error |
|---|---|---|---|---|---|---|---|
| Arm allocation; total N=500 | 100/400 | iKM | 289 | 0.009536 [0.01241] | 2.571 [3.429] | 0.01619 [0.03071] | 0.1986 [0.431] |
| Arm allocation; total N=500 | 100/400 | IPDfromKM | 300 | 0.01029 [0.01454] | 1.857 [3.143] | 0.01473 [0.03131] | 0.202 [0.4988] |
| Arm allocation; total N=500 | 100/400 | KMtoIPD | 300 | 0.009379 [0.01388] | 0.2857 [0.5714] | 0.01855 [0.03144] | 0.2286 [0.4125] |
| Arm allocation; total N=500 | 100/400 | QP | 275 | 0.01032 [0.01391] | 1.429 [2.286] | 0.0135 [0.03144] | 0.1783 [0.4411] |
| Arm allocation; total N=500 | 150/350 | iKM | 288 | 0.009258 [0.01168] | 2.286 [3.429] | 0.01001 [0.02428] | 0.1403 [0.3658] |
| Arm allocation; total N=500 | 150/350 | IPDfromKM | 300 | 0.01025 [0.0143] | 1.714 [3.143] | 0.01011 [0.02514] | 0.1665 [0.4386] |
| Arm allocation; total N=500 | 150/350 | KMtoIPD | 300 | 0.009308 [0.01397] | 0.2857 [0.5714] | 0.01194 [0.02164] | 0.1494 [0.4246] |
| Arm allocation; total N=500 | 150/350 | QP | 281 | 0.01071 [0.01489] | 1.429 [2.286] | 0.0128 [0.02963] | 0.142 [0.4215] |
| Arm allocation; total N=500 | 200/300 | iKM | 290 | 0.009388 [0.01208] | 2.286 [3.143] | 0.00865 [0.01993] | 0.119 [0.3239] |
| Arm allocation; total N=500 | 200/300 | IPDfromKM | 300 | 0.0106 [0.01437] | 1.714 [2.857] | 0.009076 [0.02289] | 0.1504 [0.4363] |
| Arm allocation; total N=500 | 200/300 | KMtoIPD | 300 | 0.009869 [0.01281] | 0.2857 [0.5714] | 0.009176 [0.01939] | 0.1287 [0.3566] |
| Arm allocation; total N=500 | 200/300 | QP | 275 | 0.01078 [0.01517] | 1.714 [2.571] | 0.01209 [0.03014] | 0.15 [0.4095] |
| Arm allocation; total N=500 | 250/250 | iKM | 282 | 0.00936 [0.01138] | 2.286 [3.429] | 0.00715 [0.01765] | 0.1123 [0.2729] |
| Arm allocation; total N=500 | 250/250 | IPDfromKM | 300 | 0.0103 [0.01375] | 1.714 [2.857] | 0.008226 [0.01924] | 0.1374 [0.4008] |
| Arm allocation; total N=500 | 250/250 | KMtoIPD | 300 | 0.009687 [0.01405] | 0.2857 [0.5714] | 0.007388 [0.01778] | 0.1249 [0.3284] |
| Arm allocation; total N=500 | 250/250 | QP | 271 | 0.01096 [0.01585] | 1.429 [2.286] | 0.01026 [0.0225] | 0.1598 [0.3889] |
| Balanced control/treatment sample size | 100/100 | iKM | 296 | 0.0103 [0.01424] | 1.429 [3.571] | 0.01302 [0.03319] | 0.1599 [0.4001] |
| Balanced control/treatment sample size | 100/100 | IPDfromKM | 300 | 0.01207 [0.01816] | 0.7143 [2.857] | 0.01383 [0.03671] | 0.223 [0.5745] |
| Balanced control/treatment sample size | 100/100 | KMtoIPD | 300 | 0.00935 [0.01434] | 0 [0.7143] | 0.008739 [0.02979] | 0.08319 [0.4164] |
| Balanced control/treatment sample size | 100/100 | QP | 292 | 0.01251 [0.01867] | 2.143 [2.857] | 0.01389 [0.03074] | 0.1896 [0.5241] |
| Balanced control/treatment sample size | 250/250 | iKM | 293 | 0.009637 [0.01185] | 2.286 [3.429] | 0.008288 [0.02094] | 0.1119 [0.3017] |
| Balanced control/treatment sample size | 250/250 | IPDfromKM | 300 | 0.01084 [0.01545] | 1.714 [3.143] | 0.007343 [0.02075] | 0.1532 [0.3895] |
| Balanced control/treatment sample size | 250/250 | KMtoIPD | 300 | 0.009793 [0.01408] | 0.2857 [0.5714] | 0.007331 [0.0179] | 0.121 [0.3419] |

| Factor | Level | Method | Valid runs | KM RMSE | Event deviation (%) | Cox logHR error | RMST contrast error |
|---|---|---|---|---|---|---|---|
| Balanced control/treatment sample size | 250/250 | QP | 275 | 0.01124 [0.01549] | 1.429 [2.286] | 0.01059 [0.02873] | 0.1742 [0.4538] |
| Balanced control/treatment sample size | 500/500 | iKM | 293 | 0.009784 [0.01117] | 3.143 [4] | 0.006408 [0.01524] | 0.09828 [0.232] |
| Balanced control/treatment sample size | 500/500 | IPDfromKM | 300 | 0.01045 [0.01323] | 2.714 [3.571] | 0.00626 [0.01567] | 0.1106 [0.3126] |
| Balanced control/treatment sample size | 500/500 | KMtoIPD | 300 | 0.01012 [0.0139] | 0.2857 [0.4286] | 0.005442 [0.01467] | 0.1129 [0.2958] |
| Balanced control/treatment sample size | 500/500 | QP | 254 | 0.009949 [0.01351] | 1.143 [1.857] | 0.01114 [0.02395] | 0.1404 [0.361] |

**Note:** Values are medians, with the 90th percentile in brackets. Cox and RMST errors are absolute. Each factor level used 300 assigned replicates; valid runs are the shared count used for all four outcomes in that row. One factor changed at a time from the design in S2.1.

**Table S4.** Summary of all Part II meta-analysis strategies

| Trial set | K | Missing (%) | Strategy | Reps | Success (%) | Mean trials retained | Bias, logHR | RMSE, logHR | Coverage (%) | Mean $\tau^2$ |
|---|---|---|---|---|---|---|---|---|---|---|
| 30 trials | 30 | 0 | Full AD oracle | 2700 | 100.0 | 30.0 | 0.0015 | 0.0355 | 94.2 | 0.0249 |
| 30 trials | 30 | 0 | AD only | 2700 | 100.0 | 30.0 | 0.0015 | 0.0355 | 94.2 | 0.0249 |
| 30 trials | 30 | 0 | Hybrid—iKM | 2700 | 100.0 | 30.0 | 0.0015 | 0.0355 | 94.2 | 0.0249 |
| 30 trials | 30 | 0 | Hybrid—IPDfromKM | 2700 | 100.0 | 30.0 | 0.0015 | 0.0355 | 94.2 | 0.0249 |
| 30 trials | 30 | 0 | Hybrid—KMtoIPD | 2700 | 100.0 | 30.0 | 0.0015 | 0.0355 | 94.2 | 0.0249 |
| 30 trials | 30 | 0 | Hybrid—QP | 2700 | 100.0 | 30.0 | 0.0015 | 0.0355 | 94.2 | 0.0249 |
| 30 trials | 30 | 0 | Reconstruction only—QP | 2700 | 100.0 | 27.4 | 0.0007 | 0.0376 | 93.9 | 0.0253 |
| 30 trials | 30 | 0 | Reconstruction only—IPDfromKM | 2700 | 100.0 | 30.0 | -0.0003 | 0.0358 | 93.9 | 0.0258 |
| 30 trials | 30 | 0 | Reconstruction only—KMtoIPD | 2700 | 100.0 | 30.0 | 0.0002 | 0.0356 | 94.4 | 0.0254 |
| 30 trials | 30 | 0 | Reconstruction only—iKM | 2700 | 100.0 | 29.1 | -0.0000 | 0.0362 | 94.1 | 0.0259 |
| 30 trials | 30 | 0 | Pooled true-IPD comparator | 2700 | 100.0 | 30.0 | 0.0021 | 0.0354 | 77.9 | |
| 30 trials | 30 | 20 | Full AD oracle | 2700 | 100.0 | 30.0 | 0.0015 | 0.0355 | 94.2 | 0.0249 |
| 30 trials | 30 | 20 | AD only | 2700 | 100.0 | 24.0 | 0.0014 | 0.0397 | 93.3 | 0.0248 |
| 30 trials | 30 | 20 | Hybrid—iKM | 2700 | 100.0 | 29.8 | 0.0012 | 0.0355 | 94.2 | 0.0251 |
| 30 trials | 30 | 20 | Hybrid—IPDfromKM | 2700 | 100.0 | 30.0 | 0.0011 | 0.0355 | 94.2 | 0.0251 |
| 30 trials | 30 | 20 | Hybrid—KMtoIPD | 2700 | 100.0 | 30.0 | 0.0012 | 0.0355 | 94.1 | 0.0250 |
| 30 trials | 30 | 20 | Hybrid—QP | 2700 | 100.0 | 29.5 | 0.0013 | 0.0358 | 93.9 | 0.0250 |
| 30 trials | 30 | 20 | Reconstruction only—QP | 2700 | 100.0 | 27.4 | 0.0007 | 0.0376 | 93.9 | 0.0253 |
| 30 trials | 30 | 20 | Reconstruction only—IPDfromKM | 2700 | 100.0 | 30.0 | -0.0003 | 0.0358 | 93.9 | 0.0258 |
| 30 trials | 30 | 20 | Reconstruction only—KMtoIPD | 2700 | 100.0 | 30.0 | 0.0002 | 0.0356 | 94.4 | 0.0254 |
| 30 trials | 30 | 20 | Reconstruction only—iKM | 2700 | 100.0 | 29.1 | -0.0000 | 0.0362 | 94.1 | 0.0259 |
| 30 trials | 30 | 20 | Pooled true-IPD comparator | 2700 | 100.0 | 30.0 | 0.0021 | 0.0354 | 77.9 | |
| 30 trials | 30 | 40 | Full AD oracle | 2700 | 100.0 | 30.0 | 0.0015 | 0.0355 | 94.2 | 0.0249 |
| 30 trials | 30 | 40 | AD only | 2700 | 100.0 | 17.9 | 0.0016 | 0.0457 | 93.3 | 0.0248 |
| 30 trials | 30 | 40 | Hybrid—iKM | 2700 | 100.0 | 29.6 | 0.0008 | 0.0358 | 94.2 | 0.0253 |
| 30 trials | 30 | 40 | Hybrid—IPDfromKM | 2700 | 100.0 | 30.0 | 0.0007 | 0.0356 | 94.1 | 0.0253 |
| 30 trials | 30 | 40 | Hybrid—KMtoIPD | 2700 | 100.0 | 30.0 | 0.0010 | 0.0355 | 94.3 | 0.0251 |

| Trial set | K | Missing (%) | Strategy | Reps | Success (%) | Mean trials retained | Bias, logHR | RMSE, logHR | Coverage (%) | Mean τ² |
|---|---|---|---|---|---|---|---|---|---|---|
| 30 trials | 30 | 40 | Hybrid—QP | 2700 | 100.0 | 28.9 | 0.0011 | 0.0363 | 94.0 | 0.0251 |
| 30 trials | 30 | 40 | Reconstruction only—QP | 2700 | 100.0 | 27.4 | 0.0007 | 0.0376 | 93.9 | 0.0253 |
| 30 trials | 30 | 40 | Reconstruction only—IPDfromKM | 2700 | 100.0 | 30.0 | -0.0003 | 0.0358 | 93.9 | 0.0258 |
| 30 trials | 30 | 40 | Reconstruction only—KMtoIPD | 2700 | 100.0 | 30.0 | 0.0002 | 0.0356 | 94.4 | 0.0254 |
| 30 trials | 30 | 40 | Reconstruction only—iKM | 2700 | 100.0 | 29.1 | -0.0000 | 0.0362 | 94.1 | 0.0259 |
| 30 trials | 30 | 40 | Pooled true-IPD comparator | 2700 | 100.0 | 30.0 | 0.0021 | 0.0354 | 77.9 | |
| 30 trials | 30 | 60 | Full AD oracle | 2700 | 100.0 | 30.0 | 0.0015 | 0.0355 | 94.2 | 0.0249 |
| 30 trials | 30 | 60 | AD only | 2700 | 100.0 | 12.0 | 0.0012 | 0.0565 | 93.6 | 0.0252 |
| 30 trials | 30 | 60 | Hybrid—iKM | 2700 | 100.0 | 29.4 | 0.0005 | 0.0359 | 94.2 | 0.0255 |
| 30 trials | 30 | 60 | Hybrid—IPDfromKM | 2700 | 100.0 | 30.0 | 0.0004 | 0.0356 | 94.1 | 0.0255 |
| 30 trials | 30 | 60 | Hybrid—KMtoIPD | 2700 | 100.0 | 30.0 | 0.0007 | 0.0355 | 94.3 | 0.0252 |
| 30 trials | 30 | 60 | Hybrid—QP | 2700 | 100.0 | 28.4 | 0.0008 | 0.0367 | 93.9 | 0.0251 |
| 30 trials | 30 | 60 | Reconstruction only—QP | 2700 | 100.0 | 27.4 | 0.0007 | 0.0376 | 93.9 | 0.0253 |
| 30 trials | 30 | 60 | Reconstruction only—IPDfromKM | 2700 | 100.0 | 30.0 | -0.0003 | 0.0358 | 93.9 | 0.0258 |
| 30 trials | 30 | 60 | Reconstruction only—KMtoIPD | 2700 | 100.0 | 30.0 | 0.0002 | 0.0356 | 94.4 | 0.0254 |
| 30 trials | 30 | 60 | Reconstruction only—iKM | 2700 | 100.0 | 29.1 | -0.0000 | 0.0362 | 94.1 | 0.0259 |
| 30 trials | 30 | 60 | Pooled true-IPD comparator | 2700 | 100.0 | 30.0 | 0.0021 | 0.0354 | 77.9 | |
| 30 trials | 30 | 80 | Full AD oracle | 2700 | 100.0 | 30.0 | 0.0015 | 0.0355 | 94.2 | 0.0249 |
| 30 trials | 30 | 80 | AD only | 2700 | 98.7 | 6.0 | 0.0008 | 0.0852 | 90.4 | 0.0258 |
| 30 trials | 30 | 80 | Hybrid—iKM | 2700 | 100.0 | 29.3 | 0.0003 | 0.0360 | 94.2 | 0.0257 |
| 30 trials | 30 | 80 | Hybrid—IPDfromKM | 2700 | 100.0 | 30.0 | 0.0001 | 0.0357 | 94.0 | 0.0257 |
| 30 trials | 30 | 80 | Hybrid—KMtoIPD | 2700 | 100.0 | 30.0 | 0.0005 | 0.0356 | 94.3 | 0.0253 |
| 30 trials | 30 | 80 | Hybrid—QP | 2700 | 100.0 | 27.9 | 0.0007 | 0.0371 | 93.7 | 0.0252 |
| 30 trials | 30 | 80 | Reconstruction only—QP | 2700 | 100.0 | 27.4 | 0.0007 | 0.0376 | 93.9 | 0.0253 |
| 30 trials | 30 | 80 | Reconstruction only—IPDfromKM | 2700 | 100.0 | 30.0 | -0.0003 | 0.0358 | 93.9 | 0.0258 |
| 30 trials | 30 | 80 | Reconstruction only—KMtoIPD | 2700 | 100.0 | 30.0 | 0.0002 | 0.0356 | 94.4 | 0.0254 |
| 30 trials | 30 | 80 | Reconstruction only—iKM | 2700 | 100.0 | 29.1 | -0.0000 | 0.0362 | 94.1 | 0.0259 |

| Trial set | K | Missing (%) | Strategy | Reps | Success (%) | Mean trials retained | Bias, logHR | RMSE, logHR | Coverage (%) | Mean $\tau^2$ |
|---|---|---|---|---|---|---|---|---|---|---|
| 30 trials | 30 | 80 | Pooled true-IPD comparator | 2700 | 100.0 | 30.0 | 0.0021 | 0.0354 | 77.9 | |
| 10/15 trials | 10 | 60 | Full AD oracle | 1200 | 100.0 | 10.0 | -0.0016 | 0.1074 | 92.0 | 0.0660 |
| 10/15 trials | 10 | 60 | AD only | 1200 | 95.7 | 4.0 | -0.0042 | 0.1781 | 87.1 | 0.0727 |
| 10/15 trials | 10 | 60 | Hybrid—iKM | 1200 | 100.0 | 9.8 | 0.0068 | 0.1063 | 92.2 | 0.0659 |
| 10/15 trials | 10 | 60 | Hybrid—IPDfromKM | 1200 | 100.0 | 10.0 | 0.0062 | 0.1061 | 92.8 | 0.0650 |
| 10/15 trials | 10 | 60 | Hybrid—KMtoIPD | 1200 | 100.0 | 10.0 | 0.0073 | 0.1057 | 93.0 | 0.0628 |
| 10/15 trials | 10 | 60 | Hybrid—QP | 1200 | 100.0 | 9.5 | -0.0083 | 0.1175 | 91.9 | 0.0757 |
| 10/15 trials | 10 | 60 | Reconstruction only—QP | 1200 | 100.0 | 9.3 | -0.0130 | 0.1234 | 92.4 | 0.0818 |
| 10/15 trials | 10 | 60 | Reconstruction only—IPDfromKM | 1200 | 100.0 | 10.0 | 0.0112 | 0.1059 | 92.5 | 0.0645 |
| 10/15 trials | 10 | 60 | Reconstruction only—KMtoIPD | 1200 | 100.0 | 10.0 | 0.0131 | 0.1050 | 92.9 | 0.0606 |
| 10/15 trials | 10 | 60 | Reconstruction only—iKM | 1200 | 100.0 | 9.8 | 0.0115 | 0.1061 | 92.2 | 0.0657 |
| 10/15 trials | 10 | 60 | Pooled true-IPD comparator | 1200 | 100.0 | 10.0 | -0.0009 | 0.1070 | 81.6 | |
| 10/15 trials | 10 | 80 | Full AD oracle | 1200 | 100.0 | 10.0 | -0.0016 | 0.1074 | 92.0 | 0.0660 |
| 10/15 trials | 10 | 80 | AD only | 1200 | 62.8 | 2.0 | 0.0016 | 0.2224 | 55.0 | 0.0816 |
| 10/15 trials | 10 | 80 | Hybrid—iKM | 1200 | 100.0 | 9.8 | 0.0088 | 0.1067 | 92.5 | 0.0658 |
| 10/15 trials | 10 | 80 | Hybrid—IPDfromKM | 1200 | 100.0 | 10.0 | 0.0088 | 0.1063 | 92.6 | 0.0646 |
| 10/15 trials | 10 | 80 | Hybrid—KMtoIPD | 1200 | 100.0 | 10.0 | 0.0103 | 0.1056 | 92.8 | 0.0617 |
| 10/15 trials | 10 | 80 | Hybrid—QP | 1200 | 100.0 | 9.4 | -0.0109 | 0.1202 | 92.6 | 0.0786 |
| 10/15 trials | 10 | 80 | Reconstruction only—QP | 1200 | 100.0 | 9.3 | -0.0130 | 0.1234 | 92.4 | 0.0818 |
| 10/15 trials | 10 | 80 | Reconstruction only—IPDfromKM | 1200 | 100.0 | 10.0 | 0.0112 | 0.1059 | 92.5 | 0.0645 |
| 10/15 trials | 10 | 80 | Reconstruction only—KMtoIPD | 1200 | 100.0 | 10.0 | 0.0131 | 0.1050 | 92.9 | 0.0606 |
| 10/15 trials | 10 | 80 | Reconstruction only—iKM | 1200 | 100.0 | 9.8 | 0.0115 | 0.1061 | 92.2 | 0.0657 |
| 10/15 trials | 10 | 80 | Pooled true-IPD comparator | 1200 | 100.0 | 10.0 | -0.0009 | 0.1070 | 81.6 | |
| 10/15 trials | 15 | 60 | Full AD oracle | 1200 | 100.0 | 15.0 | -0.0071 | 0.0913 | 92.6 | 0.0675 |
| 10/15 trials | 15 | 60 | AD only | 1200 | 99.5 | 6.0 | -0.0079 | 0.1508 | 90.4 | 0.0745 |
| 10/15 trials | 15 | 60 | Hybrid—iKM | 1200 | 100.0 | 14.8 | 0.0018 | 0.0899 | 92.8 | 0.0676 |
| 10/15 trials | 15 | 60 | Hybrid—IPDfromKM | 1200 | 100.0 | 15.0 | 0.0012 | 0.0891 | 92.3 | 0.0669 |
| 10/15 trials | 15 | 60 | Hybrid—KMtoIPD | 1200 | 100.0 | 15.0 | 0.0019 | 0.0889 | 92.4 | 0.0642 |

| Trial set | K | Missing (%) | Strategy | Reps | Success (%) | Mean trials retained | Bias, logHR | RMSE, logHR | Coverage (%) | Mean τ² |
|---|---|---|---|---|---|---|---|---|---|---|
| 10/15 trials | 15 | 60 | Hybrid—QP | 1200 | 100.0 | 14.3 | -0.0141 | 0.0987 | 92.7 | 0.0750 |
| 10/15 trials | 15 | 60 | Reconstruction only—QP | 1200 | 100.0 | 13.8 | -0.0197 | 0.1046 | 91.8 | 0.0805 |
| 10/15 trials | 15 | 60 | Reconstruction only—IPDfromKM | 1200 | 100.0 | 15.0 | 0.0061 | 0.0882 | 92.2 | 0.0665 |
| 10/15 trials | 15 | 60 | Reconstruction only—KMtoIPD | 1200 | 100.0 | 15.0 | 0.0079 | 0.0879 | 92.1 | 0.0618 |
| 10/15 trials | 15 | 60 | Reconstruction only—iKM | 1200 | 100.0 | 14.7 | 0.0070 | 0.0894 | 92.9 | 0.0676 |
| 10/15 trials | 15 | 60 | Pooled true-IPD comparator | 1200 | 100.0 | 15.0 | -0.0066 | 0.0909 | 78.5 | |
| 10/15 trials | 15 | 80 | Full AD oracle | 1200 | 100.0 | 15.0 | -0.0071 | 0.0913 | 92.6 | 0.0675 |
| 10/15 trials | 15 | 80 | AD only | 1200 | 84.2 | 3.1 | -0.0075 | 0.2054 | 75.7 | 0.0820 |
| 10/15 trials | 15 | 80 | Hybrid—iKM | 1200 | 100.0 | 14.7 | 0.0043 | 0.0898 | 92.7 | 0.0676 |
| 10/15 trials | 15 | 80 | Hybrid—IPDfromKM | 1200 | 100.0 | 15.0 | 0.0036 | 0.0887 | 92.0 | 0.0667 |
| 10/15 trials | 15 | 80 | Hybrid—KMtoIPD | 1200 | 100.0 | 15.0 | 0.0049 | 0.0883 | 92.2 | 0.0631 |
| 10/15 trials | 15 | 80 | Hybrid—QP | 1200 | 100.0 | 14.1 | -0.0171 | 0.1019 | 92.4 | 0.0778 |
| 10/15 trials | 15 | 80 | Reconstruction only—QP | 1200 | 100.0 | 13.8 | -0.0197 | 0.1046 | 91.8 | 0.0805 |
| 10/15 trials | 15 | 80 | Reconstruction only—IPDfromKM | 1200 | 100.0 | 15.0 | 0.0061 | 0.0882 | 92.2 | 0.0665 |
| 10/15 trials | 15 | 80 | Reconstruction only—KMtoIPD | 1200 | 100.0 | 15.0 | 0.0079 | 0.0879 | 92.1 | 0.0618 |
| 10/15 trials | 15 | 80 | Reconstruction only—iKM | 1200 | 100.0 | 14.7 | 0.0070 | 0.0894 | 92.9 | 0.0676 |
| 10/15 trials | 15 | 80 | Pooled true-IPD comparator | 1200 | 100.0 | 15.0 | -0.0066 | 0.0909 | 78.5 | |

**Note:** Thirty-trial rows pool the three censoring mechanisms and three heterogeneity levels at each missingness level. Rows for the 10- or 15-trial, 60%-censoring benchmark pool four censoring mechanisms for each K and missingness combination. Coverage is unconditional and failed meta-analysis fits count as noncoverage. Mean heterogeneity is not applicable to the pooled true-IPD comparator. K denotes the number of trials; Missing (%) is the percentage of unavailable reported estimates; Reps denotes simulation replicates.

**Table S5.** Auxiliary-input sensitivity performance

| Design | Method | Auxiliary inputs and controls | Success, % | KM RMSE, median (P90) | Relative event deviation, %, median (P90) | Absolute Cox logHR error, median (P90) | Absolute RMST error, median (P90) |
|---|---|---|---|---|---|---|---|
| 30% censoring: uniform | iKM | Risk table only | 99.0 | 0.0094 (0.0116) | 2.3 (3.4) | 0.0072 (0.0185) | 0.109 (0.258) |
| 30% censoring: uniform | IPDfromKM | No total-event input | 100.0 | 0.0104 (0.0140) | 1.7 (2.9) | 0.0082 (0.0199) | 0.156 (0.355) |
| 30% censoring: uniform | IPDfromKM | True total events | 100.0 | 0.0104 (0.0141) | 1.7 (2.9) | 0.0082 (0.0198) | 0.159 (0.370) |
| 30% censoring: uniform | KMtoIPD | Exact censor marks | 100.0 | 0.0096 (0.0136) | 0.3 (0.6) | 0.0069 (0.0184) | 0.124 (0.361) |
| 30% censoring: uniform | KMtoIPD | No censor marks | 100.0 | 0.0087 (0.0108) | 36.3 (40.0) | 0.0367 (0.0848) | 0.086 (0.203) |
| 30% censoring: uniform | QP | Risk table + exact censor marks; every mark enforced | 94.0 | 0.0111 (0.0153) | 1.4 (2.3) | 0.0104 (0.0267) | 0.184 (0.424) |
| 30% censoring: uniform | QP | Risk table + true total events | 96.5 | 0.0093 (0.0121) | 0.0 (0.0) | 0.0058 (0.0165) | 0.093 (0.285) |
| 30% censoring: uniform | QP | Risk table only | 99.0 | 0.0095 (0.0121) | 2.3 (3.1) | 0.0064 (0.0174) | 0.104 (0.260) |
| 30% censoring: uniform | QP | Risk table + exact censor marks as candidate locations | 92.5 | 0.0094 (0.0131) | 2.3 (3.1) | 0.0064 (0.0167) | 0.105 (0.253) |
| 30% censoring: uniform | QP | Risk table + exact censor marks + true total events | 94.0 | 0.0091 (0.0134) | 0.0 (0.0) | 0.0059 (0.0153) | 0.115 (0.342) |
| 30% censoring: uniform | QP | Initial sample size + exact censor marks + true total events | 100.0 | 0.0092 (0.0125) | 0.0 (0.0) | 0.0070 (0.0172) | 0.108 (0.274) |
| 30% censoring: uniform | QP | Initial sample size + exact censor marks | 100.0 | 0.0088 (0.0107) | 25.7 (32.3) | 0.0210 (0.0574) | 0.087 (0.223) |
| 30% censoring: informative | iKM | Risk table only | 96.0 | 0.0085 (0.0099) | 5.5 (10.1) | 0.0109 (0.0332) | 0.150 (0.517) |
| 30% censoring: informative | IPDfromKM | No total-event input | 100.0 | 0.0084 (0.0100) | 5.8 (25.0) | 0.0116 (0.0401) | 0.147 (0.510) |
| 30% censoring: informative | IPDfromKM | True total events | 100.0 | 0.0088 (0.0164) | 5.8 (14.6) | 0.0116 (0.0291) | 0.185 (1.140) |
| 30% censoring: informative | KMtoIPD | Exact censor marks | 100.0 | 0.0084 (0.0100) | 0.0 (0.6) | 0.0115 (0.0320) | 0.136 (0.448) |

| Design | Method | Auxiliary inputs and controls | Success, % | KM RMSE, median (P90) | Relative event deviation, %, median (P90) | Absolute Cox logHR error, median (P90) | Absolute RMST error, median (P90) |
|---|---|---|---|---|---|---|---|
| 30% censoring: informative | KMtoIPD | No censor marks | 100.0 | 0.0081 (0.0096) | 42.7 (48.0) | 0.0247 (0.0565) | 0.135 (0.407) |
| 30% censoring: informative | QP | Risk table + exact censor marks; every mark enforced | 90.0 | 0.0087 (0.0109) | 1.2 (2.0) | 0.0162 (0.0381) | 0.283 (0.649) |
| 30% censoring: informative | QP | Risk table + true total events | 89.0 | 0.0087 (0.0109) | 0.0 (0.0) | 0.0142 (0.0324) | 0.187 (0.838) |
| 30% censoring: informative | QP | Risk table only | 91.5 | 0.0080 (0.0095) | 4.9 (8.6) | 0.0102 (0.0290) | 0.138 (0.391) |
| 30% censoring: informative | QP | Risk table + exact censor marks as candidate locations | 90.5 | 0.0084 (0.0099) | 2.7 (4.5) | 0.0133 (0.0409) | 0.168 (0.685) |
| 30% censoring: informative | QP | Risk table + exact censor marks + true total events | 88.5 | 0.0083 (0.0102) | 0.0 (0.0) | 0.0117 (0.0426) | 0.177 (0.672) |
| 30% censoring: informative | QP | Initial sample size + exact censor marks + true total events | 100.0 | 0.0084 (0.0102) | 0.0 (0.0) | 0.0110 (0.0319) | 0.144 (0.476) |
| 30% censoring: informative | QP | Initial sample size + exact censor marks | 100.0 | 0.0082 (0.0099) | 40.2 (46.7) | 0.0215 (0.0548) | 0.136 (0.411) |
| 60% censoring: uniform | iKM | Risk table only | 100.0 | 0.0404 (0.0495) | 12.5 (15.8) | 0.0575 (0.1263) | 0.228 (0.657) |
| 60% censoring: uniform | IPDfromKM | No total-event input | 100.0 | 0.0424 (0.0562) | 10.0 (13.4) | 0.0574 (0.1341) | 0.259 (0.807) |
| 60% censoring: uniform | IPDfromKM | True total events | 100.0 | 0.0424 (0.0562) | 10.0 (13.4) | 0.0574 (0.1341) | 0.259 (0.807) |
| 60% censoring: uniform | KMtoIPD | Exact censor marks | 100.0 | 0.0425 (0.0584) | 4.2 (6.7) | 0.0599 (0.1266) | 0.243 (0.865) |
| 60% censoring: uniform | KMtoIPD | No censor marks | 100.0 | 0.0398 (0.0478) | 75.8 (98.3) | 0.1108 (0.2742) | 0.197 (0.569) |
| 60% censoring: uniform | QP | Risk table + exact censor marks; every mark enforced | 88.5 | 0.0466 (0.0660) | 8.3 (13.3) | 0.0487 (0.1160) | 0.237 (0.728) |
| 60% censoring: uniform | QP | Risk table + true total events | 91.0 | 0.0381 (0.0553) | 0.0 (0.0) | 0.0403 (0.0865) | 0.203 (0.733) |
| 60% censoring: uniform | QP | Risk table only | 93.5 | 0.0403 (0.0524) | 12.5 (15.8) | 0.0566 (0.1269) | 0.190 (0.633) |
| 60% censoring: uniform | QP | Risk table + exact censor marks as candidate locations | 87.0 | 0.0400 (0.0527) | 12.1 (15.8) | 0.0564 (0.1270) | 0.248 (0.598) |

| Design | Method | Auxiliary inputs and controls | Success, % | KM RMSE, median (P90) | Relative event deviation, %, median (P90) | Absolute Cox logHR error, median (P90) | Absolute RMST error, median (P90) |
|---|---|---|---|---|---|---|---|
| 60% censoring: uniform | QP | Risk table + exact censor marks + true total events | 86.5 | 0.0369 (0.0537) | 0.0 (0.0) | 0.0455 (0.0845) | 0.200 (0.707) |
| 60% censoring: uniform | QP | Initial sample size + exact censor marks + true total events | 100.0 | 0.0417 (0.0559) | 0.0 (0.0) | 0.0623 (0.1312) | 0.253 (0.677) |
| 60% censoring: uniform | QP | Initial sample size + exact censor marks | 100.0 | 0.0393 (0.0495) | 24.2 (35.8) | 0.0541 (0.1235) | 0.198 (0.611) |
| 60% censoring: informative | iKM | Risk table only | 93.5 | 0.0369 (0.0466) | 16.7 (24.9) | 0.0698 (0.1380) | 0.995 (2.672) |
| 60% censoring: informative | IPDfromKM | No total-event input | 100.0 | 0.0367 (0.0456) | 16.2 (102.7) | 0.0686 (0.1489) | 1.107 (2.655) |
| 60% censoring: informative | IPDfromKM | True total events | 100.0 | 0.0375 (0.0505) | 16.2 (48.8) | 0.0658 (0.1481) | 1.114 (3.073) |
| 60% censoring: informative | KMtoIPD | Exact censor marks | 100.0 | 0.0380 (0.0499) | 0.8 (2.5) | 0.0747 (0.1397) | 1.091 (2.640) |
| 60% censoring: informative | KMtoIPD | No censor marks | 100.0 | 0.0345 (0.0416) | 151.3 (175.2) | 0.0911 (0.2461) | 0.934 (2.233) |
| 60% censoring: informative | QP | Risk table + exact censor marks; every mark enforced | 87.0 | 0.0340 (0.0508) | 4.2 (8.3) | 0.0468 (0.1208) | 0.925 (2.847) |
| 60% censoring: informative | QP | Risk table + true total events | 86.5 | 0.0409 (0.0746) | 0.0 (0.0) | 0.0969 (0.1829) | 1.073 (6.601) |
| 60% censoring: informative | QP | Risk table only | 92.5 | 0.0338 (0.0431) | 15.8 (24.3) | 0.0725 (0.1542) | 0.850 (2.215) |
| 60% censoring: informative | QP | Risk table + exact censor marks as candidate locations | 87.0 | 0.0341 (0.0446) | 11.5 (15.4) | 0.0697 (0.1403) | 0.963 (2.578) |
| 60% censoring: informative | QP | Risk table + exact censor marks + true total events | 88.0 | 0.0334 (0.0469) | 0.0 (0.0) | 0.0695 (0.1418) | 0.893 (2.809) |
| 60% censoring: informative | QP | Initial sample size + exact censor marks + true total events | 100.0 | 0.0377 (0.0504) | 0.0 (0.0) | 0.0784 (0.1478) | 1.099 (2.776) |
| 60% censoring: informative | QP | Initial sample size + exact censor marks | 100.0 | 0.0350 (0.0428) | 129.9 (172.7) | 0.0727 (0.1822) | 0.951 (2.072) |

**Note:** Each design–input state used 200 assigned replicates. QP was evaluated under seven input/control configurations; exact censor marks were removed from KMtoIPD and true total-event counts were removed from IPDfromKM. Accuracy outcomes summarize valid reconstructions.

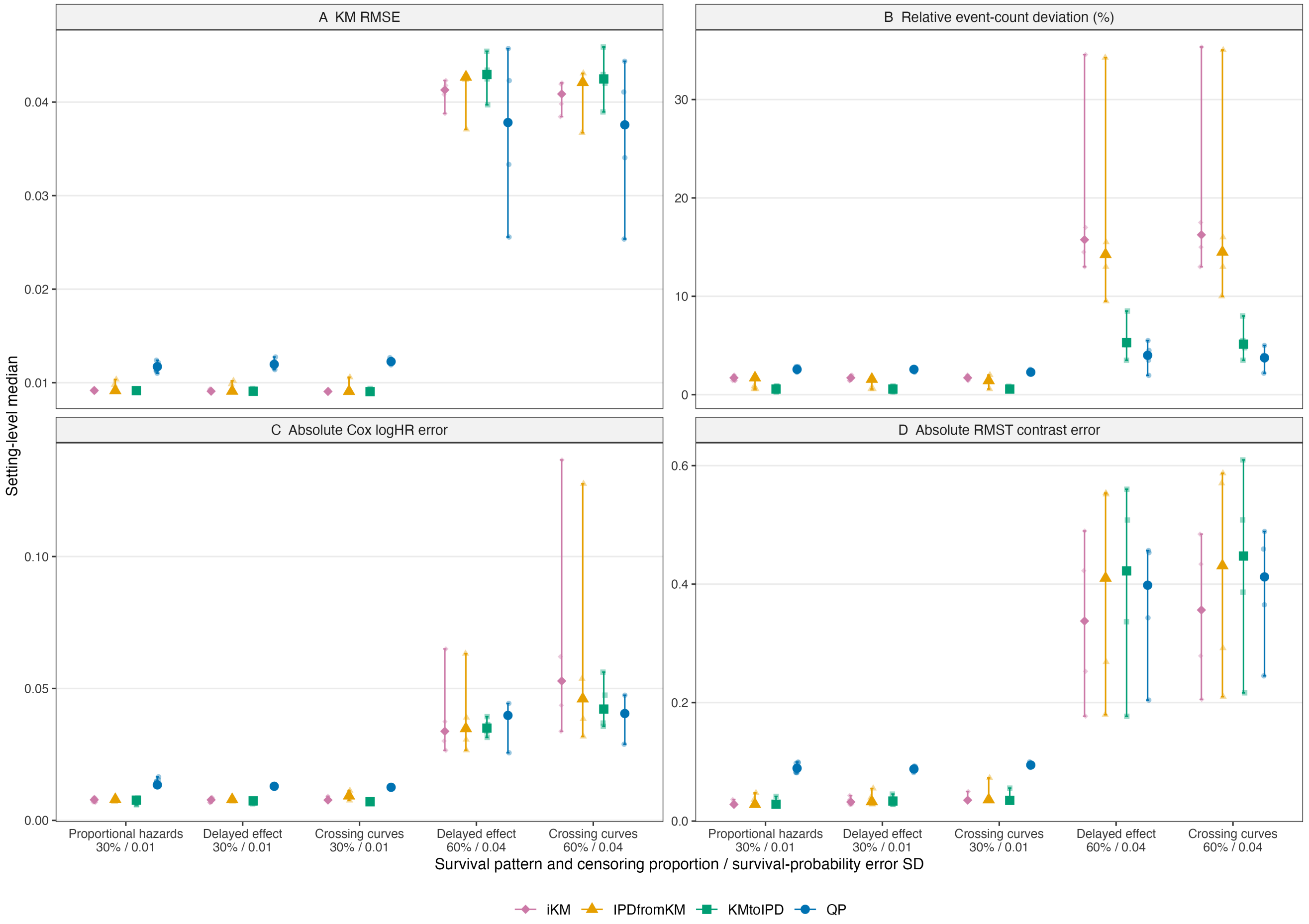

**Figure S1.** Performance across proportional-hazards, delayed-effect, and crossing-survival settings. Small points are medians over 1,000 replicates for individual design settings. Large points are medians across settings within survival-pattern and censoring/error groups; bars show the setting-level minimum and maximum. Panels report KM RMSE, relative event-count deviation, absolute Cox logHR error, and absolute RMST-contrast error; lower values indicate better recovery. Percentages on the horizontal axis denote target censoring proportions.

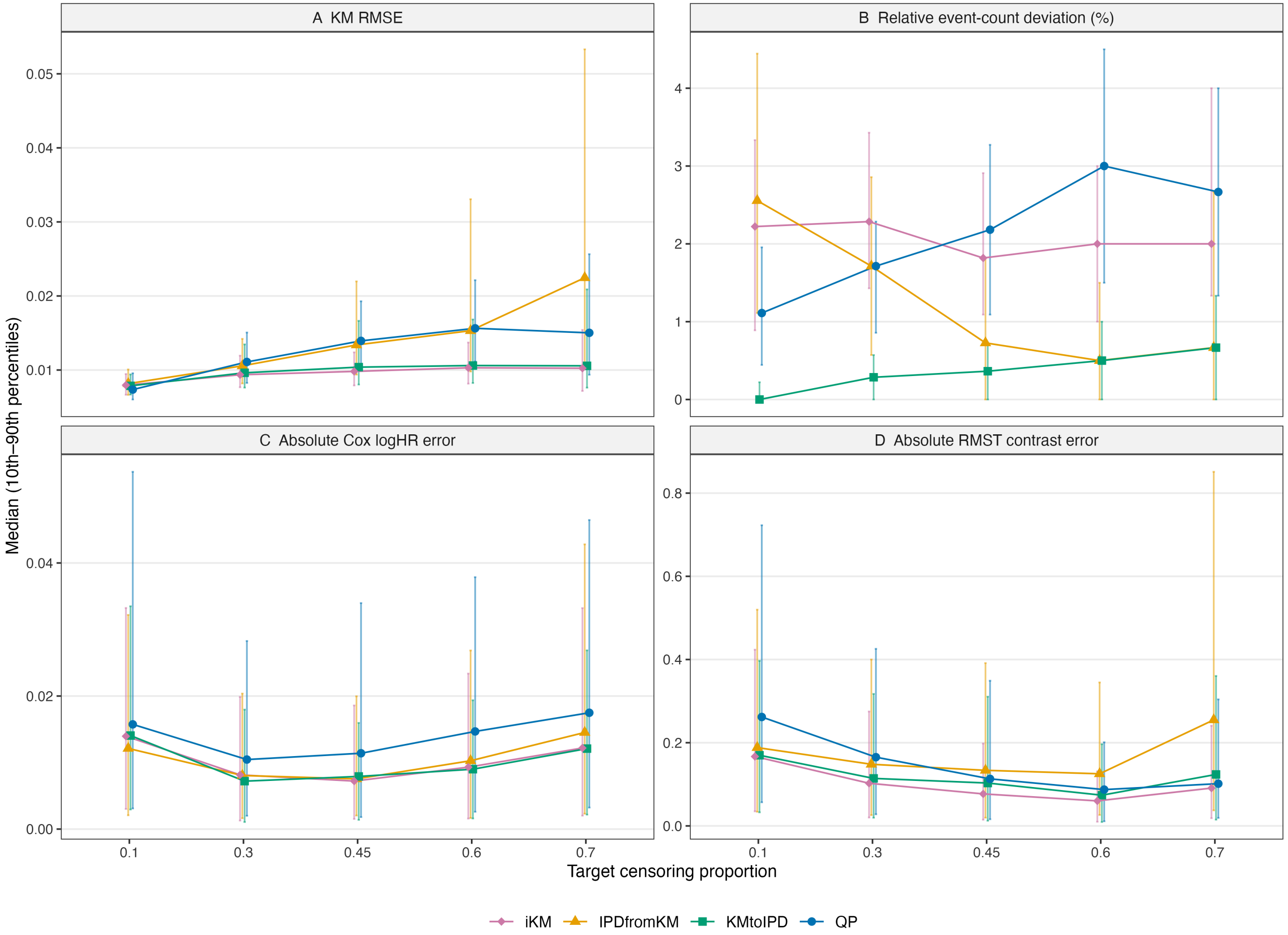

**Figure S2A.** Results from the dedicated one-factor experiment across target censoring proportions 0.10, 0.30, 0.45, 0.60, and 0.70. Points show medians and bars show P10–P90 among valid reconstructions; all other design components were fixed at the stated reference values.

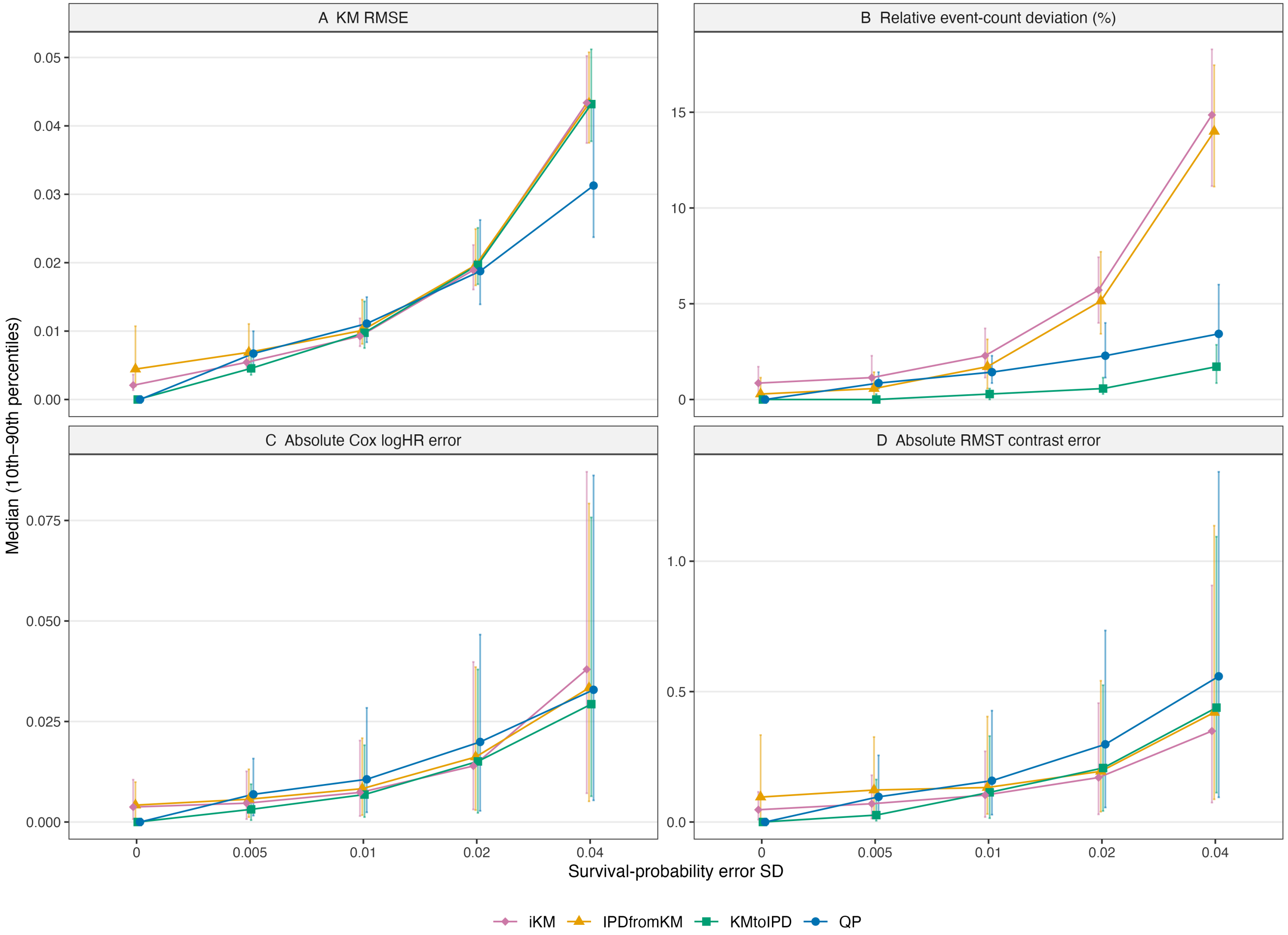

**Figure S2B.** Results from the dedicated one-factor experiment as Gaussian survival-probability error SD increased from 0 to 0.04. Time coordinates remained exact, and perturbed survival probabilities were clipped to [0,1] and monotonized. Points show medians and bars show P10–P90 among valid reconstructions.

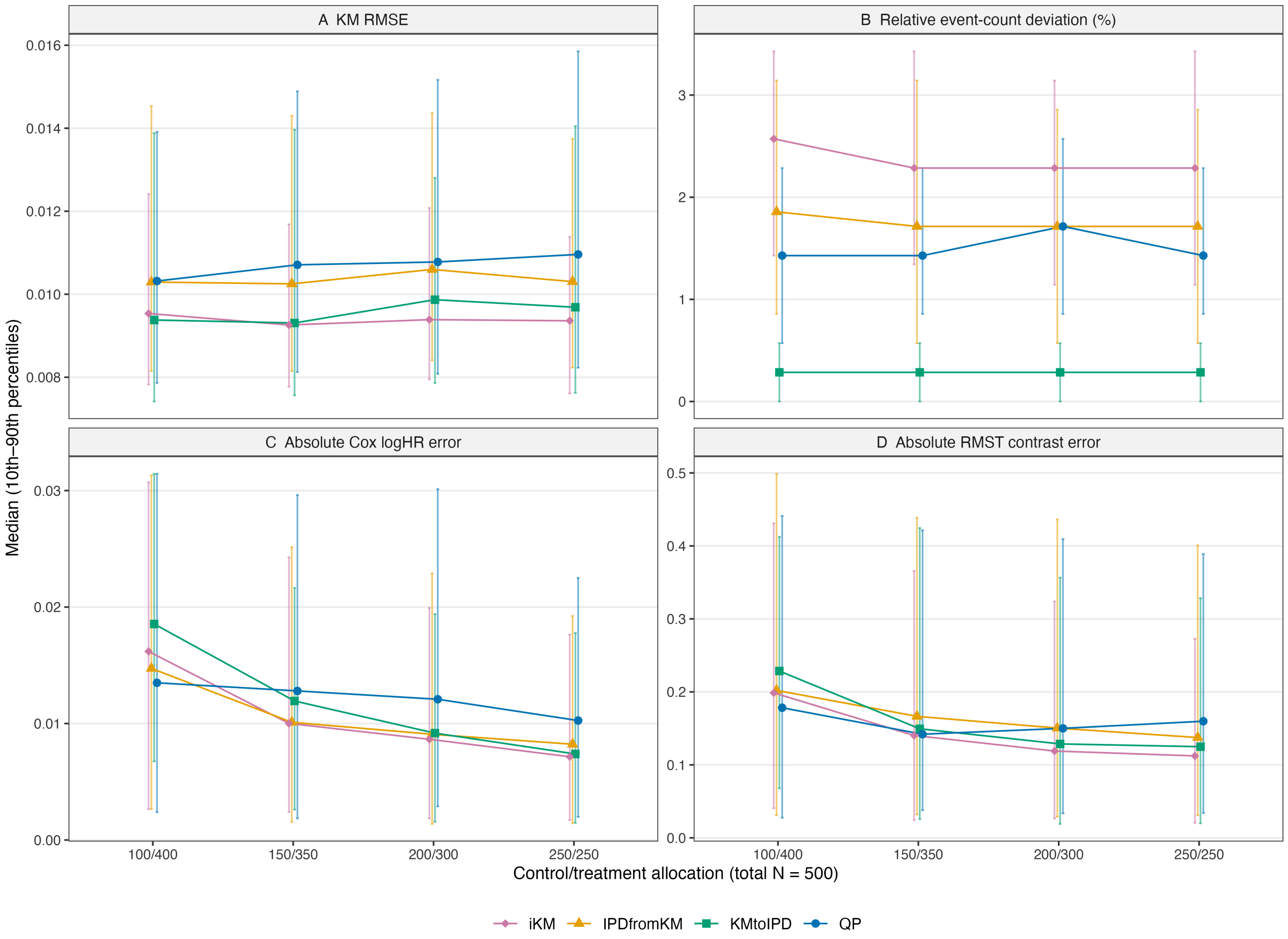

**Figure S2C.** Results from the dedicated one-factor experiment as control/treatment allocation changed from 250/250 to 100/400 while total sample size remained 500. Points show medians and bars show P10–P90 among valid reconstructions.

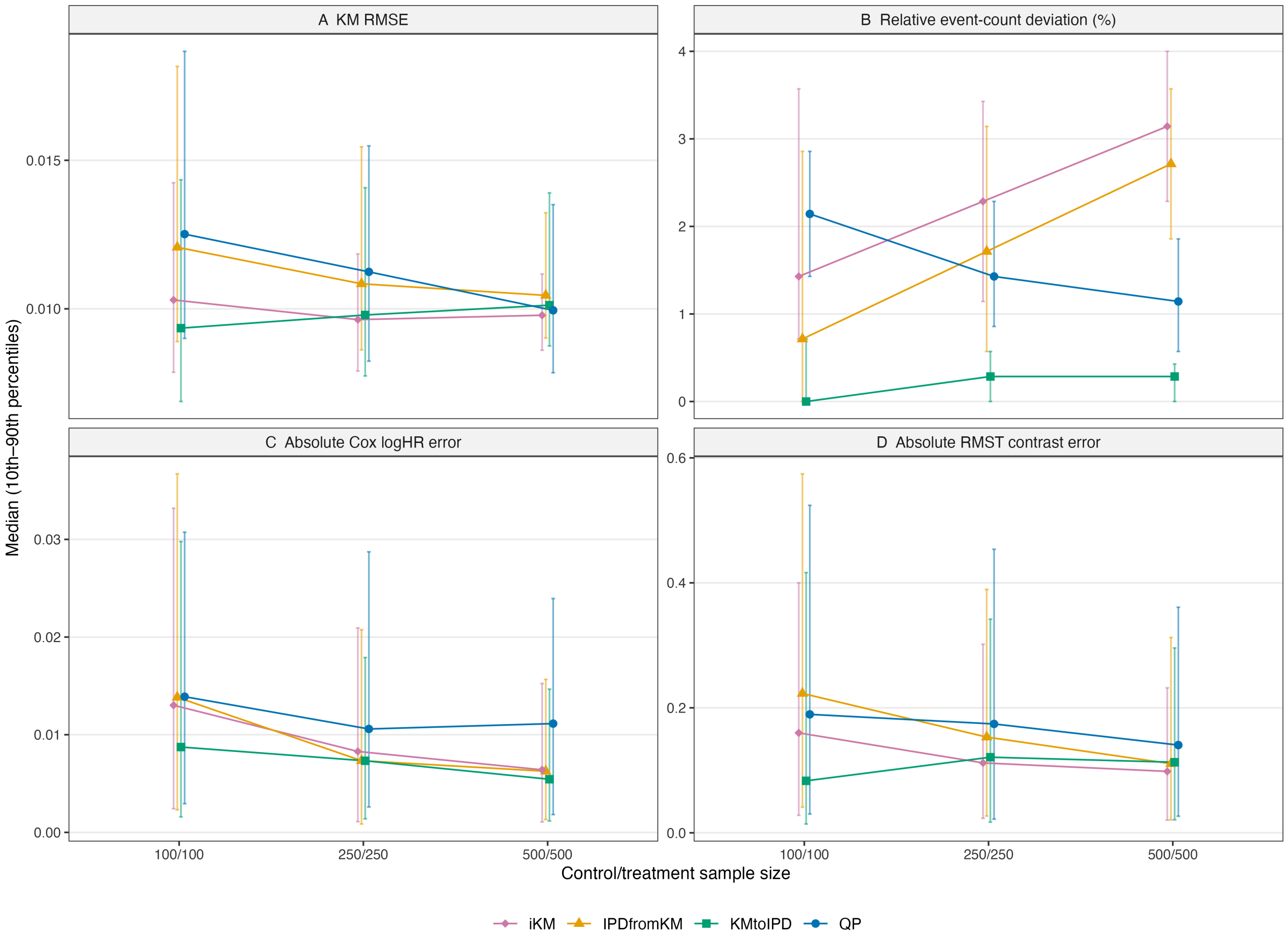

**Figure S2D.** Results from the dedicated one-factor experiment at control/treatment sample sizes of 100/100, 250/250, and 500/500. Points show medians and bars show P10–P90 among valid reconstructions.

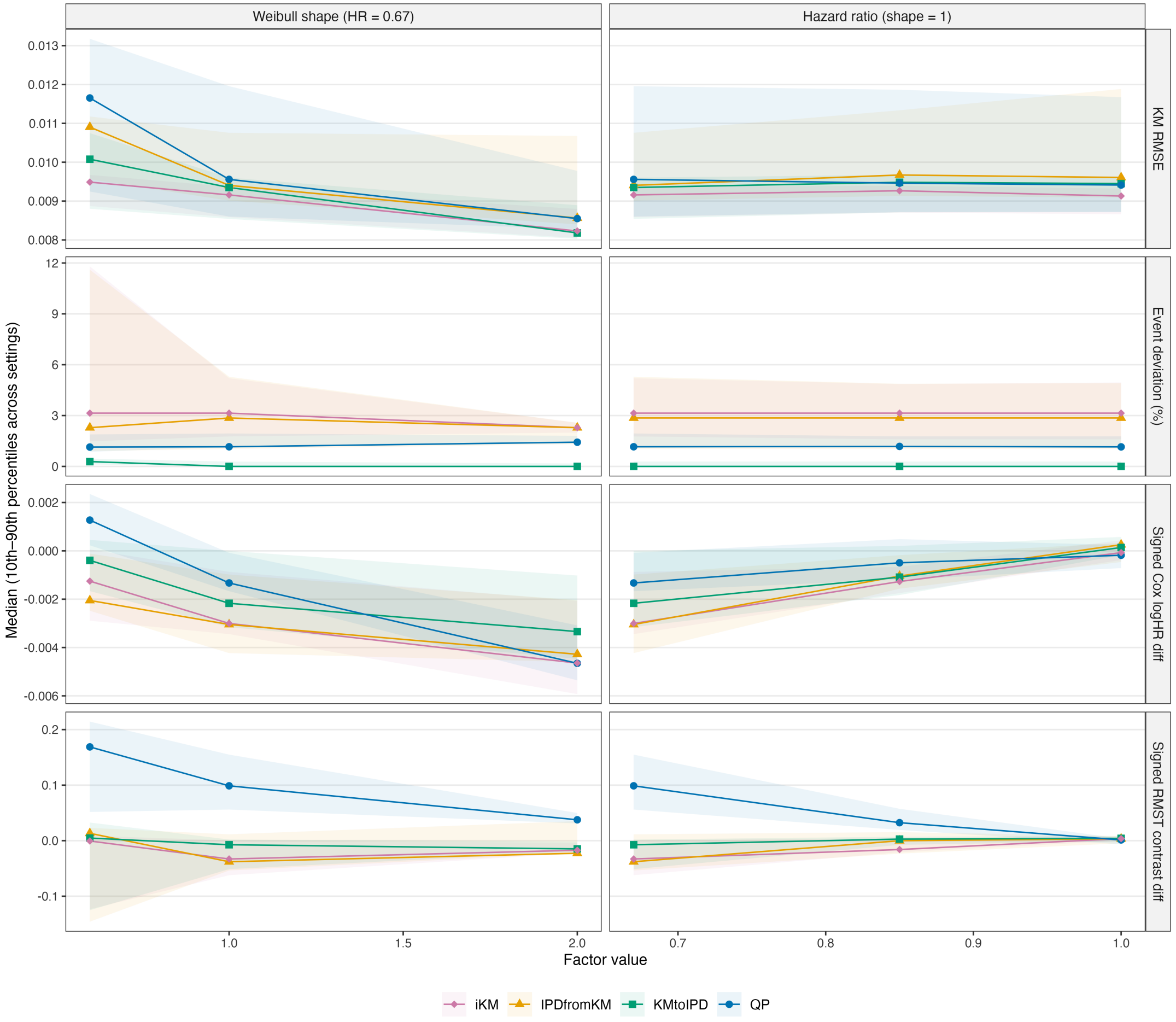

**Figure S3.** Method-specific summaries from matched 30%-censoring Weibull settings. Shape 0.6, 1, and 2 was compared at HR 0.67 within censoring mechanism; HR 0.67, 0.85, and 1.00 was compared at shape 1 within censoring mechanism. Points show medians and ribbons show P10–P90 across the matched setting-level medians. Cox and RMST panels display signed differences to show error direction.

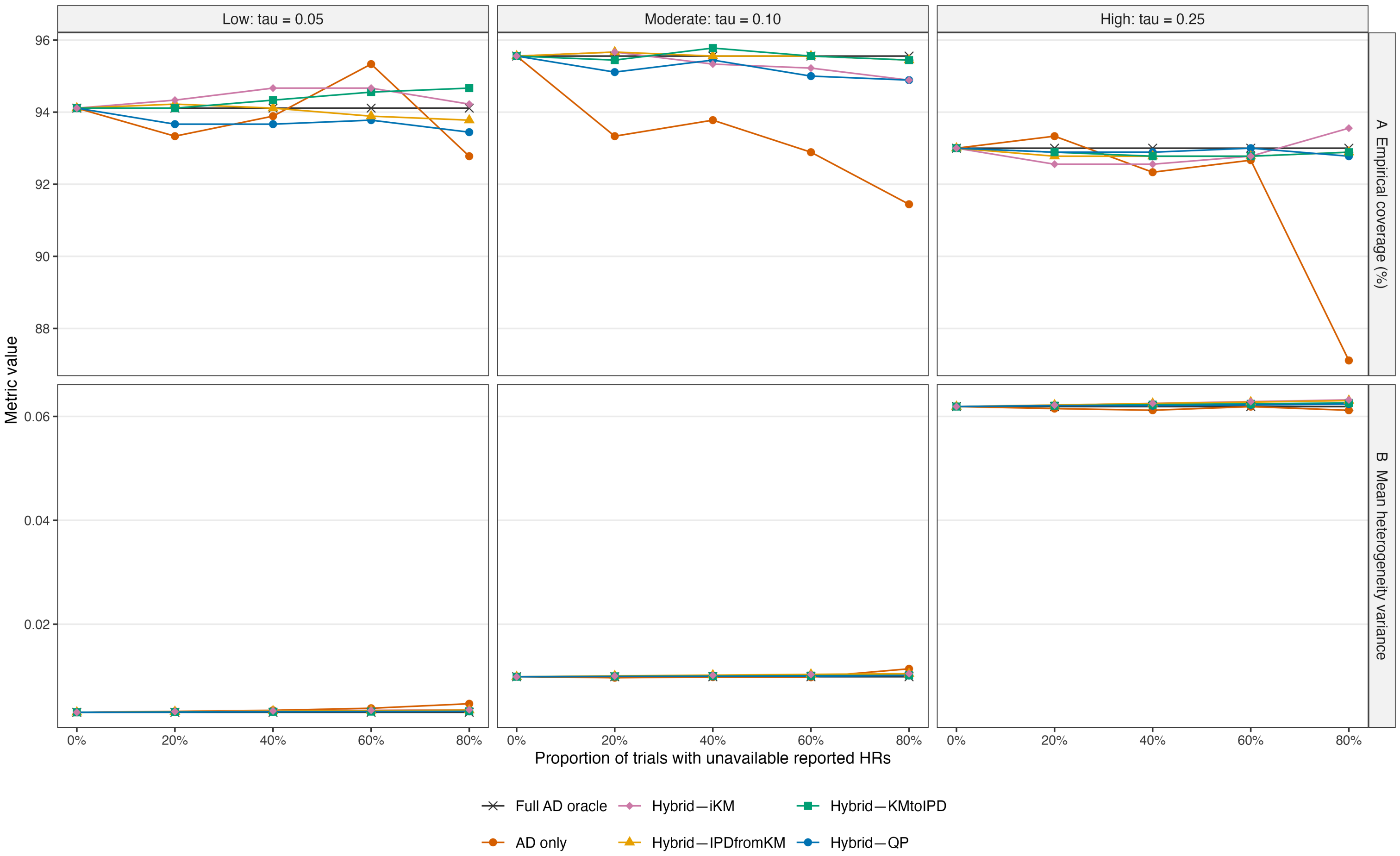

**Figure S4.** Unconditional empirical 95% coverage and mean estimated heterogeneity variance across proportions of missing reported Cox estimates and heterogeneity SDs in the 30-trial Part II benchmark. Each underlying scenario used 300 replicates, and failed meta-analysis fits counted as noncoverage.

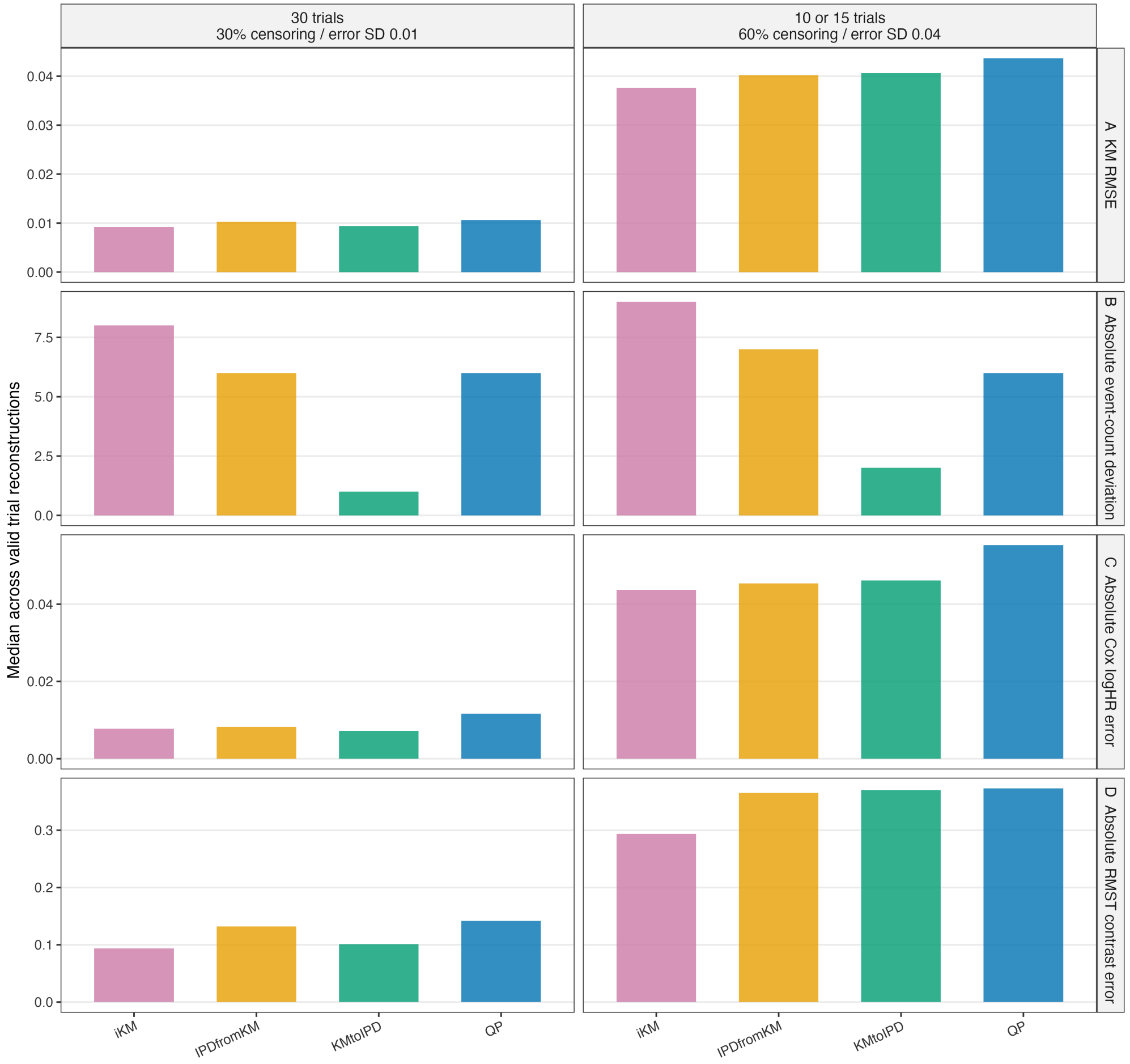

**Figure S5.** Trial-level KM RMSE, absolute event-count deviation, absolute Cox logHR error, and absolute RMST-contrast error for iKM, IPDfromKM, KMtoIPD, and QP in the 30-trial benchmark and the 10- or 15-trial benchmark with 60% censoring. Bars summarize valid reconstructed trials; Supplementary Results S2.1 reports valid-reconstruction proportions.

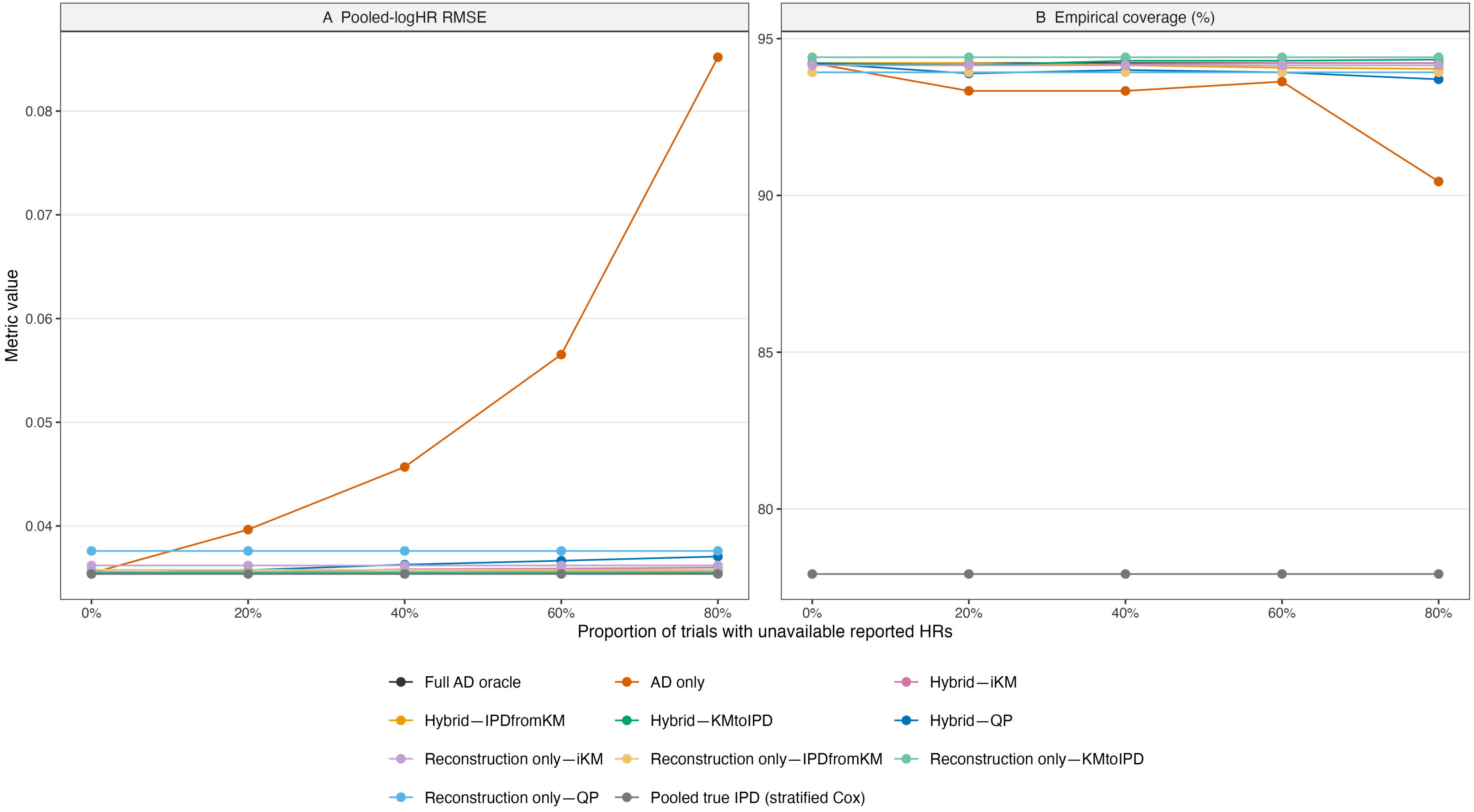

**Figure S6.** Pooled-logHR RMSE and unconditional empirical coverage across proportions of missing reported Cox estimates in the 30-trial benchmark for the Full AD oracle, AD only, four hybrid strategies, four reconstruction-only strategies, and pooled true-IPD comparator. The pooled comparator is not estimand-matched to the generating random-effects mean. Model success, trial retention, bias, and heterogeneity estimates are reported in Table S4.

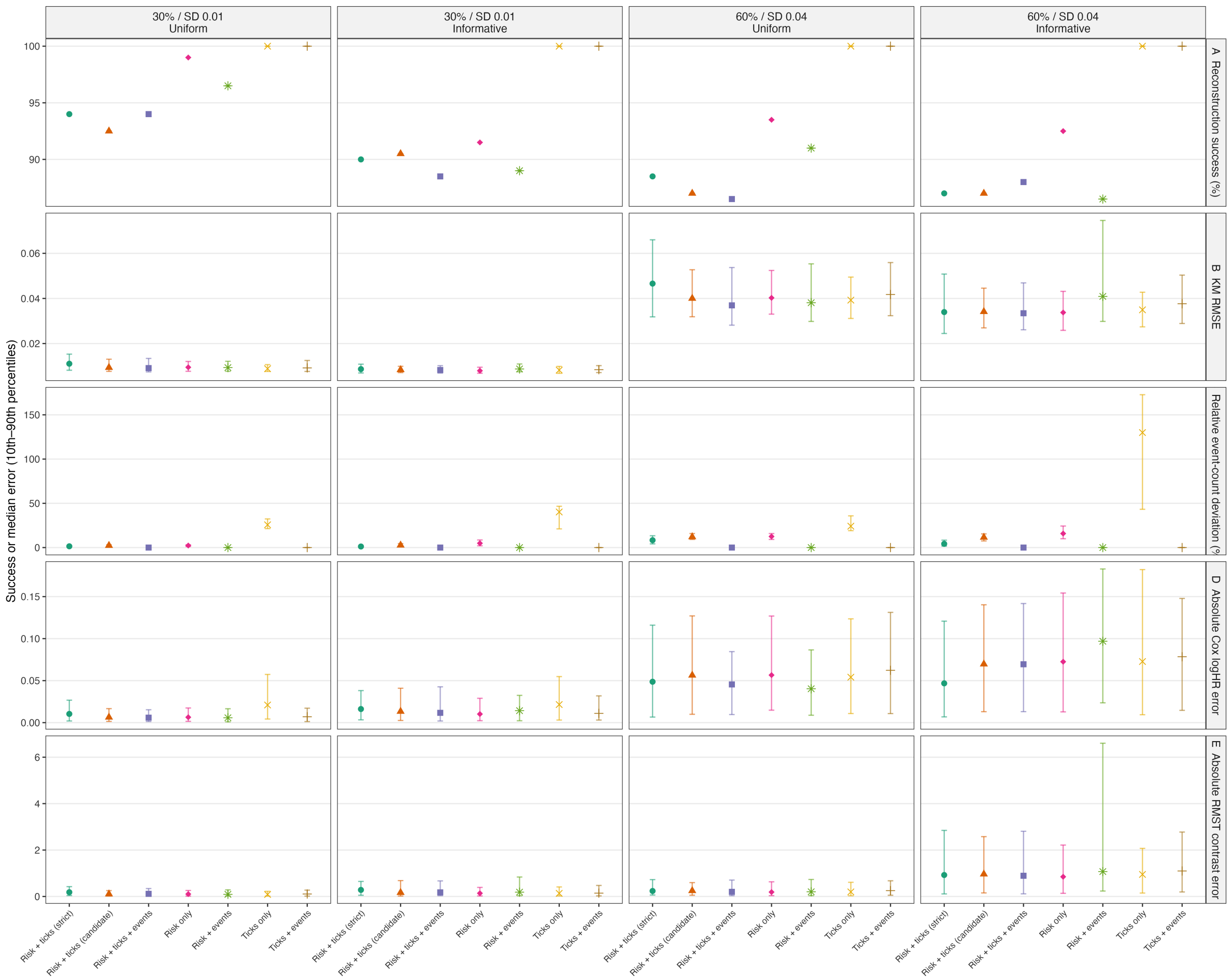

**Figure S7.** Reconstruction success and four accuracy outcomes for seven CIFresolve QP configurations across 30%- versus 60%-censoring designs with uniform or informative censoring. The configurations varied risk-table availability, exact censor marks, true total-event counts, and whether every supplied mark was enforced. Accuracy points are medians over valid reconstructions and bars show P10–P90; success uses every assigned run.

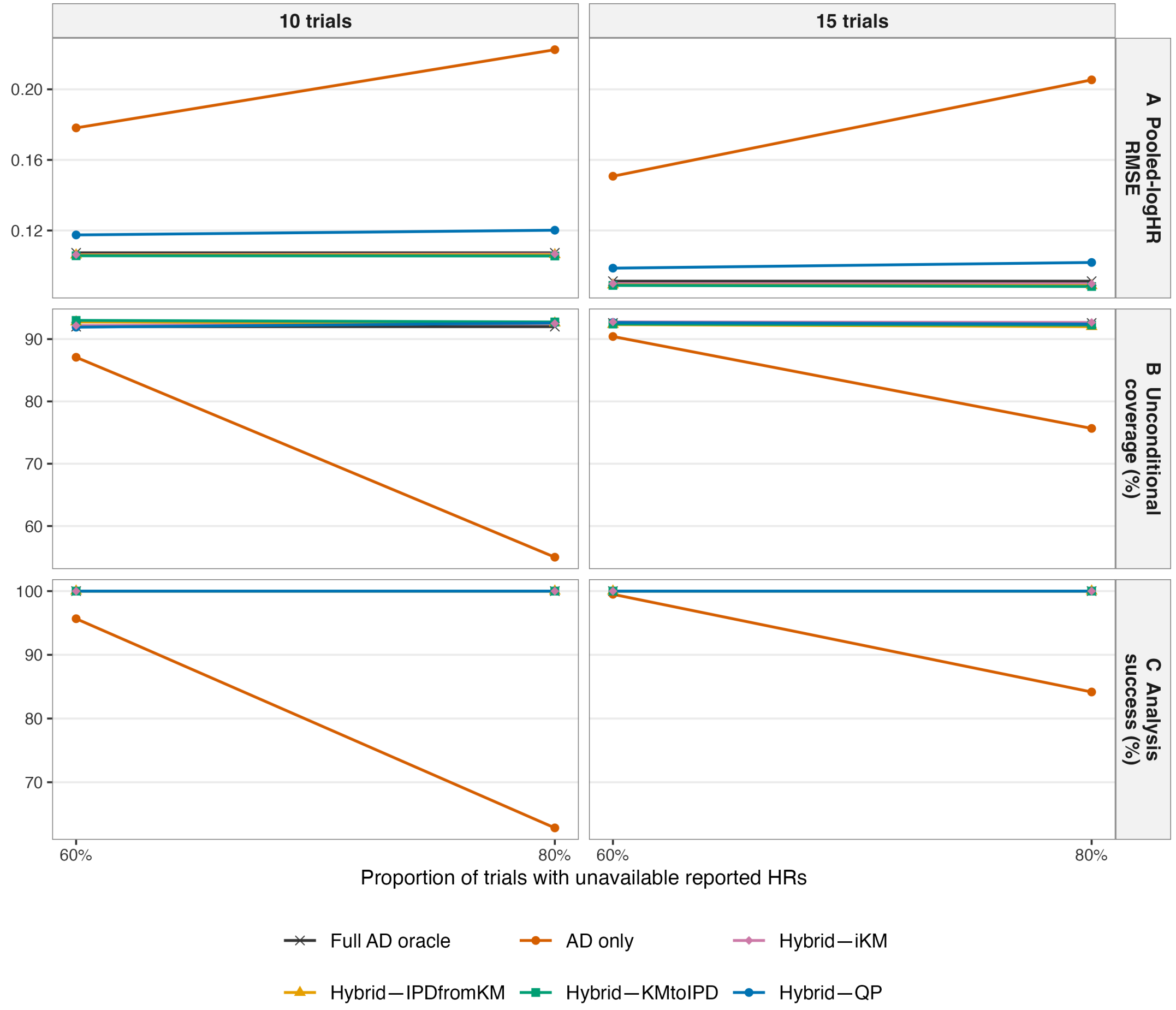

**Figure S8.** Meta-analysis benchmark with 10 or 15 trials. Scenarios used 100 participants per arm, 60% censoring, survival-probability error SD 0.04, and heterogeneity SD 0.25. Panels show pooled-logHR RMSE (A), unconditional empirical coverage (B), and analysis success (C) at 60% and 80% unavailable reported HRs. Failed fits counted as noncoverage; each scenario used 300 replicates.